\documentclass[aps,prb,twocolumn,floatfix,groupedaddress,showpacs,showkeys]{revtex4}
\pdfoutput=1

\usepackage{amsmath}
\usepackage{amsfonts}
\usepackage{bm}
\usepackage{graphicx}
\usepackage{color}
\usepackage{subfigure}

\newcommand {\Sm}{\mbox{${\mathbf{\cal{S}}}$}}

\begin{document}

\title{Evanescent channels and scattering in cylindrical nanowire 
heterostructures}

\author{P. N. Racec}
\email{racec@wias-berlin.de}
\affiliation{Weierstra\ss -Institut f\"ur Angewandte Analysis und Stochastik,
Mohrenstr. 39 10117 Berlin, Germany }
\affiliation{National Institute of Materials Physics, PO Box MG-7,
             077125 Bucharest Magurele, Romania}

\author{E.\ R.\ Racec}
\email{roxana@physik.tu-cottbus.de}
\affiliation{Institut f\"ur Physik, Technische Universit\"at Cottbus, Postfach 101344, 03013 Cottbus, Germany }
\affiliation{Faculty of Physics, University of Bucharest, PO Box MG-11,
             077125 Bucharest Magurele, Romania}

\author{H. Neidhardt}
\email{neidhard@wias-berlin.de}
\affiliation{Weierstra\ss -Institut f\"ur Angewandte Analysis und Stochastik,
Mohrenstr. 39 10117 Berlin, Germany }

\begin{abstract}
We investigate the scattering phenomena produced by a
general finite-range nonseparable potential in a
multi-channel two-probe cylindrical nanowire heterostructure.
The multi-channel current scattering matrix is efficiently 
computed using the 
R-matrix formalism extended for cylindrical coordinates. 
Considering the contribution of the evanescent channels 
to the scattering matrix, 
we are able to put in evidence the 
specific dips in the tunneling coefficient in the case of an
attractive potential.
The cylindrical symmetry cancels the "selection rules" 
known for Cartesian coordinates.
If the attractive potential
is superposed over a non-uniform potential along the nanowire,
then resonant transmission peaks appear.
We can characterize them quantitatively through 
the poles of the current scattering matrix.
Detailed maps of the localization probability density sustain
the physical interpretation of the resonances (dips and peaks). 
Our formalism is applied to a variety of model systems 
like a quantum dot, a core/shell quantum ring or a double-barrier,
embedded into the nanocylinder.
\end{abstract}



\maketitle
\section{Introduction}

In the last years, 
there is an increased interest in studying nanowire-based devices
due to their broad application area. They can be used as 
field-effect transistors (FET) \cite{Lieber_06}
or gate-all-around (GAA) FET 
\cite{Samuelson_06,suk_iedm06,cho_08}, 
nanowire resonant tunneling diodes \cite{samuelson02,wensorra},
solar cells as integrated power sources for nanoelectronic systems
\cite{Lieber_07},
lasers \cite{Lieber_08},
and also as qubits \cite{Lieber_qubit}. 
Their structural complexity has progresively increased,
and the material composition includes III-V materials 
but also, 
so attractive for semiconductor industry,
group IV materials. 

Description of electrical transport and charge distribution 
in nanowire-based devices has to be done quantum mechanically,
and the most appropriate method for such open systems 
is the scattering theory. 
Due to the confinement of the motion inside the nanowire,
the electrons are free to move only along the nanowire direction, so that 
these systems are also called quasi-one-dimensional systems.
Since many nanowires have circular cross-sectional shape
\cite{Samuelson_06,suk_iedm06,cho_08},
we present in this work a general method,
valid within the effective mass approximation,
for solving the three-dimensional (3D) Schr\"odinger equation with scattering
boundary conditions in cylindrical geometries.
The azimuthal symmetry suggests to use
cylindrical coordinates, with $z$ axis along the nanowire.
This reduces the scattering problem to two dimensions:
$r$ and $z$ directions.
Its solution is found numerically using the R-matrix formalism,
\cite{smr,wulf98,onac,roxana01,racec02,nemnes04,nemnes05,07impe,hagen08}
extended for cylindrical coordinates.

An interesting effect in a multi-channel scattering problem is that
as soon as
the potential is not anymore separable, the channels get mixed.
If furthermore the scattering potential is attractive,
then it leads to unusual scattering properties, 
like resonant dips in the transmission
coefficient just below the next channel minimum energy. 
As it was shown analytically 
for a $\delta$ scattering potential \cite{bagwell90} 
or for point scatterers \cite{Exner96}, and
later on for a finite-range scattering potential\cite{levinson93,noeckel94} 
the dips are due to quasi-bound-states splitting off from a higher 
evanescent channel. 
So that evanescent channels can not be neglected
when analyzing scattering in two- or three-dimensional quantum systems. 
These findings were recently confirmed numerically for a 
Gaussian-type scatterer \cite{gudmundsson04} and also for 
a quantum dot or a quantum ring\cite{gudmundsson05} embedded inside
nanowires tailored in two-dimensional electron gas (2DEG). 

It is the aim of this work to show that we could find the same
features in the case of a cylindrical nanowire. 
Furthermore, in cylindrical nanowires, 
due to the three-dimensional (3D) modelling, 
every magnetic quantum number defines
a two-dimensional (2D) scattering problem, 
with different structure of dips 
for the same scattering potential. 
Also, the cylindrical symmetry  
does not forbid any 
intersubband transmission, 
so that we could find dips in front of every plateau in the transmission
coefficient.
We apply our method to a variety of model systems
like quantum dot, quantum ring
or double-barrier heterostructure, embedded inside the nanowire.

\section{Model}
\label{model}

We consider a cylindrical nanowire with a constant potential on the
surface. 
Inside 
the wire the electrons are scattered by a potential of finite extend.

\subsection{Scattering problem for the cylindrical geometry}

\begin{figure}[h]
\includegraphics[width=3.25in]{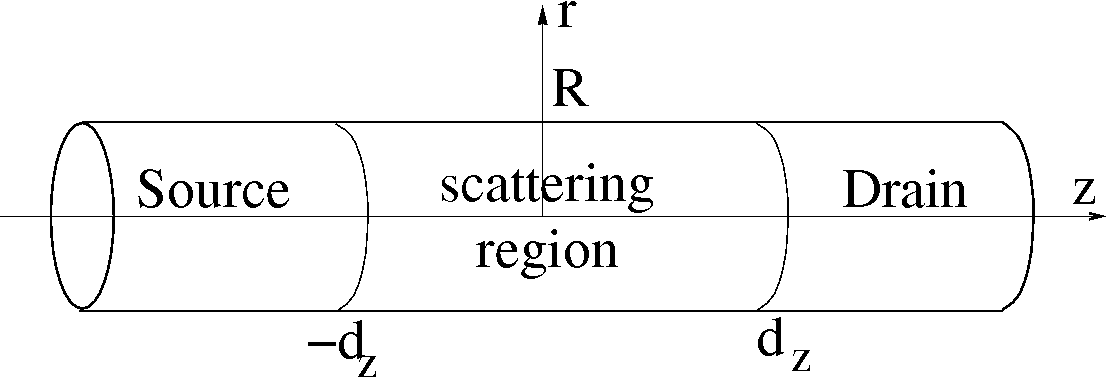}
\caption{The geometry of the 2D scattering problem.
}
\label{Schr_2D_vol}
\end{figure}

In the effective mass approximation, 
the envelope function associated to the energy $E$ 
satisfies a Schr\"odinger-type equation
\begin{equation}
\left[ -\frac{\hbar^2}{2 \mu} \Delta + V(\mathbf{r}) \right] \Psi(\mathbf{r})
   =E \Psi(\mathbf{r}).
\label{Schr_cyl}
\end{equation}
We use the symbol $\mu$ to denote the effective mass of the electrons, 
while $m$ will denote the magnetic quantum number.
As long as there are not split gates
on the surface of the nanowire,
the potential energy $V(\mathbf{r})$ 
is rotational invariant
\begin{equation}
V(\mathbf{r})=V(r,z)
\end{equation}
and nonseparable 
in a small region of the structure called scattering region. 
The $z$-axis was considered along the nanowire as shown
in Fig. \ref{Schr_2D_vol}. 

A scattering potential which does not explicitly depend on the 
azimuthal angle $\theta$ imposes the eigenfunctions of the 
orbital angular momentum operator $L_z$ as solutions of Eq. 
(\ref{Schr_cyl})
\begin{equation}
\Psi_{m}(E;r,\theta,z)=\frac{e^{im\theta}}{\sqrt{2\pi}}\psi_{m}(E;r,z),
\label{Psi_gen}
\end{equation}
where $m =0, \pm 1, \pm 2, ...$ is the magnetic quantum number.
This is an integer number due to 
the requirement that the function $e^{im\theta}$ should be single-valued.
The allowed values of the energy  $E$ and the functions $\psi_{m}(E;r,z)$
are determined from the equation
\begin{equation}
\begin{split}
\left[ -\frac{\hbar^2}{2 \mu}
             \left( \frac{\partial^2}{\partial r^2}
                   +\frac{1}{r}\frac{\partial}{\partial r}
                   -\frac{m^2}{r^2}
                   +\frac{\partial^2}{\partial z^2}
             \right)
        \right. & \left. + V(r,z)
\right] \psi_{m}(E;r,z) \\
& =  E  \psi_{m}(E;r,z).
\end{split}
\label{2D_schr}
\end{equation}
Due to the localized character of the scattering potential
it is appropriate to solve Eq. (\ref{2D_schr}) within 
the scattering theory. 
In such a way, every magnetic quantum number $m$ defines a 
\textit{two-dimensional (2D) scattering problem}. 
Furthermore, these
2D scattering problems can be solved separately
if the scattering potential is rotational invariant. 
How many of these problems have to be solved, depends on the 
specific physical quantity which has to be computed.

The potential energy which appears in Eq. (\ref{2D_schr})
has generally two components:
\begin{equation}
V(r,z) = V_\perp(r) + V_{scatt}(r,z).
\end{equation}
The first one, $V_\perp(r)$, describes the lateral confinement 
of the electrons
inside a cylinder of radius $R$ 
and is translation invariant along the nanowire.
We consider a hard wall potential
\begin{equation}
V_\perp(r) = \begin{cases}
              0, &      0 < r < R \\
              \infty, & r \ge R 
            \end{cases}
\end{equation}
suitable for modelling 
either free-standing nanowires
or nanowire transistors with no gate leakage current.
Both situations correspond to the state-of-the-art devices.

The scattering potential energy inside the nanowire, $V_{scatt}(r,z)$,
has generally a nonseparable character in a domain of finite-range
and is constant outside this domain. 
We consider here the nonseparable potential 
localized within the volume defined by the boundaries $\pm d_z$ and $R$,  
see Fig. \ref{Schr_2D_vol},
\begin{equation}
V_{scatt}(r,z)=\begin{cases}
           V_1, & r \in [0,R], z < -d_z \\
           V_s(r,z), & r \in [0,R], -d_z \le z \le d_z \\
           V_2, & r \in [0,R], z > d_z
         \end{cases}.
\end{equation}
There are not material definitions for the planes $z=\pm d_z$.
Usually, they are chosen inside the highly doped regions
of the nanowire characterized by a slowly $z$-varying potential,
practically by a constant potential. These regions play the role 
of the source and drain contacts.

\subsection{Scattering states}

In the asymptotic regions,
 $|z|> d_z$ i.e. source and drain contacts,
the potential energy
is separable in the confinement and the transport direction
($V(r,z)=V_\perp(r)+V_s$)
and Eq. (\ref{2D_schr}) can be directly solved 
using the separation of variables method
\begin{equation}
\psi_m(E;r,z)=\phi(r)\varphi(z).
\label{sepxiphi}
\end{equation}
The function $\phi(r)$ satisfies the radial equation
\begin{equation}
-\frac{\hbar^2}{2\mu}
        \left[ \frac{d^2}{dr^2}
              + \frac{1}{r} \frac{d}{dr}
              - \frac{m^2}{r^2}
              +V_\perp(r)
        \right]
 \phi(r)=E_\perp \phi(r),
\label{radeq}
\end{equation}\\
while $\varphi(z)$ satisfies the one-dimensional 
Schr\"odinger type equation
\begin{equation}
\left[ -\frac{\hbar^2}{2\mu} \frac{d^2}{dz^2} + V_s \right] \varphi(z)=
(E-E_\perp) \varphi(z),
\label{transpeq}
\end{equation}
where $s=1$ stays for the source contact ($z < -d_z$)
and $s=2$ for the drain contact ($z > d_z$).

Due to the electron confinement inside the cylinder of radius $R$,
the solutions of Eq. (\ref{radeq})
are given in terms of the Bessel functions of the first kind,
$J_m$, 
\begin{equation}
\phi_n^{(m)}(r)=\frac{\sqrt{2}}{R J_{|m|+1}(x_{mn})} J_m(x_{mn}r/R),
\quad n=1,2,...
\label{phi_channel}
\end{equation}
where $x_{mn}$ is the $n$th root of $J_m(x)$.
The eigenfunctions $\phi_n^{(m)}(r)$, called \textit{transversal modes}, 
form an orthonormal and complete system of functions.
The corresponding eigenenergies are
\begin{equation}
E^{(m)}_{\perp n}=\frac{\hbar^2}{2\mu}\left(\frac{x_{mn}}{R}\right)^2,
\quad n=1,2,...
\label{e_channel}
\end{equation}
and they depend only on the effective mass and 
the radius of the cylindrical nanowire.
It is worth to mention here that 
$|\phi_n^{(m)}(r)|^2$ and 
and $E^{(m)}_{\perp n}$
depend only on $|m|$.

Every transversal mode 
together with the associated motion on the transport direction
defines the \textit{scattering channel} 
on each side of the scattering area 
($s=1$ for the source contact and $s=2$ for the drain contact). 
The scattering channels are indexed by $(mns)$ for each $E$.

If the total energy $E$ and the lateral eigenenergy $E^{(m)}_{\perp n}$
are fixed, i.e. for every $E$, $m$ and $n$,
there are at most two linearly independent solutions of Eq.
(\ref{transpeq}). In the asymptotic region
they are given as a linear combination of
exponential functions
\begin{equation}
\varphi_{sn}^{(m)}(z) = \begin{cases}
                           A_s e^{ i k_{1nm} z}
                          +B_s e^{-i k_{1nm} z} , & z < -d_z \\
                           C_s e^{ i k_{2nm} z}
                          +D_s e^{-i k_{2nm} z} , & z > d_z
                        \end{cases}
\label{psi_plane}
\end{equation}
where $A_s$, $B_s$, $C_s$ and $D_s$ are complex coefficients 
depending on $n$, $m$ and $E$ for each value of $s=1,2$.
The wave vector is defined for each scattering channel $(mns)$ as
\begin{equation}
k_{snm}(E) = k_0 \sqrt{(E-E_{\perp n}^{(m)}-V_s)/u_0},
\label{ksn}
\end{equation}
where $k_0=\pi/2d_z$ and $u_0 = \hbar^2 k_0^2/2 \mu$.
In the case of conducting or open channels 
\begin{equation}
E-E_{\perp n}^{(m)}-V_s \ge 0,
\label{open_channels}
\end{equation}
$k_{snm}$ are positive real numbers
and correspond to propagating plane waves.
For the evanescent or closed channels 
\begin{equation}
E-E_{\perp n}^{(m)}-V_s < 0,
\label{closed_channels}
\end{equation}
$k_{snm}$  are
given from the first branch of the complex square root function,
$k_{snm}=i |k_{snm}|$,
and describe exponentialy decaying functions away from the scattering
region.
Thus, the number of the conducting channels, $N_{sm}(E)$, $s=1,2$, $m \ge 0$,
is a function of energy, and for a fixed energy $E$
this is the largest value of $n$, which satisfies the
inequality (\ref{open_channels})
for given values of $s$ and $m$.

Each conducting channel corresponds to one degree of freedom for the electron
motion through the nanowire and, consequently, there exists only one 
independent solution of Eq. (\ref{2D_schr}) for a fixed channel $(mns)$
associated with the energy $E$, $\psi^{(s)}_{nm}(E;r,z)$.
For describing further on the transport phenomena in the frame of
the scattering theory it is convenient to consider this solution
as a \textit{scattering state}, i.e. as a sum of an incoming
component on the channel $(mns)$ and a linear combination of outgoing
components on each scattering channel. 
In a convenient form\cite{roxana08}, 
the scattering function incident from the source contact ($s=1$)
is written as 
\begin{widetext}
\begin{subequations}
\label{scatt_funct}
\begin{equation}
\psi^{(1)}_{nm} (E;r,z)
=\frac{\theta(N_{1m}(E)-n)}{\sqrt{2\pi}}
 \begin{cases}
                    e^{ik_{1nm}(z+d_z)} \phi^{(m)}_n(r) +
 \sum_{n'=1}^\infty\limits \Sm^{(m)}_{1n',1n}(E) e^{-ik_{1n'm}(z+d_z)} 
                           \phi^{(m)}_{n'}(r),
  & z < - d_z \\
 \sum_{n'=1}^\infty\limits \Sm^{(m)}_{1n,2n'}(E) e^{ik_{2n'm}(z-d_z)} 
                           \phi^{(m)}_{n'}(r),
  & z > d_z
 \end{cases}
\label{scatt_left}
\end{equation}
and the scattering function incident from the drain contact ($s=2$)
as
\begin{equation}
\psi^{(2)}_{nm} (E;r,z)=\frac{\theta(N_{2m}(E)-n)}{\sqrt{2\pi}}
 \begin{cases}
 \sum_{n'=1}^\infty\limits S^{(m)}_{1n',2n}(E) e^{-ik_{1n'm}(z+d_z)} 
                           \phi^{(m)}_{n'}(r),
  & z < - d_z \\
                    e^{-ik_{2nm}(z-d_z)} \phi^{(m)}_n(r) +
 \sum_{n'=1}^\infty\limits S^{(m)}_{2n',2n}(E) e^{ik_{2n'm}(z-d_z)} 
                           \phi^{(m)}_{n'}(r),
  & z > d_z
 \end{cases}
\label{scatt_right}
\end{equation}
\end{subequations}
\end{widetext}
The step functions $\theta$ 
in the above expressions, with 
$\theta(x\ge 0)=1$ and $\theta(x< 0)=0$,
assure that the
scattering functions are defined only for conducting channels. 

The three-dimensional scattering states, solutions of Eq. (\ref{Schr_cyl}) 
for rotational invariant geometries
can be now written as
\begin{equation}
\Psi^{(s)}_{nm}(E;r,\theta,z)=\frac{e^{im\theta}}{\sqrt{2\pi}}
\psi^{(s)}_{nm}(E;r,z).
\label{Psi_gen2}
\end{equation}
Being eigenfunctions of an open system,
they are orthonormalized in the general sense \cite{07impe}
\begin{widetext}
\begin{equation}
\int_{-\infty}^\infty dz \int_0^R dr\;r \int_0^{2\pi} d\theta \;
\Psi_{mn}^{(s)}(E;r,\theta,z) \Psi_{m'n'}^{(s')}(E';r,\theta,z)^*=
\delta_{mm'} \delta_{ss'} \delta_{nn'} \frac{\delta(E-E')}{g_{snm}(E)},
\end{equation}
\end{widetext}
where $g_{snm}(E)$ is the 1D density of states, 
$g_{snm}(E)=\mu/ [\hbar^2 k_{snm}(E)]$. 

The physical interpretation of the expressions (\ref{scatt_funct}) is
that,
due to the nonseparable character of the scattering potential,
a plane wave incident onto the scattering domain is reflected on every
channel - open or closed for transport - on the same side of the system
and transmitted on every channel - open or closed for transport -
on the other side. The reflection and transmission amplitudes are described
by the complex coefficients $S^{(m)}_{sn',sn}$ and $S^{(m)}_{s'n',sn}$ with 
$s \ne s'$, respectively, and all of them should be nonzero. 
These coefficients define a matrix with $N_{1m}(E)+N_{2m}(E)$ 
infinite columns. For an elegant solution of the scattering problem we
extend $S^{(m)}(E)$ to an infinite square matrix and set at zero
the matrix elements without physical meaning,
$S^{(m)}_{s'n',sn}(E)=0$, $n > N_{sm}(E)$, $s=1,2$.
In this way we define the \textit{wave transmission matrix}
\cite{bagwell90} or the \textit{generalized scattering matrix} \cite{schanz95}.
This is not the well-known scattering matrix (current transmission
matrix) whose unitarity reflects the current conservation. 
The generalized scattering matrix is a non-unitary matrix,
which has the big advantage that it allows for a description of the
scattering processes not only in the asymptotic region but also inside the
scattering area.

For the sake of simplicity and also 
considering that for rotational invariant potentials the 2D scattering problems
generated by every magnetic quantum number $m$ can be solved separately,
the index $m$ will be omitted in the following subsections.

\subsection{R-matrix formalism for cylindrical geometry}

The scattering functions inside the scattering region are determinated
using the R-matrix formalism, i.e. they are 
expressed in terms of the eigenfunctions 
corresponding to the closed counterpart of the scattering
problem 
\cite{smr,wulf98,onac,roxana01,racec02,nemnes04,nemnes05,07impe}.
In our opinion this is a more appropriate method than the common mode 
space approach which implies the expansion of the scattering functions 
inside the scattering area in the basis of the transversal modes
$\phi_n(r)$. 
As it is shown in Ref. [\onlinecite{noeckel94},\onlinecite{luisier_06}]
the mode space approach has limitations
for structures with abrupt changes in the potential or
sudden spatial variations in the widths of the wire;
it breaks even down for coupling operators that are not
scalar potentials, like in the case of an external magnetic field.
In the R-matrix formalism the used basis contains all the information 
about the scattering potential, and this type of difficulties 
can not appear. 

Thus, the scattering functions inside the scattering region are given as
\begin{eqnarray}
\psi_{n}^{(s)}(E;r,z)&=&\sum_{l=1}^\infty a_{ln}^{(s)}(E) \chi_{l}(r,z),
\label{varphi_expan} 
\end{eqnarray}
with $r\in[0,R]$ and $z\in[-d_z,d_z]$. 

The so-called Wigner-Eisenbud functions, $\chi_{l}(r,z)$,
firstly used in the nuclear physics \cite{wigeis,lane},
satisfy the
same equation as $\psi_{n}^{(s)}(r,z)$,
Eq. (\ref{2D_schr}), but with different  boundary conditions
in the transport direction: Since the scattering function
$\psi_{n}^{(s)}(r,z)$ satisfies energy
dependent boundary conditions derived from Eq. (\ref{scatt_funct})
due to the continuity of the scattering function and its derivative at
$z = \pm d_z$, the Wigner-Eisenbud function
$\chi_{l}(r,z)$ has to satisfy Neumann boundary
conditions at the interfaces between the scattering region and contacts,
$\left. \partial \chi_{l}/\partial z \right|_{z=\pm d_z}=0$, $l \ge 1$.
The infinite potential outside the nanowire requires
Dirichlet boundary condition on the cylinder surface
for the both functions,
$\psi_{n}^{(s)}(R,z)=0$ and $\chi_{l}(R,z)=0$.
The functions $\chi_{l}$, $l \ge 1$, built a basis which verifies the
orthogonality-
and the closure-  
relation.
The corresponding eigenenergies to $\chi_{l}$ are denoted by $E_l$
and are called Wigner-Eisenbud energies.
Since the Wigner-Eisenbud problem is defined on a closed volume
with self-adjoint boundary conditions, the 
eigenfunctions $\chi_{l}$ and the eigenenergies $E_l$ can be choosen 
as real quantities.
The Wigner-Eisenbud problem is, thus, the closed counterpart of the scattering
problem. 

In the case of the one-dimensional system
it was recently proven mathematically rigorous that 
the R-matrix formalism allows for a proper expansion of the scattering matrix
on the real energy axis\cite{hagen08}.
In this section we present an extension of 
the R-matrix formalism for 2D scattering problem with cylindrical symmetry.

To calculate the expansion coefficients $a_{ln}^{(s)}(E)$
we multiply Eq. (\ref{2D_schr}) by $\chi_{l}(r,z)$ and the equation satisfied
by the Wigner-Eisenbud functions by $\psi_{n}^{(s)}(E;r,z)$. The difference
between the resulting equations is integrated over
$[-d_z,\;d_z] \times [0,R]$, and one obtains on the right hand side 
the coefficient $a_{ln}^{(s)}(E)$. 
After an integration by parts in the
kinetic energy term and using the boundary conditions
one finds $a_{ln}^{(s)}$ and feed in it into
Eq. (\ref{varphi_expan}).
So,
\textit{the scattering functions inside the scattering region}
($z \in [-d_z,d_z],\; r \in [0,R]$)
are obtained in terms of their derivatives at the edges of this
domain,
\begin{widetext}
\begin{equation}
\begin{split}
\psi_{n}^{(s)}(E,r,z)=\frac{2d_z}{\pi} \int_0^{R} dr' \; r'
   \left[ R(E,r',-d_z,r,z)
   \left.
   \frac{\partial \psi_{n}^{(s)}(E,r',z')}{\partial z'} \right|_{z'=-d_z}
   \right. 
  -\left. R(E,r',d_z,r,z)
   \left.
   \frac{\partial \psi_{n}^{(s)}(E,r',z')}{\partial z'} \right|_{z'=d_z}
   \right], 
\end{split}
\label{psi-inside}
\end{equation}
\end{widetext}
where the R-function is defined as
\begin{equation}
R(E,r,z,r',z') \equiv \frac{\hbar^2}{2\mu}\sum_{l=1}^\infty
  \frac{\chi_{l}(r,z) \chi_{l}(r',z')}{E-E_l}\frac{\pi}{2d_z}.
\end{equation}
The functions 
$\partial \psi_{n}^{(s)}/\partial z$ at $z=\pm d_z$
are calculated from the asymptotic form (\ref{scatt_funct})
based on the continuity
conditions  for the derivatives of the scattering functions
on the interfaces
between the scattering region and contacts.

With these results the scattering functions
inside the scattering domain
are expressed in terms of the
wave transmission matrix $\mathbf{S}$
\begin{equation}
\vec{\Psi}(E;r,z) = \frac{i}{\sqrt{2 \pi}}  
                       \mathbf{\Theta}(E) [{\bf 1} - \mathbf{S}^T(E) \\
                   \mathbf{K}(E) \vec{R}(E;r,z)],
\label{psi3}
\end{equation}
where the component $(sn)$ of the vector $\vec{\Psi}$ is the
scattering function $\psi_{n}^{(s)}(E;r,z)$, $n \ge 1$, $s=1,2$
and $\mathbf{S}^T$ denotes the matrix transpose.
The diagonal matrix $\mathbf{K}$ has on its diagonal the wave
vectors (\ref{ksn}) of each scattering channel 
\begin{equation}
\mathbf{K}_{sn,s'n'}(E)= \frac{k_{sn}(E)}{k_0} \, 
                               \delta_{nn'} \delta_{ss'},
\label{K-matrix}
\end{equation}
$n,n' \ge 1$, $s,s'=1,2$,
and the vector $\vec{R}(E;r,z)$ as
\begin{equation}
\vec{R}(E;r,z)= \frac{u_0}{\sqrt{k_0}}
                     \sum_{l=1}^{\infty}
                       \frac{\chi_{l}(r,z) \vec{\chi}_{l}}
                            {E-E_l},
\label{R-vector}
\end{equation}
where $\vec{\chi}_{l}$ is a vector with the components
\begin{equation}
(\vec{\chi}_{l})_{sn} = \frac{1}{\sqrt{k_0}}
                       \int_{0}^{R} \; \chi_{l}(r,(-1)^{s} d_z) 
                       \phi_n(r) r dr ,
\label{chi-vector}
\end{equation}
$n \ge 1, s=1,2$. 
The diagonal $\mathbf{\Theta}$-matrix,
$\mathbf{\Theta}_{sn,s'n'}(E) = \theta(N_{s}(E)-n) \,
                                 \delta_{ss'} \,
                                 \delta_{nn'}$, $n \ge 1$, $s=1,2$,
assures non-zero values only for the
scattering functions corresponding to the conducting channels.

Using further the continuity of the scattering functions on the surface of
the scattering area and expanding $\vec{R}(E;\pm d_z,r)$ in the basis
$\left\{\phi_n(r)\right\}_{n \ge 1}$ we find the relation between
matrixes $\mathbf{S}$ and $\mathbf{R}$
\begin{equation}
\mathbf{S}(E) = \left[ \mathbf{1} - 2 \left( \mathbf{1} + i \mathbf{R}(E) 
                                                        \mathbf{K}(E)
                                  \right)^{-1}
                \right] \mathbf{\Theta}(E),
\label{R-Srelation}
\end{equation}
with the $\mathbf{R}$-matrix given by means of a dyadic product
\begin{equation}
\mathbf{R}(E)= u_0 \sum_{l=1}^{\infty}
                     \frac{\vec{\chi}_{l} \, \vec{\chi}_{l}^T}
                          {E-E_l}.
\label{R-matrix}
\end{equation}
According to the above relation, $\mathbf{R}$ is an infinite-dimensional 
symmetrical real matrix. The above form allows for a very efficient
numerical implementation for computing the $\mathbf{R}$-matrix. 

The expression (\ref{R-Srelation}) of the $\mathbf{S}$-matrix in terms of the
$\mathbf{R}$-matrix is the key relation for solving
2D scattering problems using only the eigenfunctions
and the eigenenergies of the closed system. 
On the base of  Eq. (\ref{R-Srelation}) the
wave transmission matrix
is calculated and after that the scattering functions in each point of the
system are obtained using
Eqs. (\ref{scatt_funct}) and (\ref{psi3}). 

To come back to the dependence on $m$, we point out that 
the Wigner-Eisenbud functions and energies are $m$-dependent, so
that 
the matrixes $\mathbf{R}$, $\mathbf{K}$ and 
$\mathbf{\Theta}$ and the vector $\vec{\Psi}(E;r,z)$ are $m$-dependent
in relations (\ref{R-Srelation}) and (\ref{psi3}).

\subsection{Reflection and transmission coefficients}

Using the density current operator 
\begin{equation}
\mathbf{j}(\mathbf{r})=\frac{\hbar}{2i\mu}
  \Bigl(  \Psi(\mathbf{r}) \nabla \Psi(\mathbf{r})^*
        - \Psi(\mathbf{r})^*\nabla \Psi(\mathbf{r})
  \Bigr),
\end{equation}
one can define, as usually, the transmission and reflection probabilities.
\cite{buett85}
$\Psi(\mathbf{r})^*$ denotes the complex conjugate of the 
scattering wave function (\ref{Psi_gen2}).

The $r$-component of the density current $j_r(r,\theta,z)$ is zero in leads,
because $\phi_n(r)$ are real functions.
The component $\theta$ of the incident density current 
is $m$ dependent,
\begin{displaymath}
(j_{inc}(r,\theta,z))_\theta=\frac{\hbar^2}{\mu} \frac{1}{(2\pi)^2}
                             \frac{1}{r} m |\phi_n(r)|^2,
\end{displaymath}
so that if one sums over all $m$ values,
then they cancel each other. 
This is also valid for the reflected and transmitted current fluxes.
What remains is the $z$-component of the particle density current
$j_z(r,\theta,z)$,
which integrated over the cross section of the nanocylinder 
with the corresponding measure, $rdr$, provides the very well known
relations for the transmission and reflection probabilities.
The probability for an electron incident
from source, $s=1$, on channel $n$  to be reflected back into source
on channel $n'$ is 
\begin{equation}
R_{nn'}^{(1)}=\frac{k_{1n'}}{k_{1n}} |S_{1n,1n'}^t|^2,
\end{equation}
and the probability to be tranmitted into drain, $s=2$, on channel $n'$ is 
\begin{equation}
T_{nn'}^{(1)}=\frac{k_{2n'}}{k_{1n}} |S_{1n,2n'}^t|^2.
\end{equation}
The reflection and transmission probabilities for evanescent (closed)
channels are zero.
The total transmission and reflection coefficients for 
an electron incident from reservoir $s=1$ are defined as
\begin{equation}
T^{(1)}=\sum_{n,n'} T^{(1)}_{nn'}, \quad R^{(1)}=\sum_{n,n'} R^{(1)}_{nn'}.
\end{equation}
More detailed properties of the many-channel tunneling and reflection
probabilities are given in Ref. [\onlinecite{buett85}], 
but note that our indexes are interchanged with respect to the definitions 
used there. Of course, all these coefficients are $m$-dependent.

\subsection{Current scattering matrix}
Further, we define the energy dependent \textit{current scattering matrix} as
\begin{equation}
{\tilde{\mathbf{S}}}(E) = 
\mathbf{K}^{1/2}(E) \mathbf{\Theta}(E) \mathbf{S}(E) \mathbf{K}^{-1/2}(E),
\label{Stilde}
\end{equation}
so that its elements give directly the reflection and transmission probabilities
\begin{align}
|\tilde{S}_{1n',1n}(E)|^2 & =  R_{nn'}^{(1)}(E), &
|\tilde{S}_{2n',2n}(E)|^2 & =  R_{nn'}^{(2)}(E), \label{refl_amplt_stilde}\\
|\tilde{S}_{2n',1n}(E)|^2 & =  T_{nn'}^{(1)}(E), &
|\tilde{S}_{1n',2n}(E)|^2 & =  T_{nn'}^{(2)}(E). \label{transm_amplt_stilde}
\end{align}

The diagonal $\mathbf{\Theta}$-matrix assures that the matrix elements of
$\tilde{\mathbf{S}}$ are nonzero only for the conducting channels,
for which the transmitted flux is nonzero.
Using the $\mathbf{R}$-matrix representation of $\mathbf{S}$, Eq.
(\ref{R-Srelation}), we find from the above relation
\begin{equation}
{\tilde{\mathbf{S}}}(E) = \mathbf{\Theta}(E)
                \left[ \mathbf{1} - 2 (\mathbf{1} + i \mathbf{\Omega}(E))^{-1}
                \right]
                \mathbf{\Theta}(E),
\label{Stilde2}
\end{equation}
with the infinite dimensional matrix $\mathbf{\Omega}$
\begin{equation}
\mathbf{\Omega}(E) = u_0 \sum_{l=1}^{\infty}
                      \frac{\vec{\alpha}_l \, \vec{\alpha}^T_l}
                           {E-E_l}
                =\mathbf{K}^{1/2}(E)\mathbf{R}(E)\mathbf{K}^{1/2}(E)
\label{omega}
\end{equation}
and the column vector
\begin{equation}
\vec{\alpha}_l(E) =\mathbf{K}^{1/2}(E) \, \vec{\chi}_l,
\label{alpha}
\end{equation}
with $l \ge 1$. 
According to the definition (\ref{omega}) the matrix
$\mathbf{\Omega}$ is a symmetrical one,
$\mathbf{\Omega}=\mathbf{\Omega}^T$,
and from Eq. (\ref{Stilde2}) it follows that ${\tilde{\mathbf{S}}}$ 
also has this
property, ${\tilde{\mathbf{S}}}={\tilde{\mathbf{S}}}^T$. On this basis one can
demonstrate that the tunneling coefficient characterizes one pair of
open channels 
irrespective of the origin of the incident flux
$T^{(1)}_{nn'}=\left|{\tilde{S}}_{2n',1n}\right|^2
       =\left|{\tilde{S}}_{1n,2n'}\right|^2 = T^{(2)}_{n'n}$.
This is a well-known property of the transmission through a scattering
system and it shows that the current scattering matrix used here is 
properly defined.
The restriction of $\tilde{\mathbf{S}}$-matrix to the open channels is the well
known current scattering matrix \cite{wulf98,roxana01,racec02},
commonly used in the
Landauer-B\"uttiker formalism. For a given energy $E$ this is a
$(N_1+N_2) \times (N_1+N_2)$  matrix
which has to satisfy the unitarity
condition, 
according to the flux conservation.

The relation (\ref{Stilde2}) is the starting point for a resonance theory
\cite{roxana01,roxana08},
which allows for an explicit analytical expression for the transmission peak
as a Fano resonance with a complex asymmetry parameter. 

In the numerical computations, 
the matrixes $\mathbf{S}$, $\mathbf{R}$, $\mathbf{\Omega}$, 
$\tilde{\mathbf{S}}$ and $\mathbf{\Theta}$
have the dimension $ 2N\times 2N$,
and the vectors $\vec{\chi}_l$, $\vec{\alpha}_l(E)$ 
have $2N$ components, where $N$ is the number of scattering channels
(open and closed) taken numerically into account. 
The number of the Wigner-Eisenbud
functions and energies computed numerically 
establishes the maximum value for the index $l$.

\section{Cylindrical nanowire heterostructure model systems}

The formalism presented above
is general and can be applied to a
variety of the cylindrical nanowire heterostructures.
We consider a series of heterostructures embedded in an infinite cylindrical 
nanowire of radius $R=5$nm 
and effective mass $\mu=0.19m_0$ (corresponding to transverse mass in Silicon). 
We set in all our computations $d_z=16$nm, see Fig. \ref{Schr_2D_vol}, 
and the total number of channels (open and closed)
$N=8$. 
In our calculations, the results do not change if more channels are added.
\begin{figure}[h]
           \includegraphics[width=3.25in]{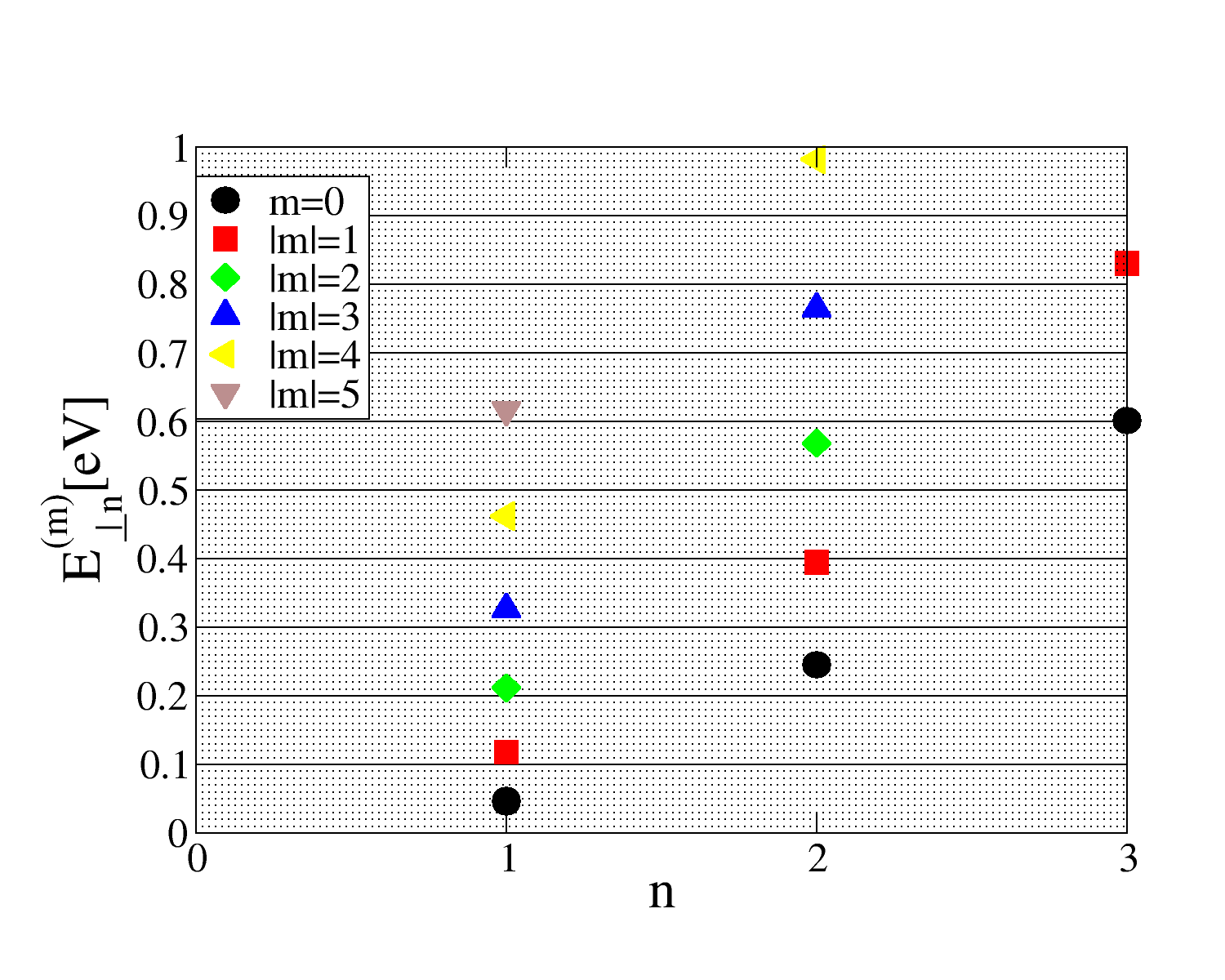}
\caption{(color online)
Energies $E^{(m)}_{\perp,n}$ of the transversal modes 
for a cylinder with $R=5$nm and $\mu=0.19m_0$.
}
\label{Epmn}
\end{figure}
In Fig. \ref{Epmn} are plotted
the energies of the transversal modes $E^{(m)}_{\perp,n}$ until $1eV$, 
for different magnetic quantum numbers $m$,
according to (\ref{e_channel}).
The difference between two successive energies of the transversal modes
is $m$-dependent, 
due to the roots $x_{mn}$ of the Bessel functions $J_m$.

\subsection{Quantum dot embedded into the nanocylinder}
\subsubsection{Same radius as the host cylinder}

\begin{figure}[h]
\subfigure[]{\label{1QW_v_sketch}
           \includegraphics[width=3.00in]{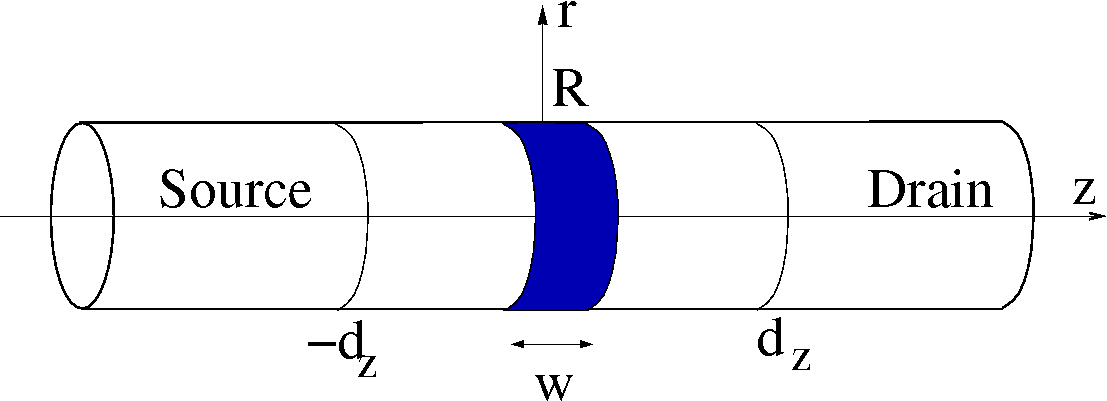}}
\subfigure[]{\label{1QW_v_pot}
           \includegraphics[width=3.00in]{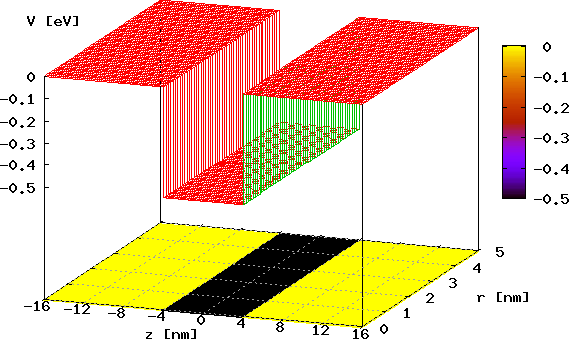}}
\caption{(color online) (a) Sketch of a cylindrical quantum dot embedded
into a nanowire with the same radius.
We consider that the dot yields an attractive potential $V(r,z)$, represented
in (b) by a rectangular quantum well of depth $W_b=-0.5eV$ and width $w=8$nm.
}
\label{1QW_v}
\end{figure}

In Fig. \ref{1QW_v_sketch} is sketched
a cylindrical quantum dot embedded
into a cylindrical nanowire with the same radius.
This kind of structures and even compositionally 
modulated, called also "nanowire superlattice" \cite{lieber02},
are already realized technologically on
different materials basis, 
as is summarized in a recent review article \cite{shen08}. 

Depending on the 
band-offsets between the dot material and the host material the potential 
produced by the dot can be repulsive, yielding a quantum barrier, or 
attractive, yielding a quantum well. 
As it is mentioned in Ref. [\onlinecite{lieber02}] 
the interfaces between the dot and the host material may be considered
sharp for nanowires with diameter less than $20$nm. 
We consider here 
that the dot yields an attractive potential $V(r,z)$, represented
in Fig. \ref{1QW_v_pot} by a rectangular quantum well of depth $W_b=-0.5eV$ and width $w=8$nm.

\begin{figure}[hb!]
\includegraphics[width=3.25in]{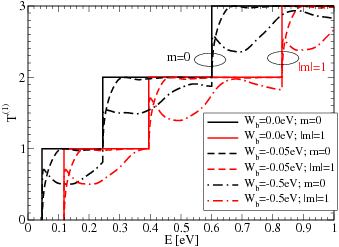}
\caption{(color online) Total tunneling coefficient 
as a function of incident energy $E$
for the scattering potential represented in Fig. \protect\ref{1QW_v_pot},
for different magnetic quantum numbers 
$m$ 
and various values of the well depth $W_b=0.0eV$ (continuous line), 
$W_b=0.05eV$ (dashed line), and $W_b=0.5eV$ (dot-dashed line).
}
\label{QW_wb_R5nm_T1}
\end{figure}

The total tunneling coefficient $T^{(1)}$ versus the incident energy $E$
is plotted in Fig. \ref{QW_wb_R5nm_T1}, for different magnetic
quantum numbers $m$ and different quantum well depths $W_b$.
In the absence of the quantum well, $W_b=0.0eV$, one can recognize the 
abrupt steps\cite{vanwees88} in the tunneling coefficient. 
The transmission increases with a unity, every time
a new channel $E^{(m)}_{\perp,n}$ becomes available for transport, i.e. 
becomes open. The length of the plateaus is given by the difference between
two successive transversal mode energies, which differ for different $m$ values.
Due to the almost square dependence of the transverse energy levels on the 
channel number, the length of the plateaus increases.
Increasing the depth of the well,
deviations from the step-like transmission appear. 
There are no effects due to the influence of the evanescent 
channels\cite{gudmundsson05}, 
because the scattering potential for this configuration remains 
furthermore separable in the confinement and the transport direction, 
$V(r,z)=U(r)+W(z)$.

The spectral representation of the 2D Hamiltonian in 
this situation is a superposition of the spectrum of each channel, 
without being perturbed by channel mixing.
Considering an attractive potential in $z$-direction, 
there is always at least one bound state \cite{simon76,klaus77} below the
continuum spectrum for every channel $n$, see Fig. \ref{evan_sketch_n}(a).
In turn, the bound states of the higher channels $n$
get embedded in the continuous part of the lower channels, forming bound states
in continuum (BIC). Since the potential is separable, there is no mix of
states, and the BIC states can not be seen as scattering states.

\subsubsection{Surrounded by host material}

\begin{figure}[h]
\subfigure[]{\label{1QW_v_R1nm_sketch}
           \includegraphics[width=3.00in]{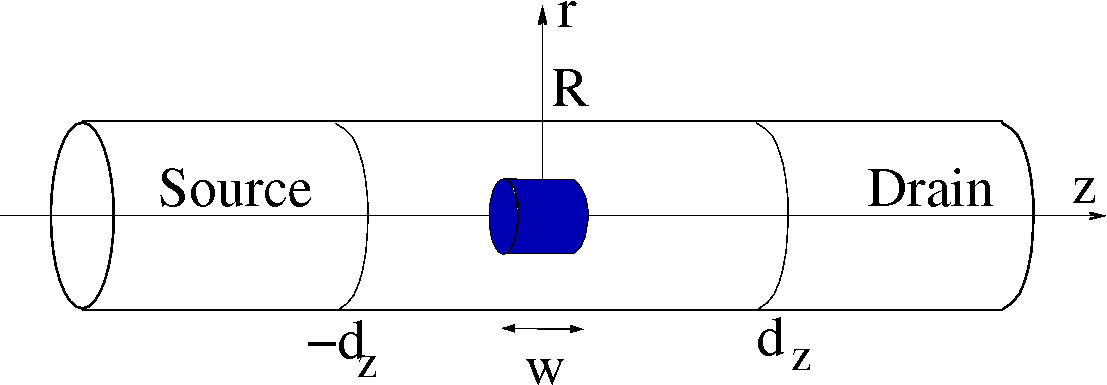}}
\subfigure[]{\label{1QW_v_R1nm_pot}
           \includegraphics[width=3.00in]{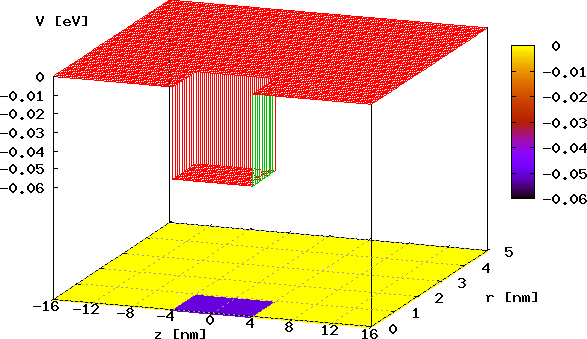}}
\caption{(color online) a) Sketch of a cylindrical quantum dot 
embedded into the nanocylinder and surrounded by the host material.  
We consider that the dot yields an attractive potential $V(r,z)$,
represented in b) by a rectangular quantum well of depth $W_b=-0.05eV$
and width $w=8$nm.
The radius of the dot is $R'=1$nm.
}
\label{1QW_v_R1nm}
\end{figure}

\begin{figure}[htb]
\subfigure[$m=0$]{\label{1QW_R1nm_T1_m0}
                  \includegraphics[width=3.25in]{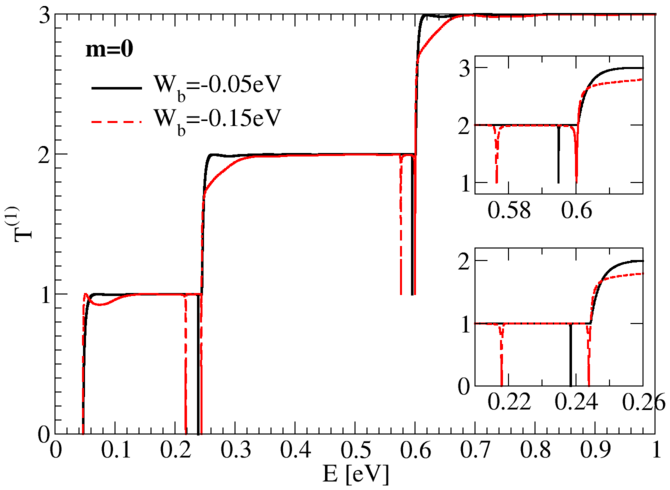}}
\subfigure[$|m|=1$]{\label{1QW_R1nm_T1_m1}
                  \includegraphics[width=3.25in]{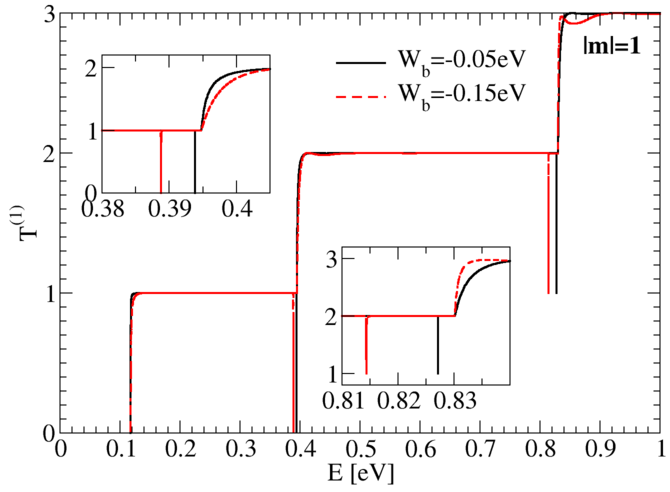}}
\caption{(color online)
Total tunneling coefficient as a function 
of total incident energy $E$ 
for different magnetic quantum numbers $m$
for a cylindrical dot surrounded by the host material as in 
Fig. \protect\ref{1QW_v_R1nm}. 
A detailed view around the channel minima is presented in the insets.
}
\label{1QW_R1nm_T1}
\end{figure}

Further we study a cylindrical dot embedded into the nanocylinder,
but whose radius $R'$ is smaller than the cylinder radius $R$, 
so that the dot gets surrounded by the host material, 
see Fig. \ref{1QW_v_R1nm_sketch}. We consider here again the case
that the dot yields an attractive potential,
i.e. a rectangular quantum well, plotted in 
Fig. \ref{1QW_v_R1nm_pot}. 
Even we have chosen a small value for the depth of the quantum well, 
$W_b=-0.05eV$, there are significant deviations in the tunneling coefficient 
from the step like characteristic, see Fig. \ref{1QW_R1nm_T1}.
Just before a new channel gets open, below $E^{(m)}_{\perp,n}$, 
there is a dip, i.e. sharp drop,  in the tunneling coefficient.
These dips are owing to modification of the tunneling coefficient due to the
evanescent (closed) channels\cite{bagwell90}. This is a multichannel effect
that was until now studied only in Cartesian 
coordinates for quantum wires tailored in two-dimensional electron gas
\cite{bagwell90,levinson93,noeckel94,gudmundsson04,gudmundsson05}. 

The dips can be understood considering 
the simple couple mode model \cite{bagwell90,levinson93,noeckel94}.
\begin{figure}[htb]
\subfigure[ For a channel $n$]{\label{evan_sketch_n}
          \includegraphics[width=3.00in]{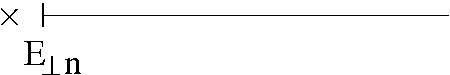}}

\subfigure[ For 2D Hamiltonian, for the first two channels]
           {\label{evan_sketch_s12}
          \includegraphics[width=3.00in]{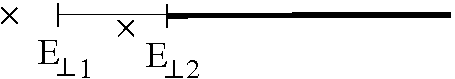}}
\caption{Sketch of the energy spectrum.
}
\label{evan_sketch}
\end{figure}
For a dot surrounded by host material, 
the scattering potential $V(r,z)$ is not anymore separable,
so that the scattering mixes the channels \cite{bagwell90,levinson93,noeckel94}.
As soon as the scattering potential is attractive, 
the diagonal coupling matrix element 
\begin{equation}
V_{nn}(z)=\int_0^R \phi_n(r)V(r,z)\phi_n(r) r dr < 0
\label{eff_attr}
\end{equation}
acts for every channel $n$ as an 
effective one-dimensional (1D) attractive potential \cite{levinson93}, 
which always allows for at least one bound state 
\cite{simon76,klaus77}
below the threshold of the continuum spectrum. 
We have sketched in Fig. \ref{evan_sketch_n} the energy spectrum
of a channel $n$: the continuous part represented by continuous line
is real and starts at $E^{(m)}_{\perp,n}$; the bound state represented by
a cross (we consider for simplicity only one), 
is also real but just below the threshold. 
By mixing the channels, this bound state becomes a 
\textit{quasi-bound state} or resonance, 
i.e with complex energy, whose real part gets embedded into continuum 
spectrum of the lower channel, and the imaginary part describes
the width of the resonance. 
The spectrum of the 2D scattering
problem is a superposition of the above discussed spectra and is sketched
in Fig. \ref{evan_sketch_s12} for spectra corresponding to 
channels $1$ and $2$.
These resonances can be seen now as dips in the tunneling coefficient.
The energy difference between the position of the dips 
and the next subband minima $E^{(m)}_{\perp,n}$ 
gives the quasi-bound state energy. 
The positions of the dips, i.e. the quasi-bound state energy, 
depend on the channel number
$n$ and on the magnetic quantum number $m$ and, of course, on the detailed
system parameters.
In Cartesian coordinates the specific symmetry of the channels (odd and even) 
do not allow for dips in the first plateau\cite{gudmundsson04}. 
In the cylindrical geometry
this symmetry is broken, so that we obtain a dip in front of every
plateau.
Our numerical method allows for high energy resolution in computing the
tunneling coefficient, so that 
we were able to find
the dips also in front of higher-order plateaus.

\begin{figure}[hb!]
\subfigure[$E=0.238eV$, $n=1$]{\label{1QW_R1nm_psi1_m0_n1}
          \includegraphics[width=2.5in]{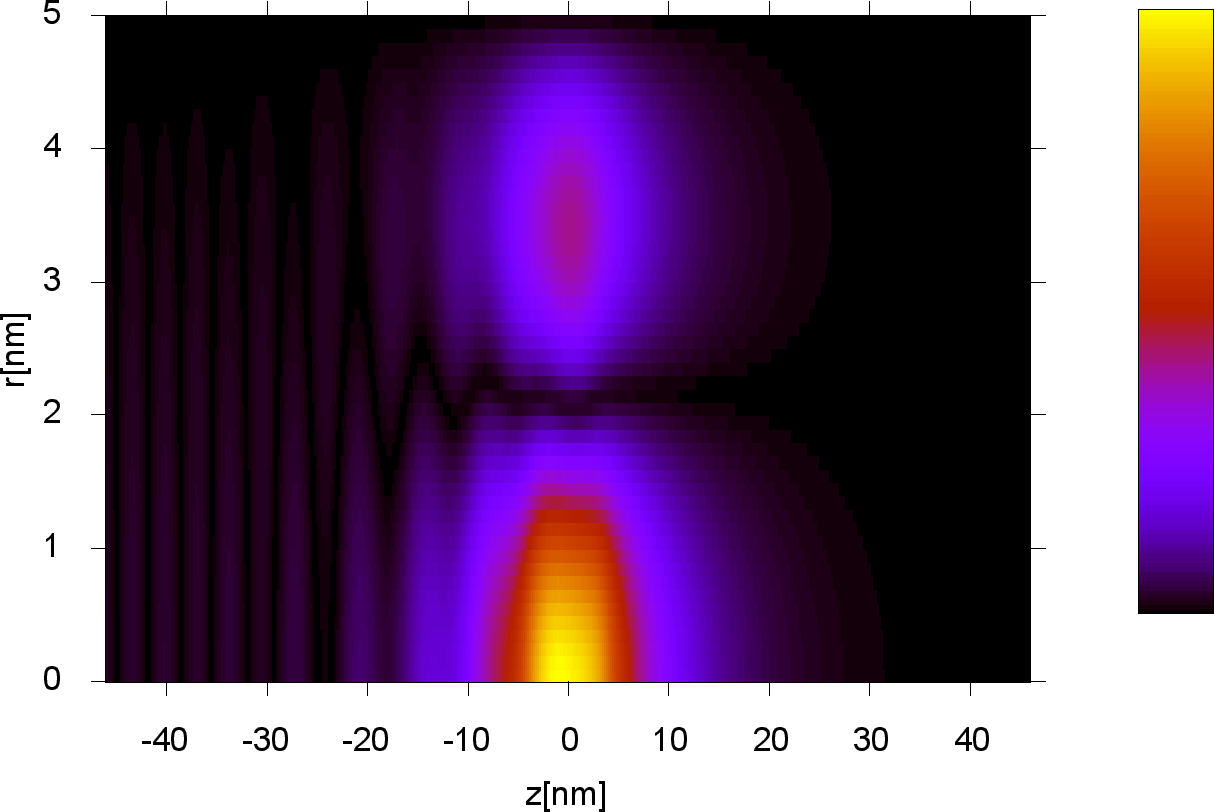}}

\subfigure[$E=0.594eV$, $n=2$]{\label{1QW_R1nm_psi1_m0_n2}
          \includegraphics[width=2.5in]{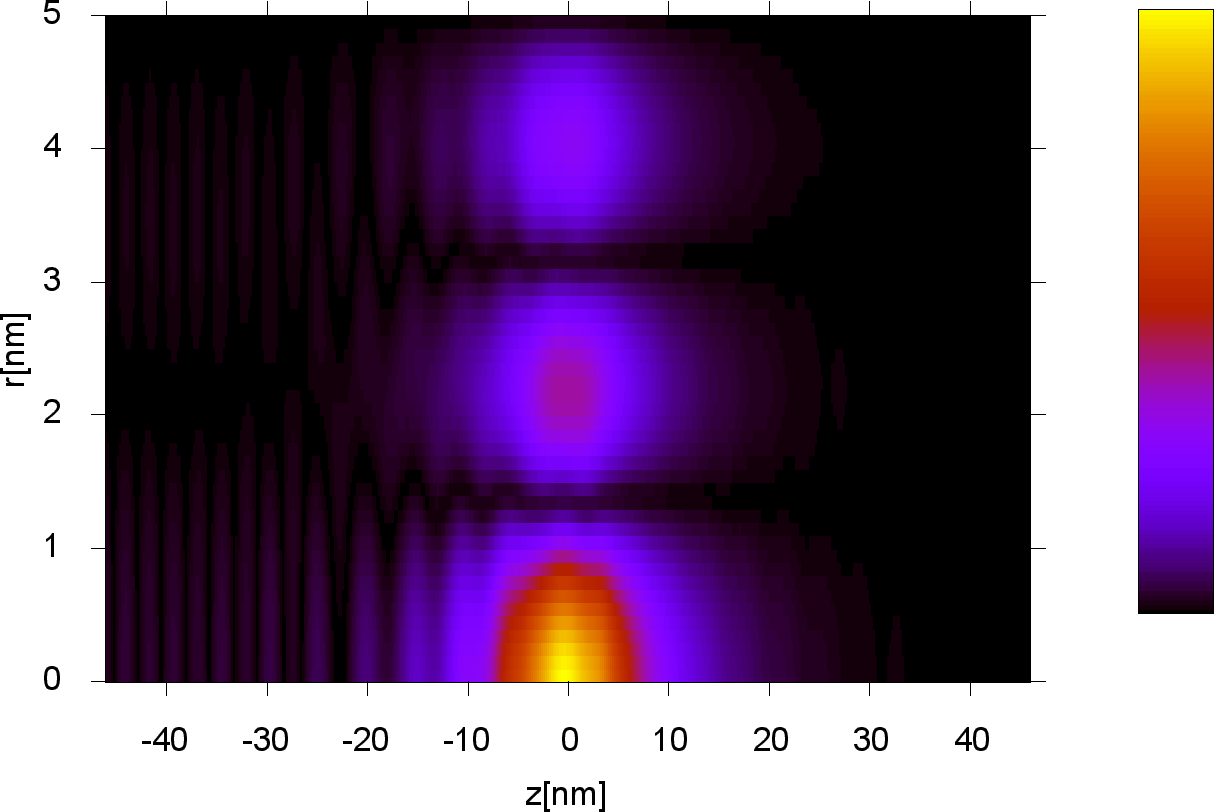}}
\caption{(color online) 
Localization probability density, $|\psi^{(1)}_n(E,r,z)|^2$,
for an electron with $m=0$,
incident from reservoir $s=1$ into channel $n$ and with total energy $E$,
both indicated in the captions. 
The energies are the dips in Fig. \protect\ref{1QW_R1nm_T1_m0}. 
}
\label{1QW_R1nm_psi1_m0}
\end{figure}

Further insight about the quasi-bound states of the evanescent
channels can be gained looking at the wave functions,
whose square absolute value $|\psi^{(s)}_n(E,r,z)|^2$ gives the 
\textit{localization probability density}.
Considering that the scattering states are orthonormalized in the general 
sense, the more appropriate quantity to analize would be 
the \textit{local density of states}
\begin{equation}
g(E,r)=|\psi^{(s)}_n(E,r,z)|^2 g_{snm}(E),
\label{ldos}
\end{equation}
which differs from the localization probability density just by
1D density of states.
For this reason we plot the localization probability density
in arbitrary units.

\begin{figure}[h]
\subfigure[$E=0.393eV$, $n=1$]{\label{1QW_R1nm_psi1_m1_n1}
           \includegraphics[width=2.5in]{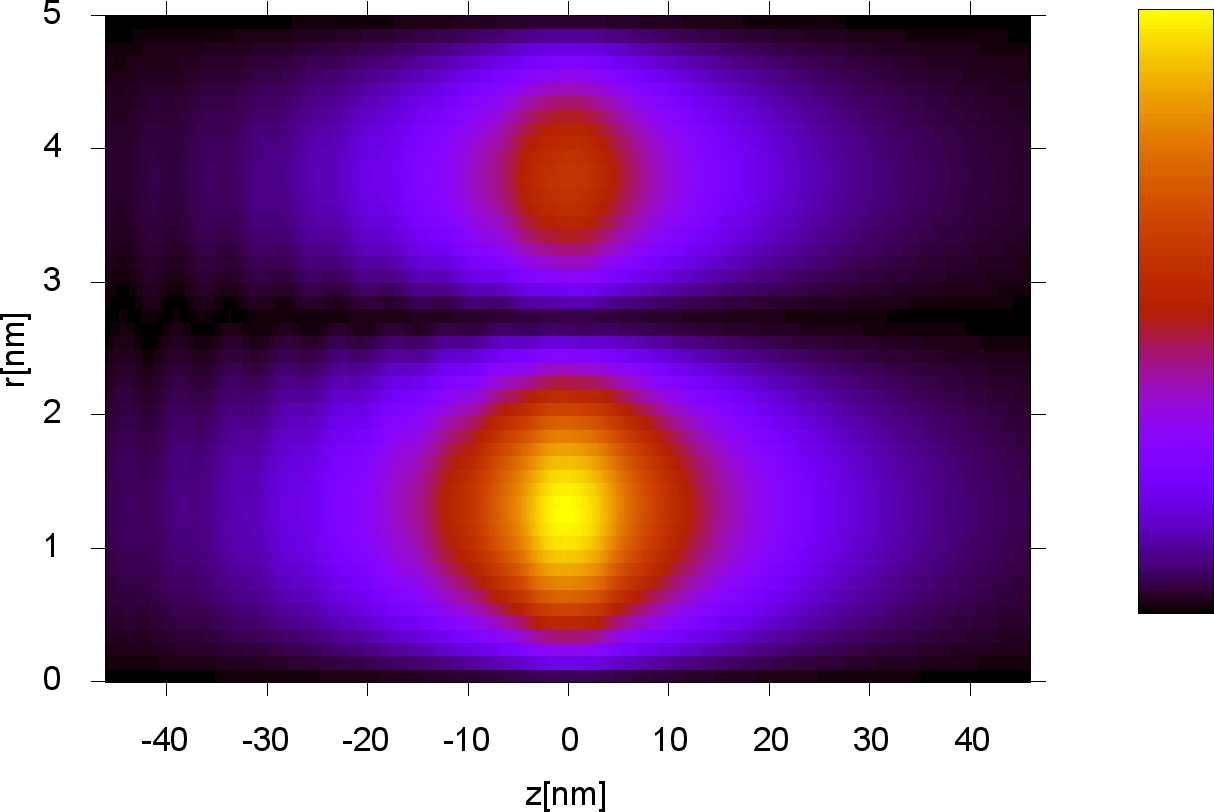}}

\subfigure[$E=0.827eV$, $n=2$]{\label{1QW_R1nm_psi1_m1_n2}
           \includegraphics[width=2.5in]{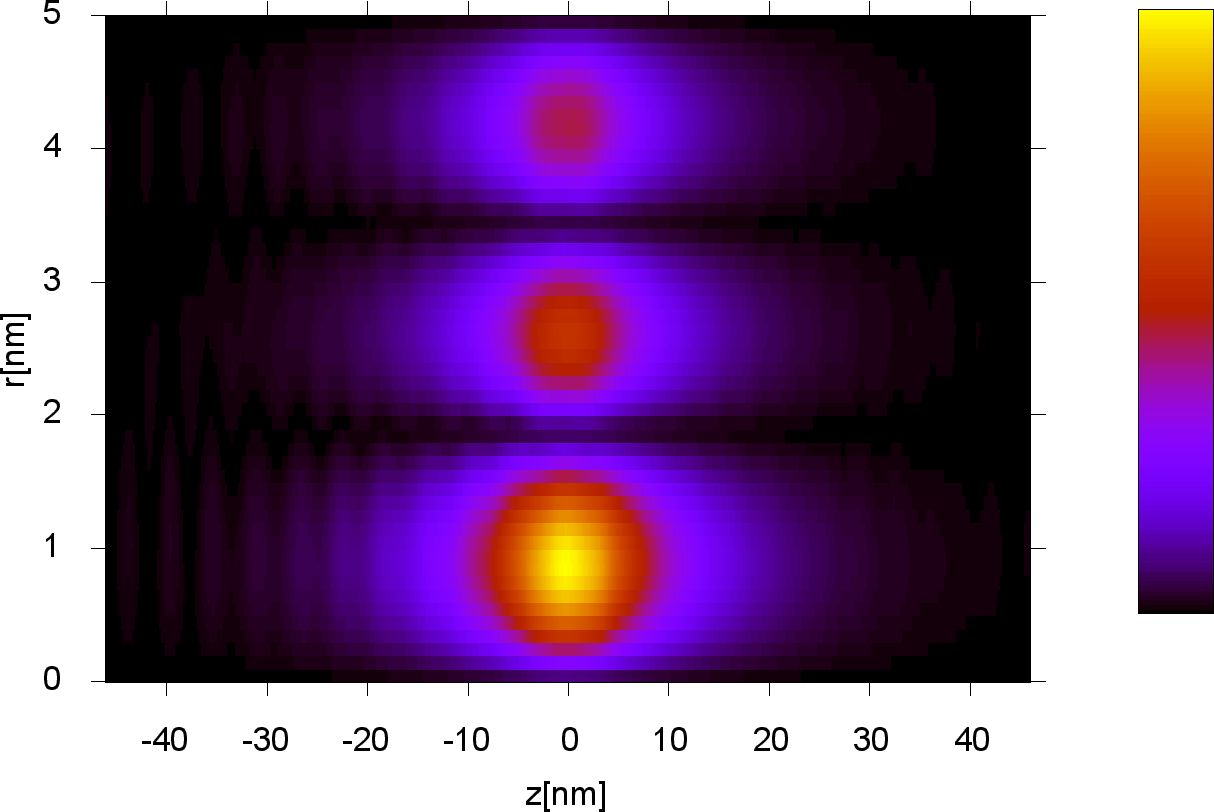}}
\caption{(color online) 
Localization probability density $|\psi^{(1)}_{nm}(E,r,z)|^2$
for an electron, 
with $|m|=1$,
incident from reservoir $s=1$ 
into channel $(nm)$ and with total energy $E$,
indicated in the captions.
The energies are the dips in Fig. \protect\ref{1QW_R1nm_T1_m1}
for $W_b=-0.05eV$.
}
\label{1QW_R1nm_psi1_m1}
\end{figure}

Our numerical implementation based on the R-matrix formalism
allows us to produce high resolution maps of the
wave functions 
inside the scattering region, see Eq. (\ref{psi3}).
In Figs. \ref{1QW_R1nm_psi1_m0}, \ref{1QW_R1nm_psi1_m1}
is represented the localization 
probability density $|\psi^{(1)}_n(E,r,z)|^2$ 
of an electron,  
incident from source ($s=1$) and has a total energy corresponding
to the dips in Fig. \ref{1QW_R1nm_T1}. The total energy $E$ 
and the channel $n$, on which the electron is incident, are specified at every
plot. 
Let discuss Fig. \ref{1QW_R1nm_psi1_m0_n1}. 
The total energy $E=0.238eV$ is less than
the energy of the second transversal mode, $E_{\perp,2}^{(0)}=0.244eV$,
so that only first channel is open,
thus the incident wave from the source contact 
is node-less in $r$-direction. 
But, as it can be seen in Fig. \ref{1QW_R1nm_psi1_m0_n1}, 
the scattering wave function inside the scattering region 
has a node in the $r$-direction, i.e. position in $r$ where the
wave function is zero. 
This means that the wave function 
corresponds to the quasi-bound state splitting off 
from the second transversal mode,
which is an evanescent one.
The quasi-bound state is
reachable now in a scattering formulation due to channel mixing.  
The wave function has a pronounced peak around the scattering potential,
i.e. $z\in [-4,4]$nm,
which decreases exponentially to the left and to the right.
To the left of the scattering potential one observes the interference
pattern produced by the incident wave and the reflected one, while
to the right only the transmitted part exists. 

The wave function considered in Fig. \ref{1QW_R1nm_psi1_m0_n2} 
has the energy less than the third transversal channel,
$E^{(0)}_{\perp,3}=0.6006eV$,
so that the incident part of the scattering state 
on the second mode $n=2$ 
has one node in $r$-direction. But the scattering function shows 
inside the scattering region two nodes in
the $r$-direction, so it corresponds
to a quasi-bound state splitting off from the above evanescent channel,
the third one.
One gets similar pictures for all $m$-values, with the
difference that for $m\not=0$ the wave functions
are zero for $r=0$, 
like it is shown in Fig. \ref{1QW_R1nm_psi1_m1} for the case $|m|=1$. 
In Figs. \ref{1QW_R1nm_psi1_m0} and \ref{1QW_R1nm_psi1_m1}
one can observe that the transmitted part of the scattering wave function 
is zero, in agreement with the resonant backscattering specific
to the quasi-bound states of the evanescent channels 
\cite{bagwell90,levinson93}.

The extension of the quasi-bound state of an evanescent channel
is given outside the scattering region
by the exponential decaying functions
$\exp[\kappa_{1nm}(z+d_z)]$ for $z > -d_z$ and 
$\exp[-\kappa_{2nm}(z-d_z)]$ for $z < d_z$, where 
$\kappa_{snm}=i|k_{snm}|$ and $k_{snm}$ is defined in (\ref{ksn}).
This means, the closer the resonance to the threshold of the evanescent channel 
$E^{(m)}_{\perp,n}$, the slower the exponential function decreases,
yielding long exponential tails into the leads. 
This can be clearly seen for the quasi-bound state represented 
in Fig. \ref{1QW_R1nm_psi1_m1_n1},
whose quasi-bound state is just $0.92meV$ below the subband minimum
$E^{(1)}_{\perp,2}=0.394eV$, see Fig. \ref{1QW_R1nm_T1_m1}. 
Since the localization probability density enters the 
quantum calculation of the charge distribution
one gets difficulties in setting the correct boundaries for the 
Hartree calculations, i.e. Schr\"odinger-Poisson system. 
This has to be studied in a future work.

\begin{figure}[htbp!]
\subfigure[$E=0.217eV$, $n=1$]{\label{QW_wb015_m0_ev1_j1}
             \includegraphics[width=2.5in]{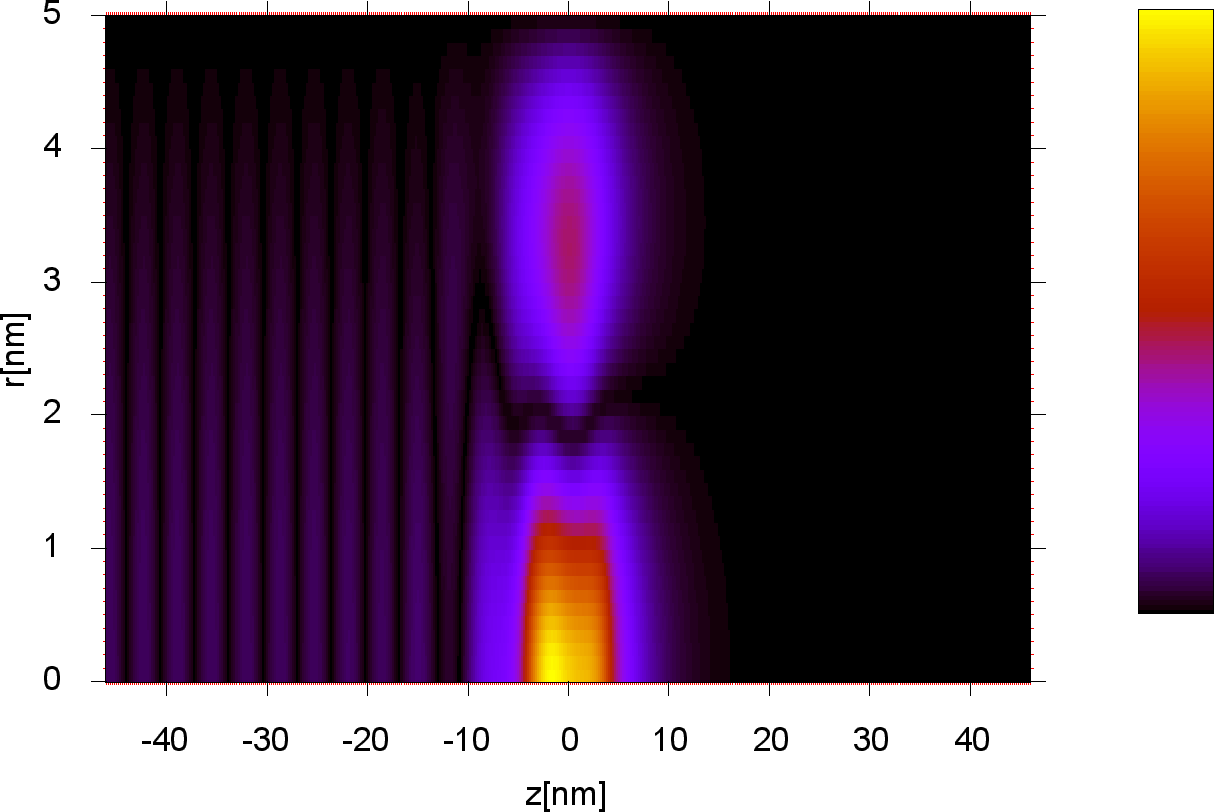}}

\subfigure[$E=0.243eV$, $n=1$]{\label{QW_wb015_m0_ev2_j1}
             \includegraphics[width=2.5in]{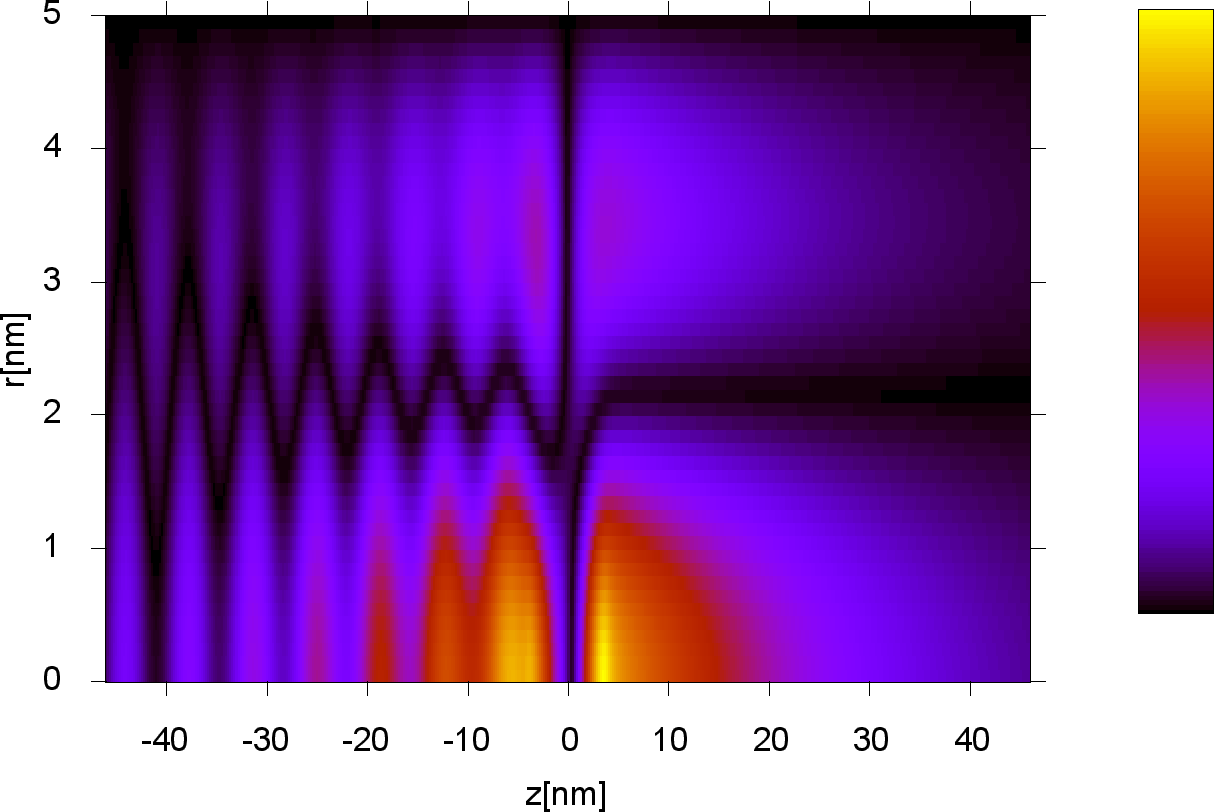}}

\subfigure[$E=0.576eV$, $n=2$]{\label{QW_wb015_m0_ev3_j2}
             \includegraphics[width=2.5in]{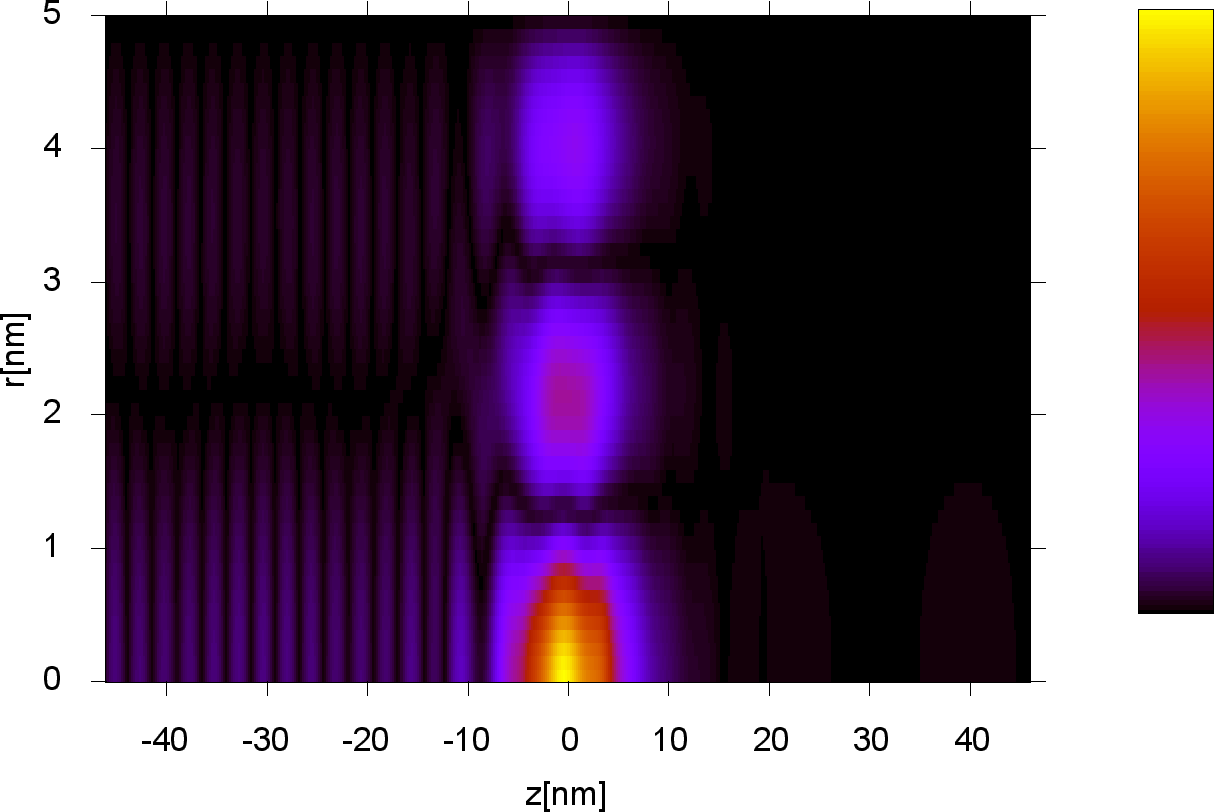}}

\subfigure[$E=0.600eV$, $n=2$]{\label{QW_wb015_m0_ev4_j2}
             \includegraphics[width=2.5in]{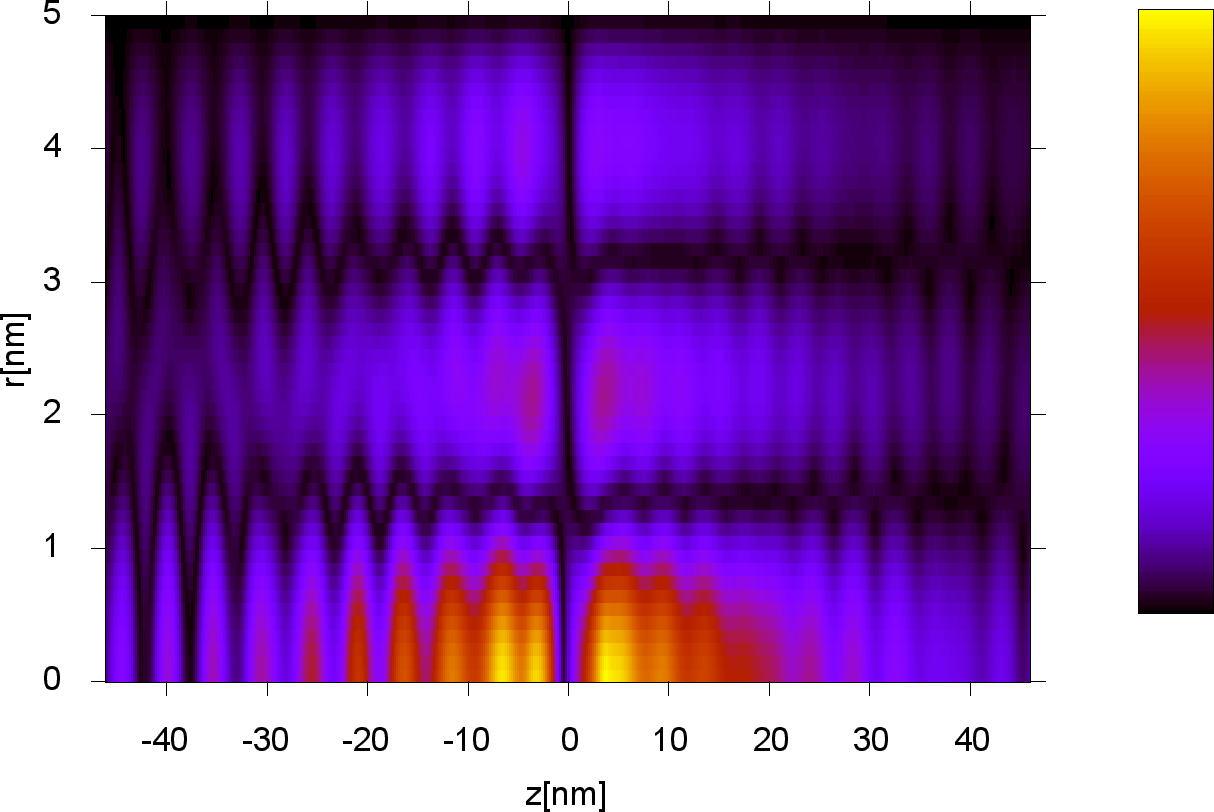}}
\caption{(color online)
Localization probability density $|\psi^{(1)}_{nm}(E,r,z)|^2$
for an electron with $m=0$,
incident from reservoir $s=1$ into channel $n$ and with energy $E$,
both indicated  in the captions.
The energies correspond to the dips in Fig. \protect\ref{1QW_R1nm_T1}
for $W_b=-0.15eV$.
}
\label{1QW_R1nnm_wb015_psi_m0}
\end{figure}

Increasing the strength of the attractive potential
to $W_b=-0.15eV$ one can see more dips\cite{gudmundsson04}
in the tunneling coefficient 
in Fig. \ref{1QW_R1nm_T1}. Interesting is that there are two dips
in the first and second plateau for $m=0$,
while for $|m|=1$ there is only one dip in every plateau.
This can be easily understood
if one thinks at the effective attractive potential $V_{nn}$,
Eq. (\ref{eff_attr}), created for every subband $n$. In the case of $m\not=0$,
the transversal modes $\phi^{(m)}_n(r)$ 
are zero on the cylinder axis, $r=0$, so that
the effective potential for every subband is weakened.
To confirm that the dips correspond to higher-order quasi-bound states, 
we plot in Fig. \ref{1QW_R1nnm_wb015_psi_m0}
the probability density of the scattering states at the energies
corresponding to the two dips in every plateau for $m=0$.
For the dips on the first plateau, Figs. \ref{QW_wb015_m0_ev1_j1} and
\ref{QW_wb015_m0_ev2_j1}, both scattering wave functions
have a node in $r$-direction corresponding to the transversal channel $n=2$.
For the dips on the second plateau, 
Figs. \ref{QW_wb015_m0_ev3_j2} and \ref{QW_wb015_m0_ev4_j2}, 
the wave functions 
have 
two nodes in $r$-direction
corresponding to the transversal channel $n=3$.
But looking in the $z$-direction,
the scattering state for the lower energy dip in every plateau
is node-less, 
while the one for higher-energy dip in every plateau
has a node at $z=0$,
which is evidence of the second quasi-bound states of the next 
evanescent channel.

\subsection{Core/shell quantum ring}

\begin{figure}[ht!]
\subfigure[]{\label{QRing_sketch}
           \includegraphics[width=3.00in]{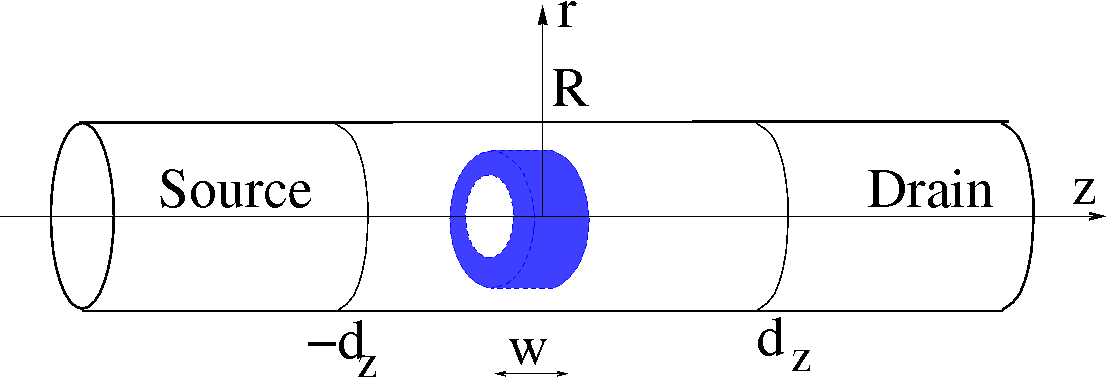}}
\subfigure[]{\label{QRing_pot}
           \includegraphics[width=3.00in]{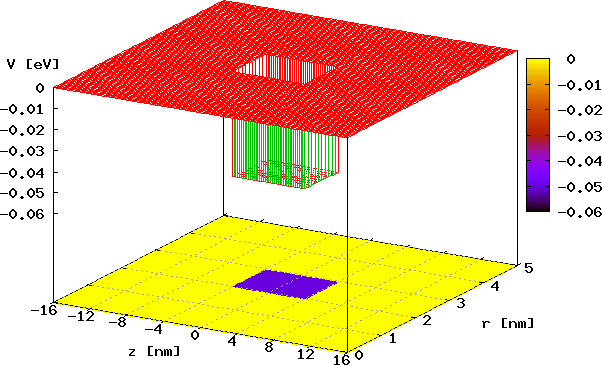}}
\caption{(color onine) 
a) Sketch of a quantum ring surrounded by the host 
material. We consider that the ring yields and attractive potential $V(r,z)$
represented in b) by an off-centered rectangular quantum well
of depth $W_b=-0.05eV$. 
The thickness of the ring is 1nm, between $R_1=2$nm and $R_2=3$nm,
and the width of the ring is $8$nm.
The radius of the cylinder is $R=5$nm. 
}
\label{QRing_vb}
\end{figure}

Now, consider the same rectangular quantum well but off-centered. 
This would corespond to  \textit{a quantum ring}
embedded into the nanocylinder, as sketched in Fig. \ref{QRing_sketch} 
and could be realised in a 
core-shell heterostructure with suplimentar structuring along
the nanowire.

\begin{figure}[ht!]
\includegraphics[width=2.75in]{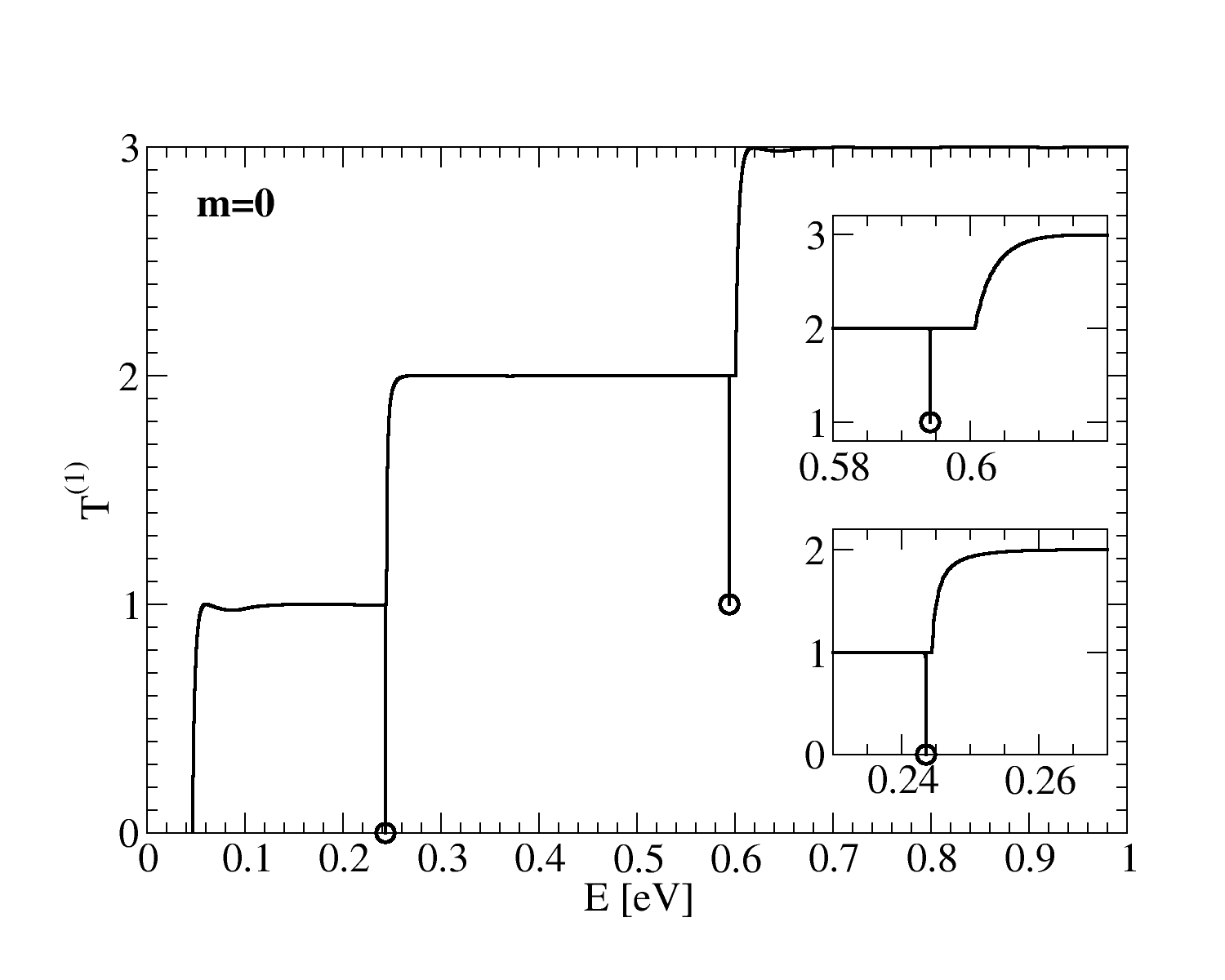}
\caption{Total tunneling coefficient as function of incident energy $E$
for magnetic quantum numbers $m=0$
for a ring surrounded by the host material.
The structure parameters are as in Fig. \protect\ref{QRing_vb}.
The symbols show the energies, at which the wave functions are analyzed
in the next graphs.
}
\label{QRing_R2_3nm_T1}
\end{figure}

\begin{figure}[h!]
\subfigure[$E=0.243eV$, $n=1$]{\label{QR_psim_m0_min1}
             \includegraphics[width=2.5in]{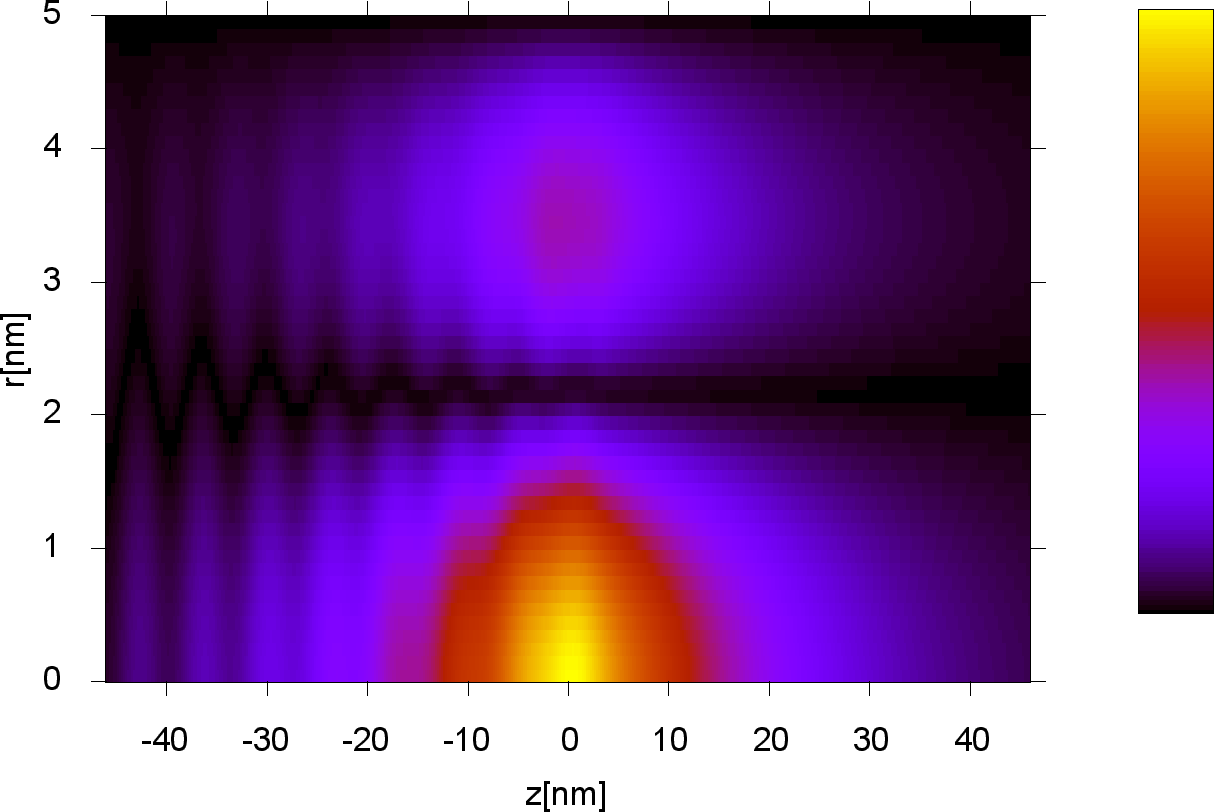}}

\subfigure[$E=0.594eV$, $n=2$]{\label{QR_psim_m0_min2}
             \includegraphics[width=2.5in]{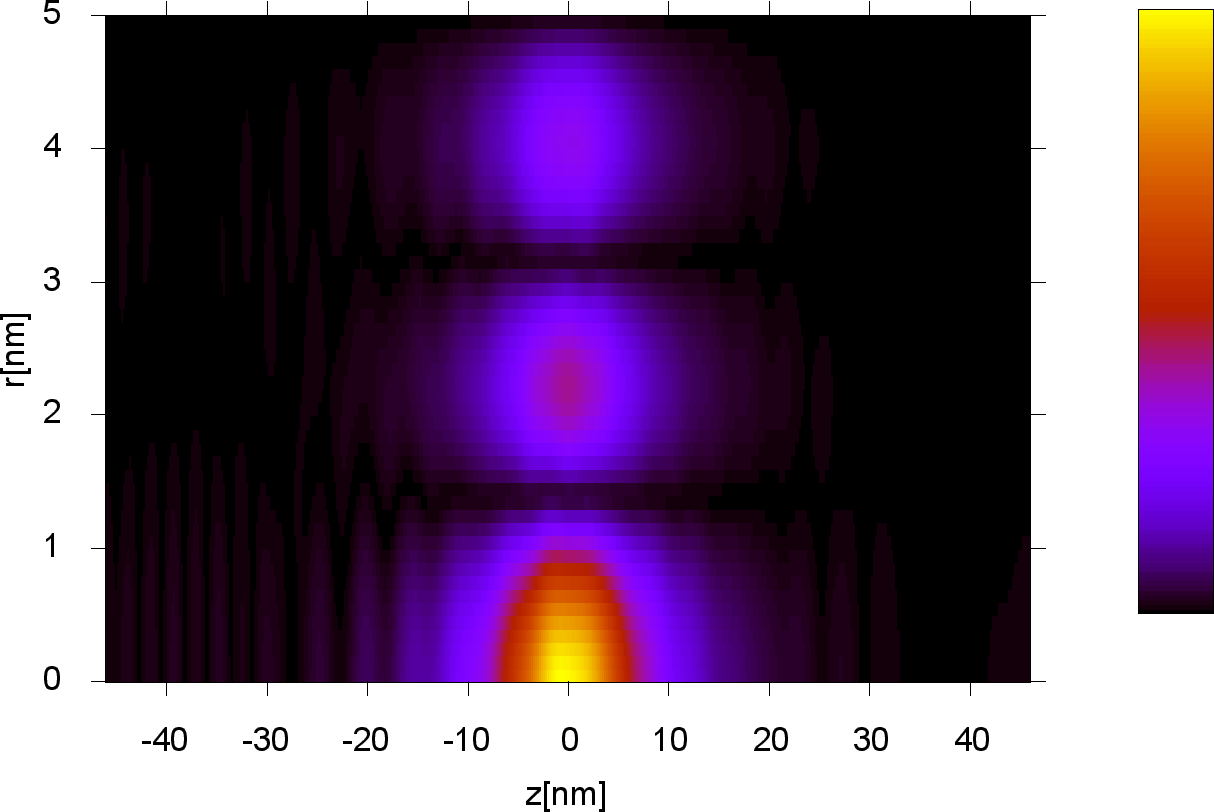}}
\caption{(color online)
Localization probability density $|\psi^{(1)}_{nm}(E,r,z)|^2$ 
for an electron with $m=0$, 
incident from reservoir $s=1$ into channel $n$ and with energy $E$ 
indicated  in the captions.
The energies correspond to the symbols in Fig. \protect\ref{QRing_R2_3nm_T1}.
}
\label{QR_psim_m0}
\end{figure}

\begin{figure}[htb!]
\includegraphics[width=3.00in]{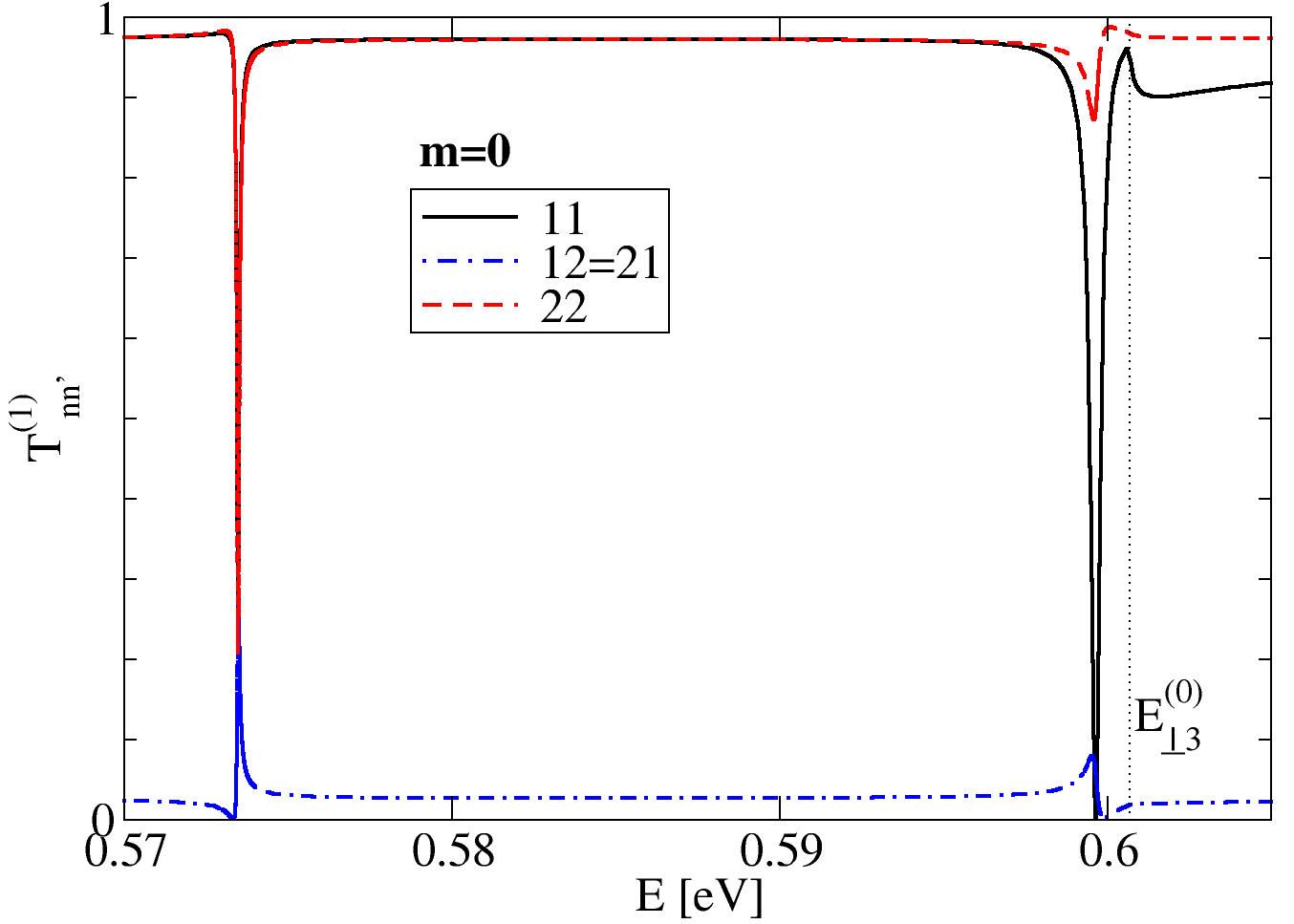}
\caption{(color online)
Intrasubband and intersubband transmission probabilities
for a quantum ring 
(Fig. \protect\ref{QRing_vb}) with $W_b=-0.15eV$ and for $m=0$.
The curves for $T^{(1)}_{12}$ and $T^{(1)}_{21}$ coincide.
The vertical doted line shows the third subband minimum $E^{(0)}_{\perp 3}$.
}
\label{QRing_R2_3nm_wb015_T1_ij_m0}
\end{figure}

The tunneling coefficient $T^{(1)}(E)$ for $m=0$ is plotted in 
Fig. \ref{QRing_R2_3nm_T1}, showing the characteristic dips due
to the quasi-bound states of the evanescent channels.
The localization probability density for the energies marked with
symbols in Fig. \ref{QRing_R2_3nm_T1} are plotted in Fig. \ref{QR_psim_m0}.
By shifting the potential from the cylinder axis and keeping the same
parameters as for the quantum dot surrounded by the host material,
one can recognize the same behavior of the wave functions corresponding 
to the quasi-bound states as in the previous case. 
This means that the quasi-bound states of an evanescent channel
extend over the whole width of the nanowire,
independent where the scattering potential is located in the lateral 
direction. Similar results hold for $m\not=0$.
\begin{figure}[ht!]
\includegraphics[width=3.00in]{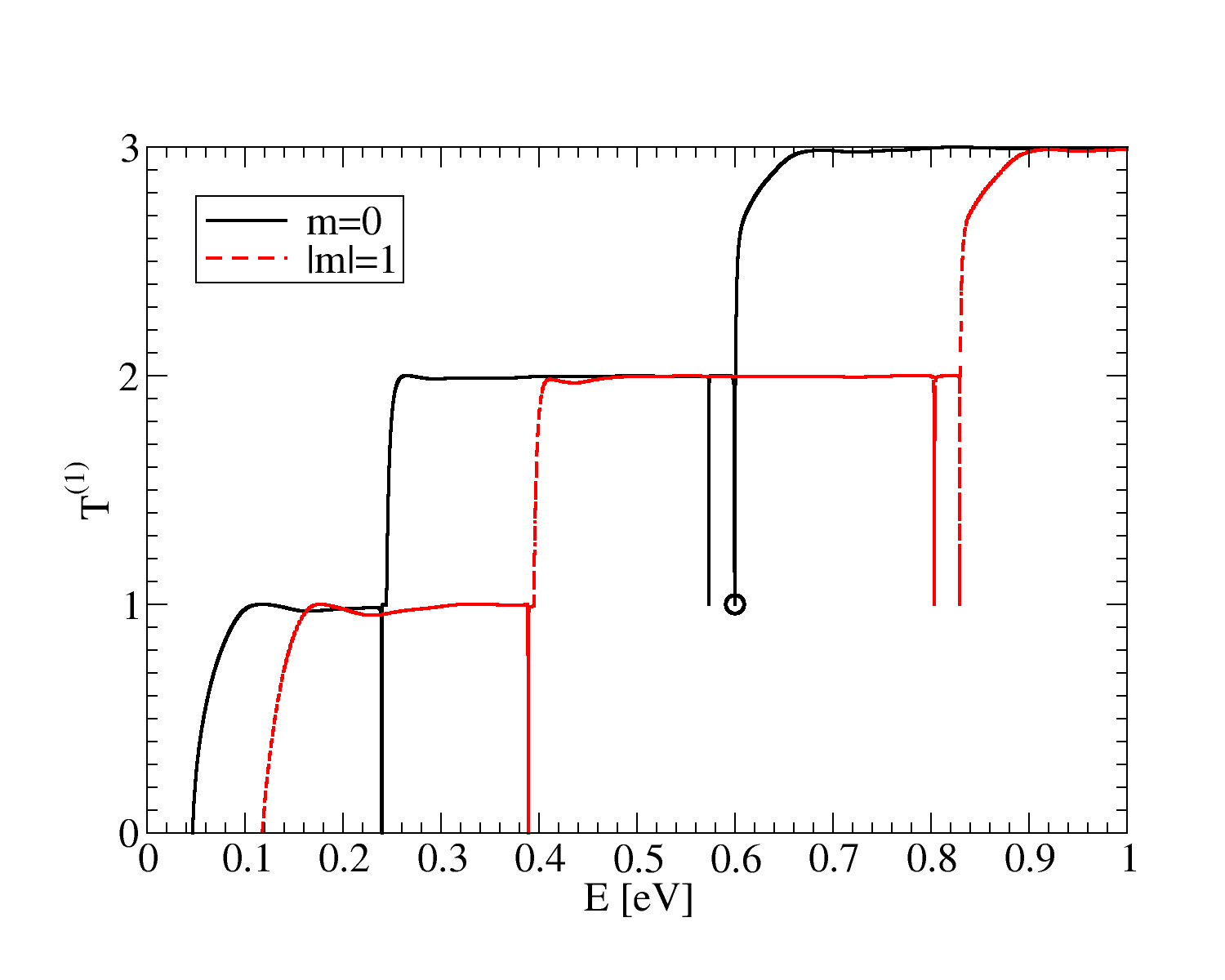}
\caption{(color online)
The tunneling coefficient $T^{(1)}$
for different magnetic quantum numbers $m$
for a ring surrounded by the host material.
The parameters are as in Fig. \protect\ref{QRing_vb}, but $W_b=-0.15eV$.
The symbols show the energies, at which the wave functions are analyzed
in the next graphs.
}
\label{QRing_R2_3nm_wb015_T1}
\end{figure}
Considering a deeper quantum well
one is surprised to see in Fig. \ref{QRing_R2_3nm_wb015_T1} that
for $m=0$ only one dip appears in the first plateau but two dips in
the second plateau. This can be understood considering that
if the transversal channels $\phi^{(m)}_n(r)$ has a node
at the off-centered position of the scattering potential, then $V_{nn}$, 
Eq. (\ref{eff_attr}), is being weakened 
allowing for less quasi-bound states.
\begin{figure}[hb!]
\subfigure[$E=0.599eV$, $n=1$]{\label{QR_wb015_psim_m0_j1}
          \includegraphics[width=2.25in]{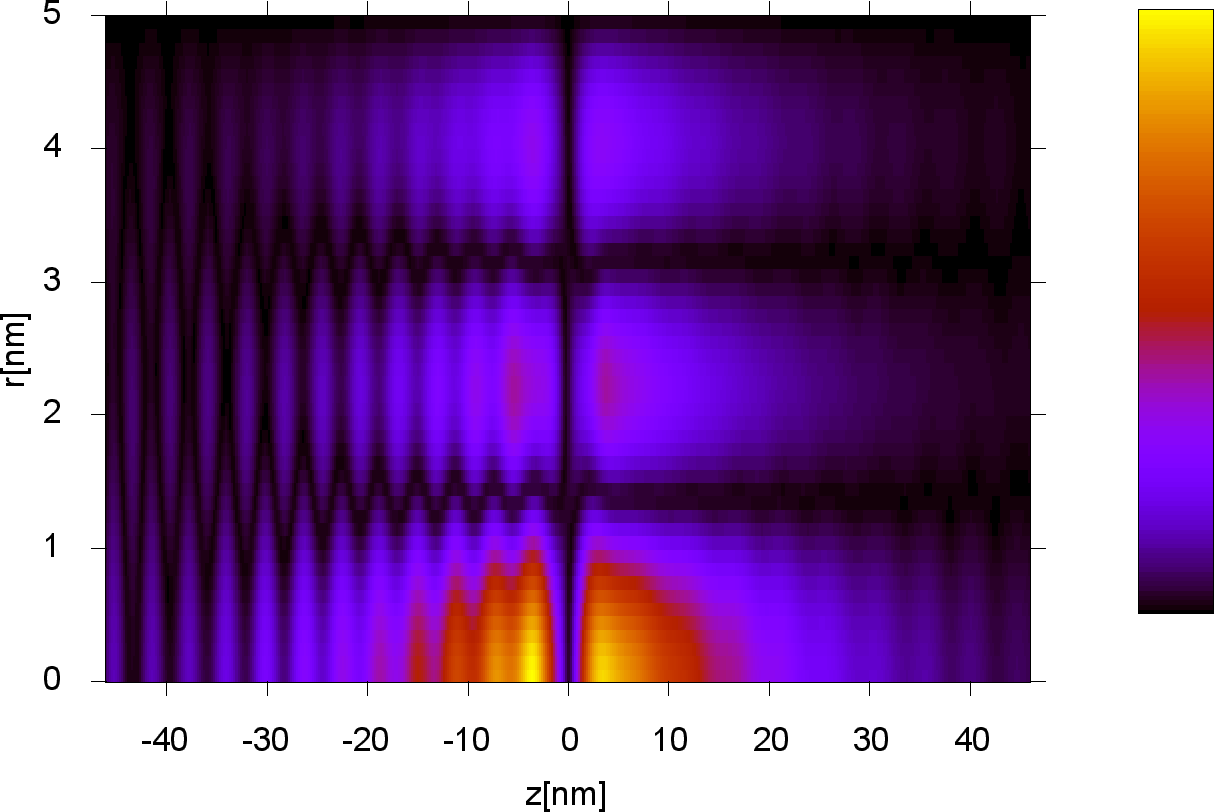}}

\subfigure[$E=0.599eV$, $n=2$]{\label{QR_wb015_psim_m0_j2}
          \includegraphics[width=2.25in]{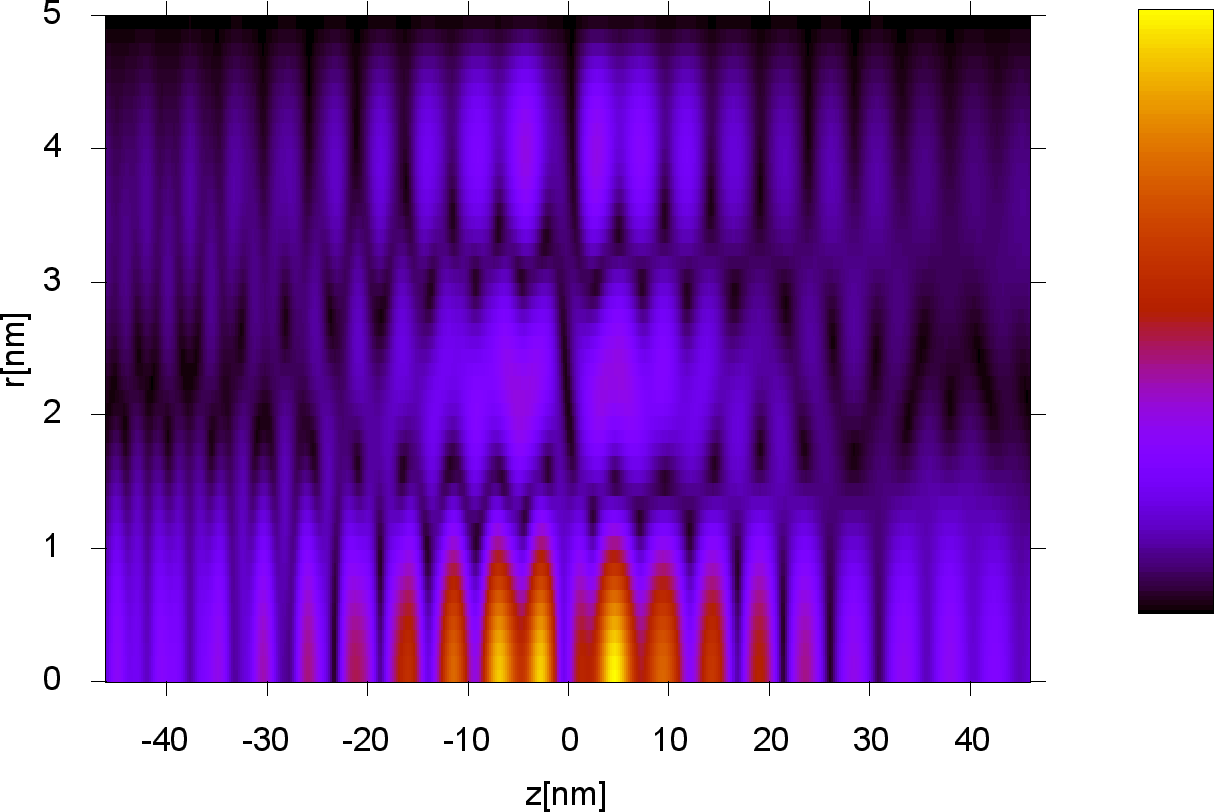}}
\caption{(color online)
Localization probability density $|\psi^{(1)}_n(E,r,z)|^2$
for an electron with $m=0$, 
incident from left into both open channels $n=1$ (a) and $n=2$ (b)
and with energy $E$ 
corresponding to the second quasi-bound state on the second plateau
in Fig. \protect\ref{QRing_R2_3nm_wb015_T1}.
}
\label{QR_wb015_psim_m0}
\end{figure}
Interesting is to take a closer look at the scattering states corresponding
to the second quasi-bound state on the second plateau,
marked with a symbol
in Fig. \protect\ref{QRing_R2_3nm_wb015_T1}.
For this energy, there are two
open channels, and we have represented in Fig. \ref{QR_wb015_psim_m0}
the localization probability density for both of them. 
One can recognize
immediately the structure of the wave function
with two nodes in $r$-direction 
corresponding to the third evanescent channel and 
one node in $z$-direction specific to the second
quasi-bound state. Specific for the quasi-bound states of an 
evanescent channel is also the exponential decaying far from the scattering 
potential. 
But the interference patterns on the left and on the right
of the scattering potential are quite 
different for these two scattering wave functions.
This can be explained by looking at intrasubband  and intersubband
transmission probabilities represented in 
Fig. \ref{QRing_R2_3nm_wb015_T1_ij_m0},
and which give detailed information about channel mixing. 
The intrasubband transmission $T^{(1)}_{11}$ is stronger
influenced by channel mixing, showing two pronounced dips, 
while the intrasubband transmission $T^{(1)}_{22}$ shows only one dip
and a second asymmetric Fano line \cite{noeckel94,roxana01}.
Both intersubband transmission probabilities $T^{(1)}_{12}$ and $T^{(1)}_{21}$
coincide and show asymmetric Fano lines with zero minima.
Now it is clear that there is no interference pattern 
to the right of the quantum ring
in Fig. \ref{QR_wb015_psim_m0_j1}
because both transmission probabilities
$T^{(1)}_{11}$ and $T^{(1)}_{12}$ are zero for the second quasi-bound state.
One recognizes in Fig. \ref{QR_wb015_psim_m0_j1} for the first channel
strong interference pattern between the incident part and
the reflected part.
For the scattering wave function incident on the second channel
$T^{(1)}_{22}$ has values close to $1$, and also $T^{(1)}_{21}$ has a maximum.
In such a way one sees in Fig. \ref{QR_wb015_psim_m0_j2}
right to the scatterer 
an interference pattern between the transmitted
wave in the first channel and the one transmitted in the second channel. 
One can recognize 
far from the scattering potential the
structure of the second channel with a node in the $r$-direction.

\subsection{Double-barrier heterostructure along the nanowire}

\begin{figure}[htb]
\subfigure[]{\label{NWTDB_sketch}
                    \includegraphics[width=3.25in]{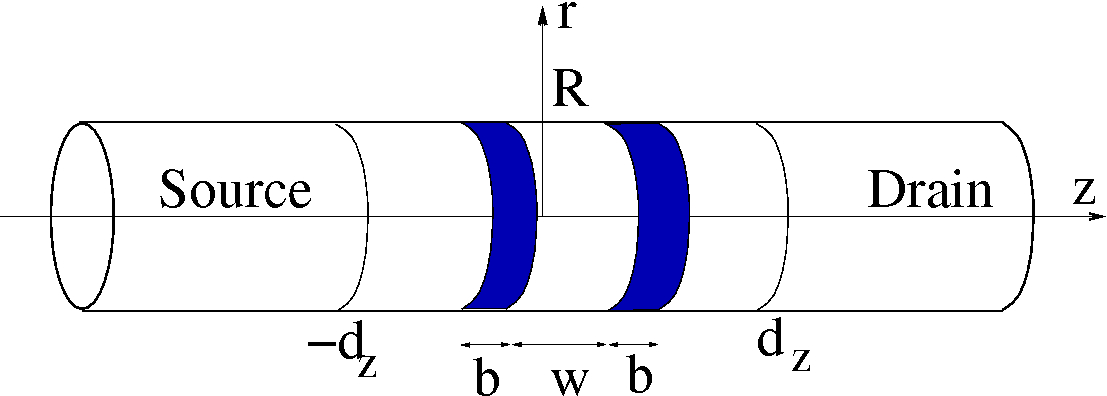}}
\subfigure[]{\label{NWTDB_pot}
                    \includegraphics[width=3.25in]{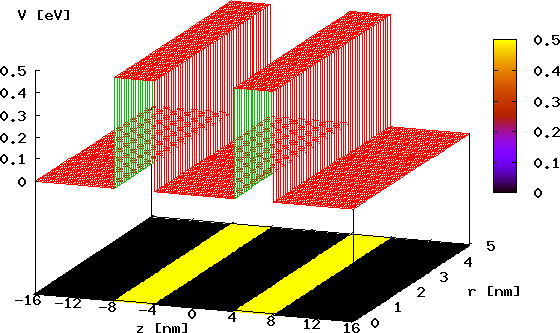}}
\caption{(color online)
a) The sketch of a double-barrier heterostructure along the nanowire 
and b) the scattering potential $V(r,z)$. 
The height of the barriers is $V_b=0.5eV$, 
the widths of the barriers is $b=4$nm, 
the width of the quantum well is $w=8$nm and the radius of the nanowire
is $R=5$nm. 
}
\label{NWTDB_vb}
\end{figure}

\begin{figure}[h]
\subfigure[linear scale]{\label{NWTDB_T1_m0_li}
                    \includegraphics[width=3.in]{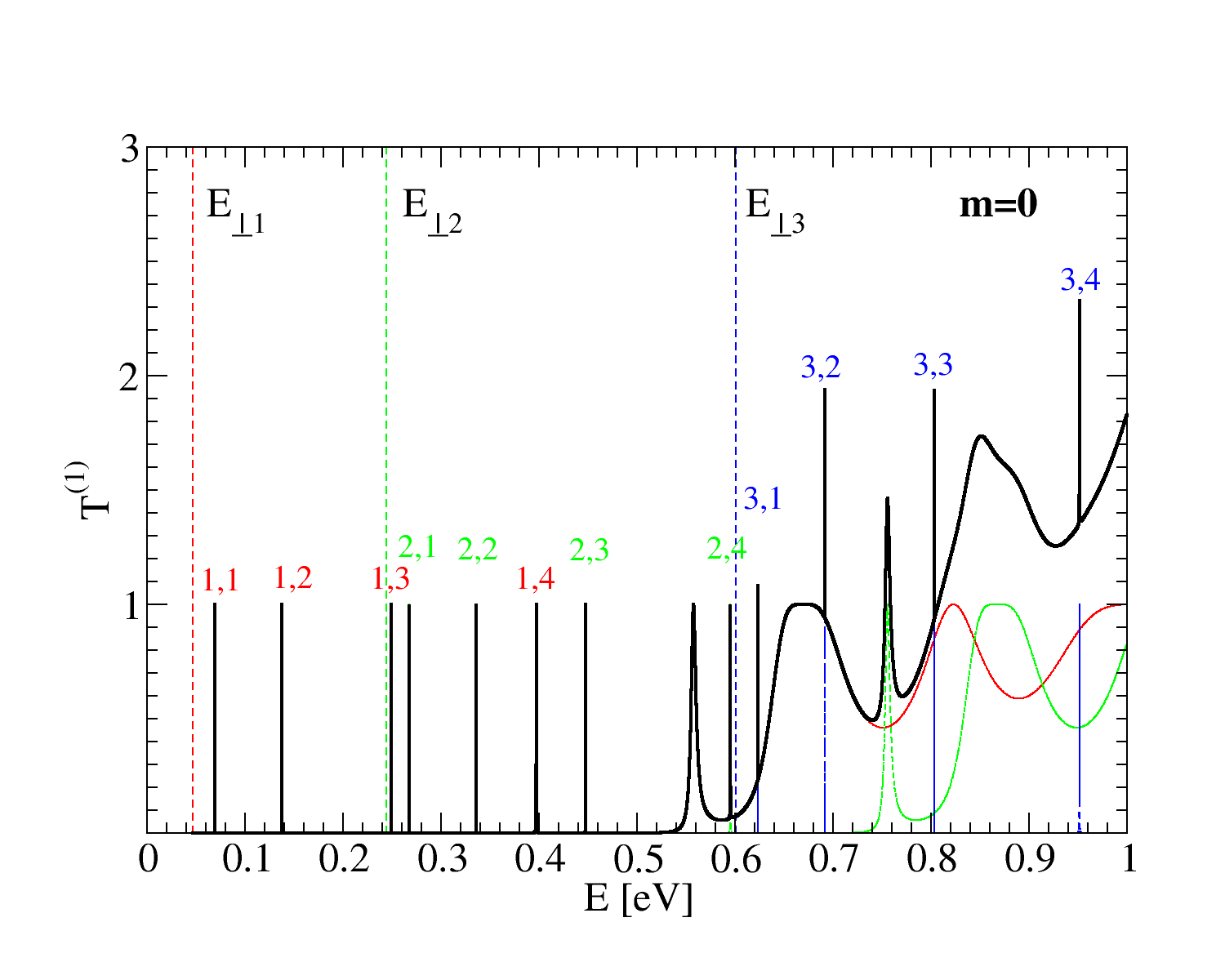}}

\subfigure[logarithmic scale]{\label{NWTDB_T1_m0_ln}
                    \includegraphics[width=3.25in]{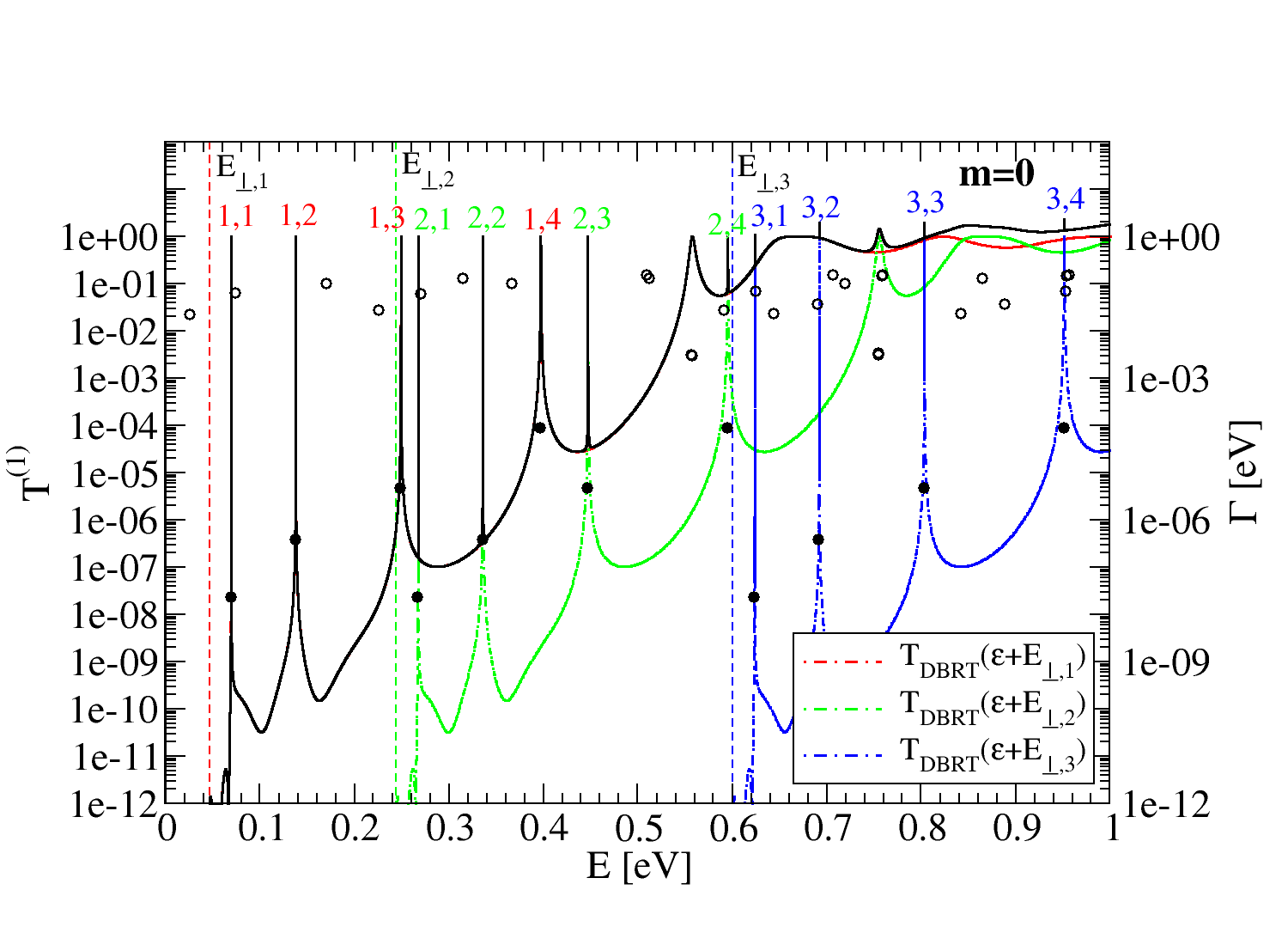}}
\caption{(color online)
Total tunneling coefficient 
in linear scale (a) and logarithmic scale (b),
as function of incident energy $E$ for
a double-barrier heterostructure along the nanowire,
as depicted in Fig. \protect\ref{NWTDB_vb}. The symbols represents 
the poles: their real part on x-axis and the imaginary part on right y-axis.
The peaks are indexed by $(n,i)$, where $n$ denotes the incident channel
and $i$ denotes the resonance between the barriers.
The same indexes are used in Table \protect\ref{NWTDB_psim_m0}.
}
\label{NWTDB_T1_m0}
\end{figure}

In Fig. \ref{NWTDB_sketch} is sketched 
a double-barrier heterostructure along the cylindrical
nanowire. 
Such systems with sharp interfaces between the layers
are realized experimentally based on InAs/InP \cite{samuelson02} 
or on GaAs/AlGaAs \cite{wensorra}. We consider rectangular 
barriers of height $V_b=0.5eV$,
widths $b=4$nm, and the width of the rectangular quantum well $w=8$nm, 
as is plotted in Fig. \ref{NWTDB_pot}.
The total tunneling coefficient $T^{(1)}(E)$ for $m=0$ is plotted in 
Fig. \ref{NWTDB_T1_m0_li} in linear scale. 
The barriers supprese the transmission, 
except for a series of sharp peaks due to the quasi-bound states
between the barriers.
With vertical dashed lines we have represented the energies of the
first three transversal channels $E^{(0)}_{\perp,n}$, $n=1,2,3$.
One can observe that the tunneling coefficient can reach values higher than 1, 
if there are more channels open. 
In case of no applied bias between source and drain contact,
the scattering potential is separable and has variations only in 
$z$-direction $V(r,z)=V(z)$, where $V(z)$ describes
an 1D double-barrier potential. The scattering does not mix the
channels, so that the total tunneling coefficient is given by summation of
the intrasubband transmission probabilities for every open channel
$T^{(1)}(E)=\sum_n T^{(1)}_{nn}(E)$. 
The intrasubband transmission probability 
on every open channel is the transmission
through a double-barrier structure, but shifted with the transversal energy of
the channel $E^{(m)}_{\perp,n}$. So that it can be also computed
as for a double-barrier resonant tunneling (DBRT) diode,
$T^{(1)}_{nn}(E)=T_{DBRT}(\epsilon+E^{(m)}_{\perp,n})$. 

This identity can be used as a verification of the numerical implementation
of our method, because the first quantity is computed with the 2D code,
while the second quantity is computed with the 1D code\cite{racec02}.
This is explicitly illustrated in a logarithmic plot
of the tunneling coefficient in Fig. \ref{NWTDB_T1_m0_ln}. 
We represent here by vertical dashed lines the positions
of the first three transversal channels $E^{(0)}_{\perp,n}$,  $n=1,2,3$. 
By dot-dashed
lines we have represented the tunneling coefficient through the DBRT,
but the energies are
shifted with the transversal channel energy, as is written in the legend.
The curve for the first channel (red dotted line) 
is just under the total transmission curve (black continuous line).
One can observe that the 1D double-barrier potential allows for four
resonances (quasi-bound states) 
between the barriers, i.e. every dashed curve has four
peaks below the barrier height. 

On the same plot we have plotted with symbols the poles of the current
scattering matrix $\tilde{\mathbf{S}}$, 
computed using the method presented in 
Ref.  [\onlinecite{roxana01}] 
recently developed for 2D geometries \cite{roxana08}.
The real part is on the axis of abscissae, while the imaginary part 
is on the right axis of ordinates.
One can see that among the poles there exist resonant ones, 
marked by filled symbols, 
with very low widths, i.e. $\Gamma < 10^{-4} eV$,
and which are very well separated from the others. 
Using the same axis of abscissae for the real part and for
the tunneling coefficient, one can directly see that to every
resonant pole corresponds a transmission peak. 
Increasing the energy, but keeping
the same channel, the widths of the poles increase, and so the widths
of the transmission peaks. 

This physical interpretation of the tunneling coefficient peaks
allows us to label them in Fig. \ref{NWTDB_T1_m0} 
by a pair of numbers $(n,i)$, where $n$ describes
the incident channel and 
$i$ describes the resonance (quasi-bound state) between the barriers.

\begin{table}[t!]
\begin{tabular}{c|c|c|c|c|}
(n,i) & i=1 & i=2 & i=3 & i=4 \\
\hline
 n=1 &\includegraphics[width=0.7in]{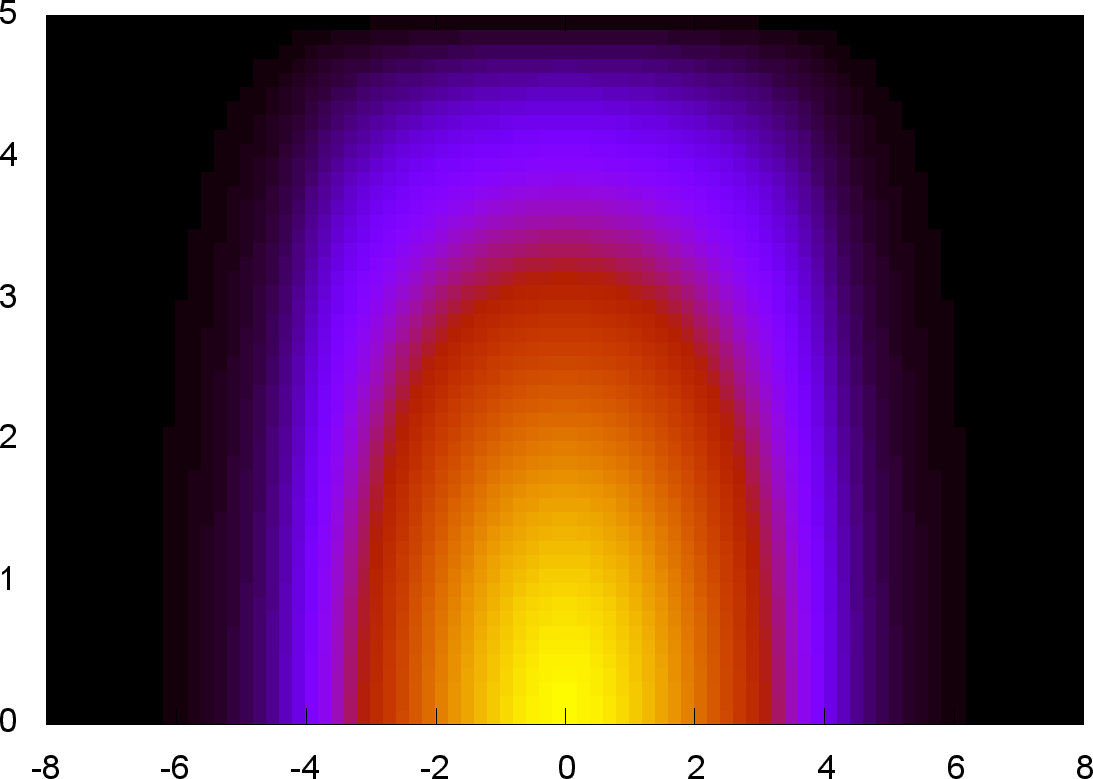}
     &\includegraphics[width=0.7in]{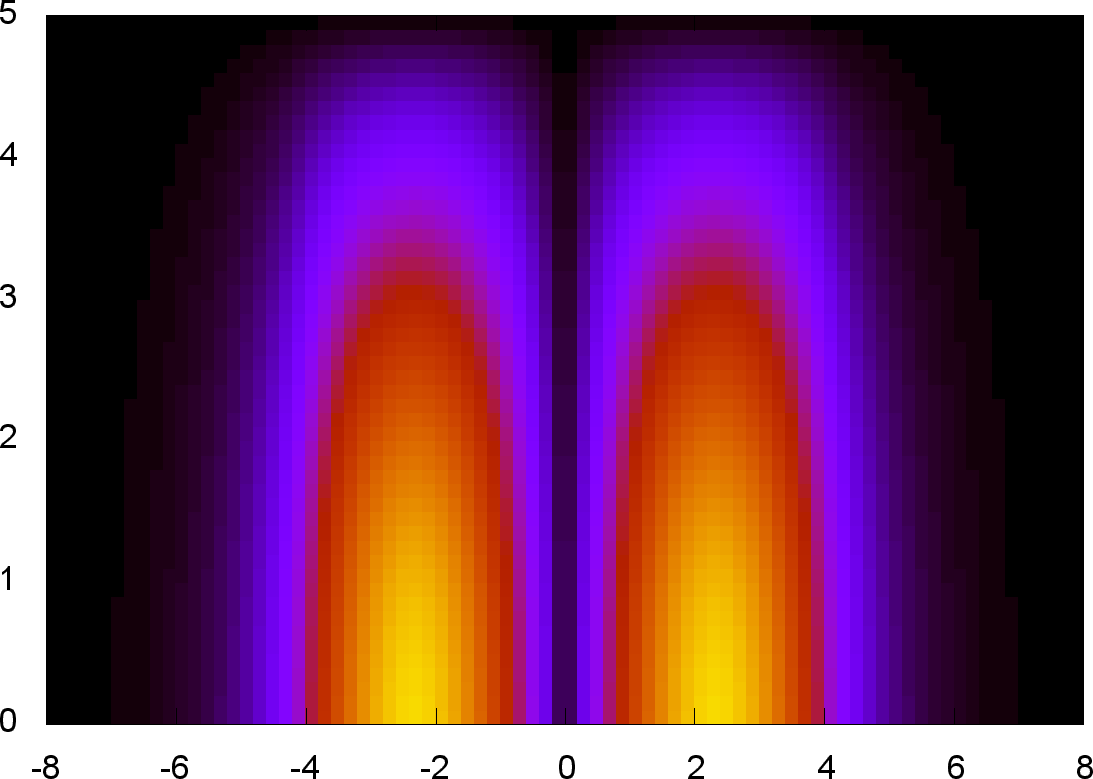}
     &\includegraphics[width=0.7in]{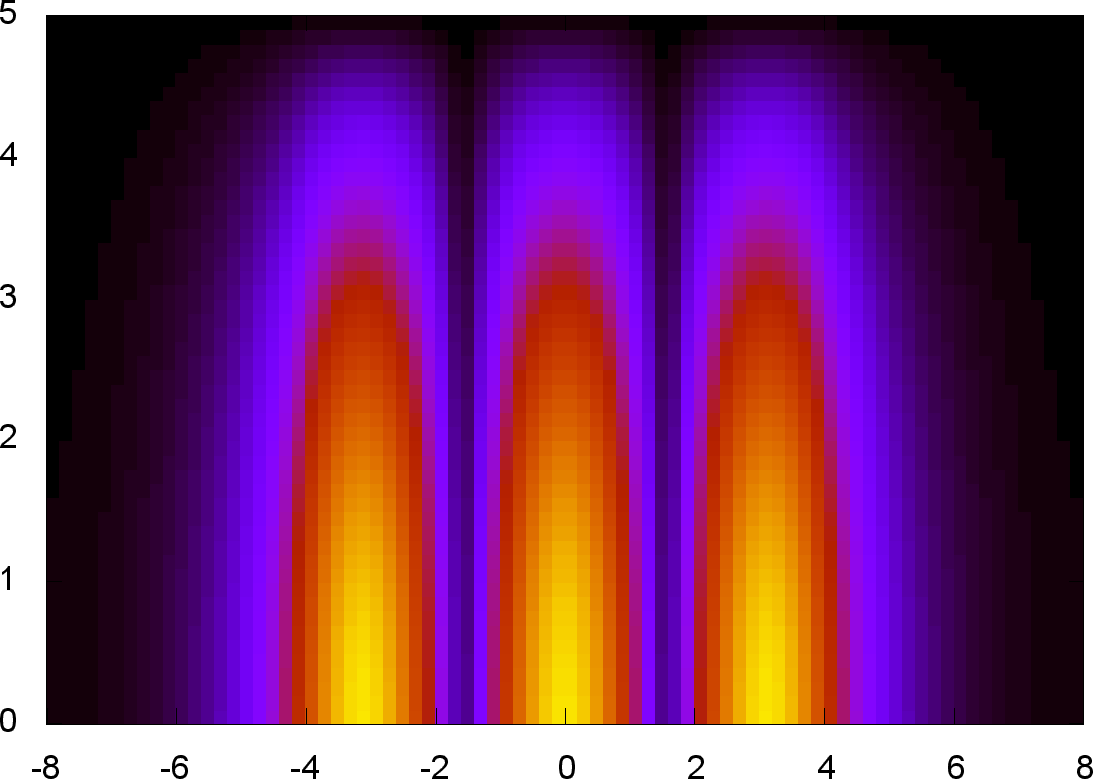}
     &\includegraphics[width=0.7in]{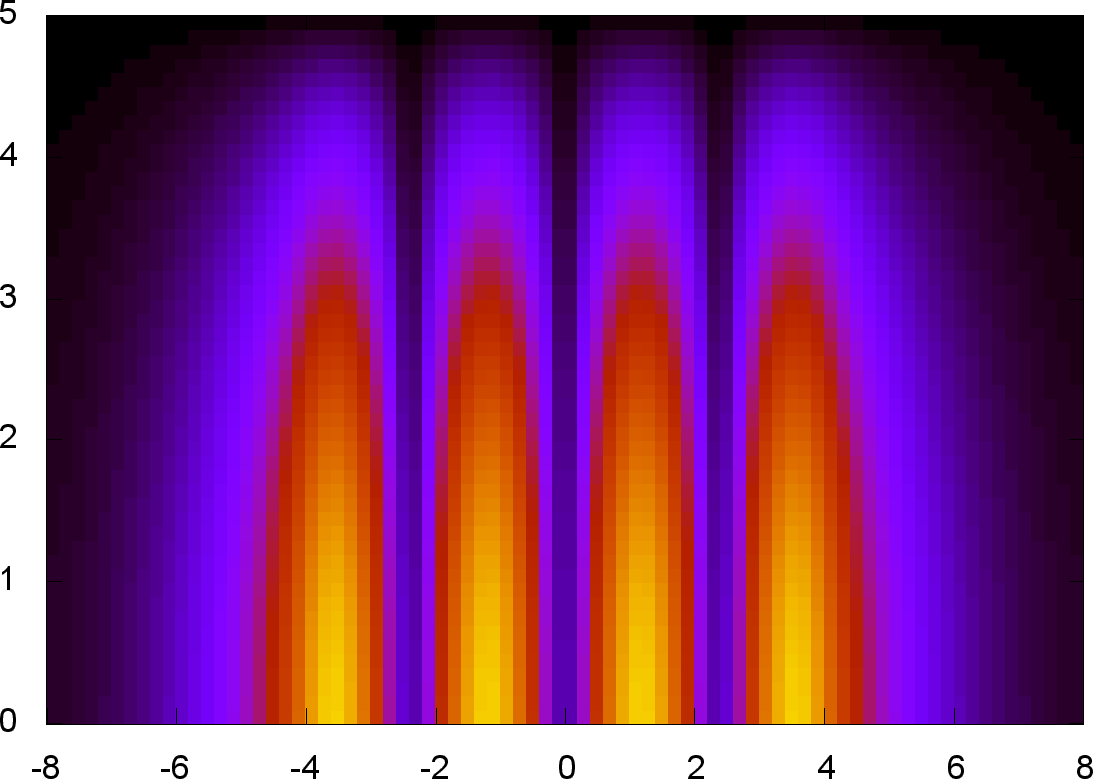} \\
\hline
 n=2 &\includegraphics[width=0.7in]{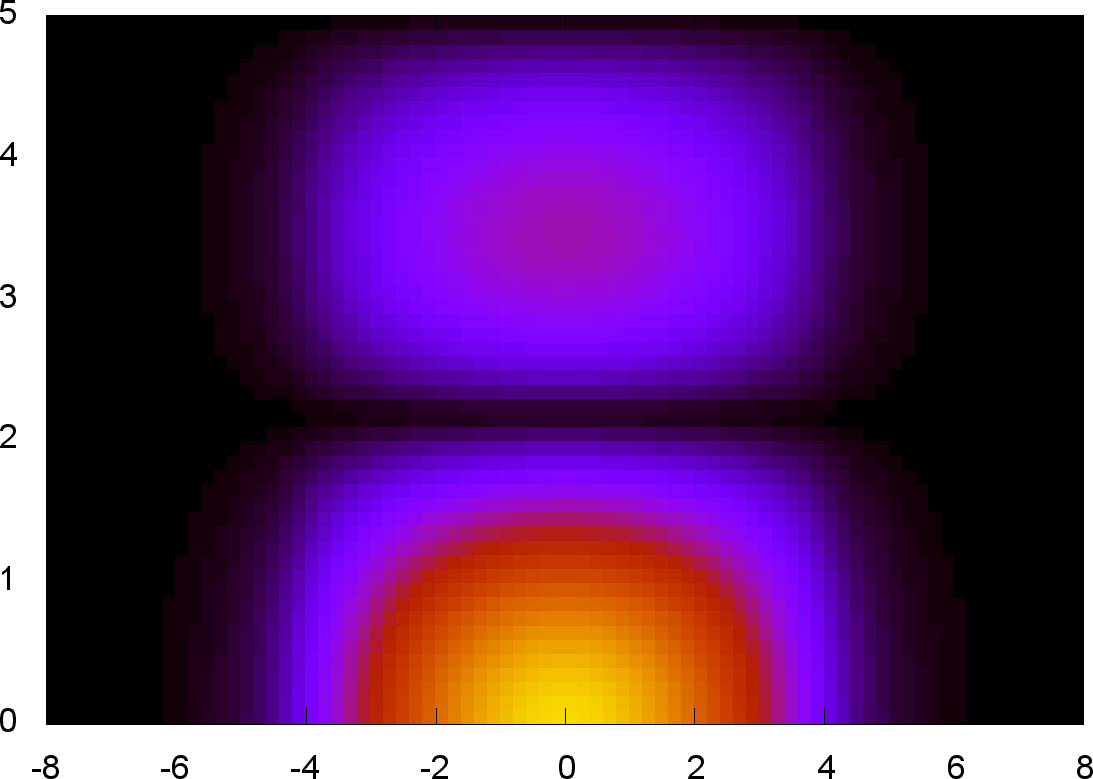}
     &\includegraphics[width=0.7in]{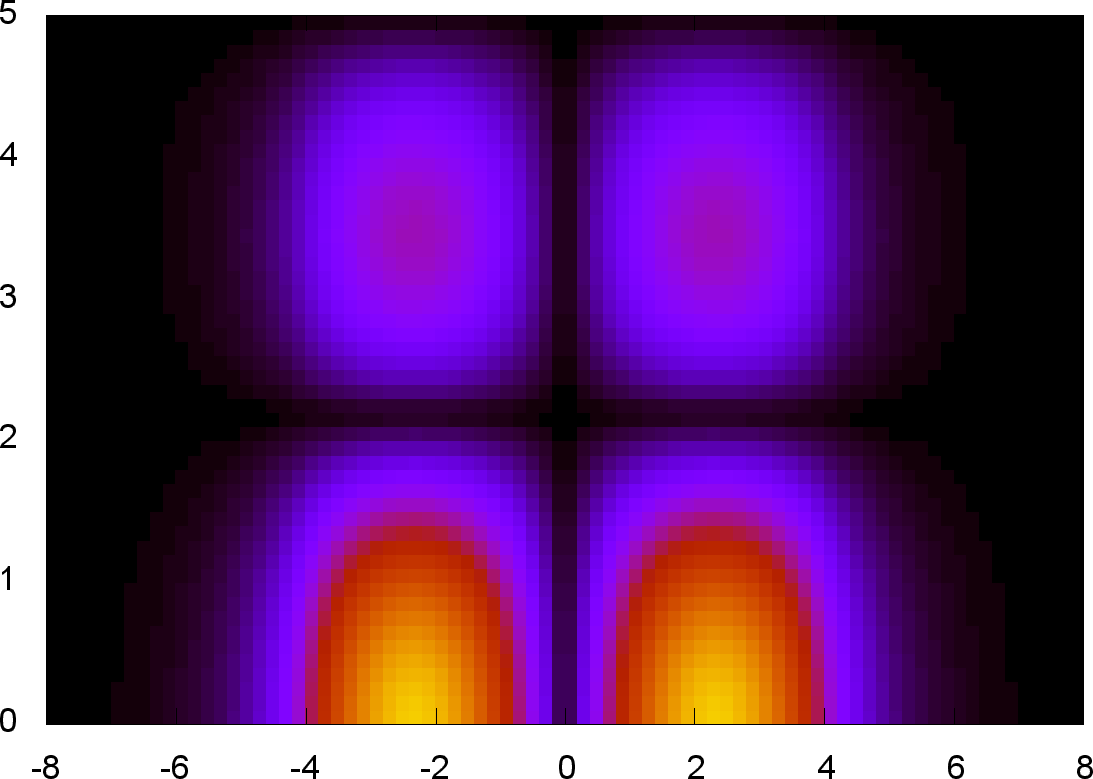}
     &\includegraphics[width=0.7in]{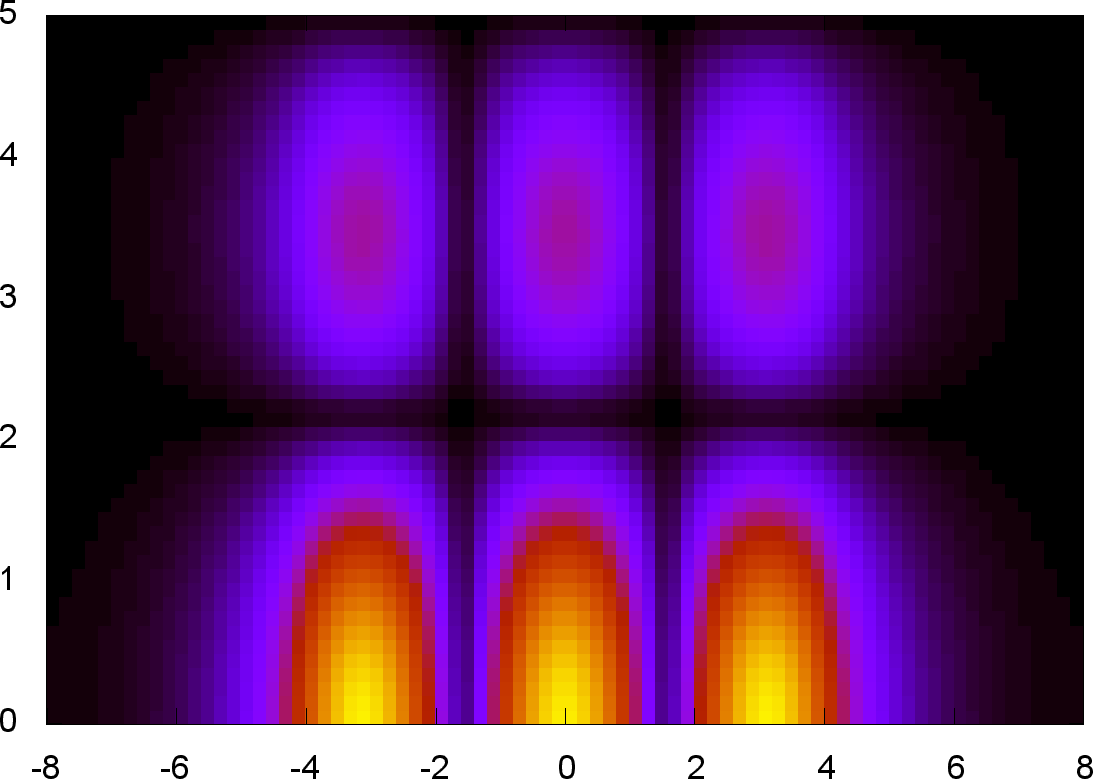}
     &\includegraphics[width=0.7in]{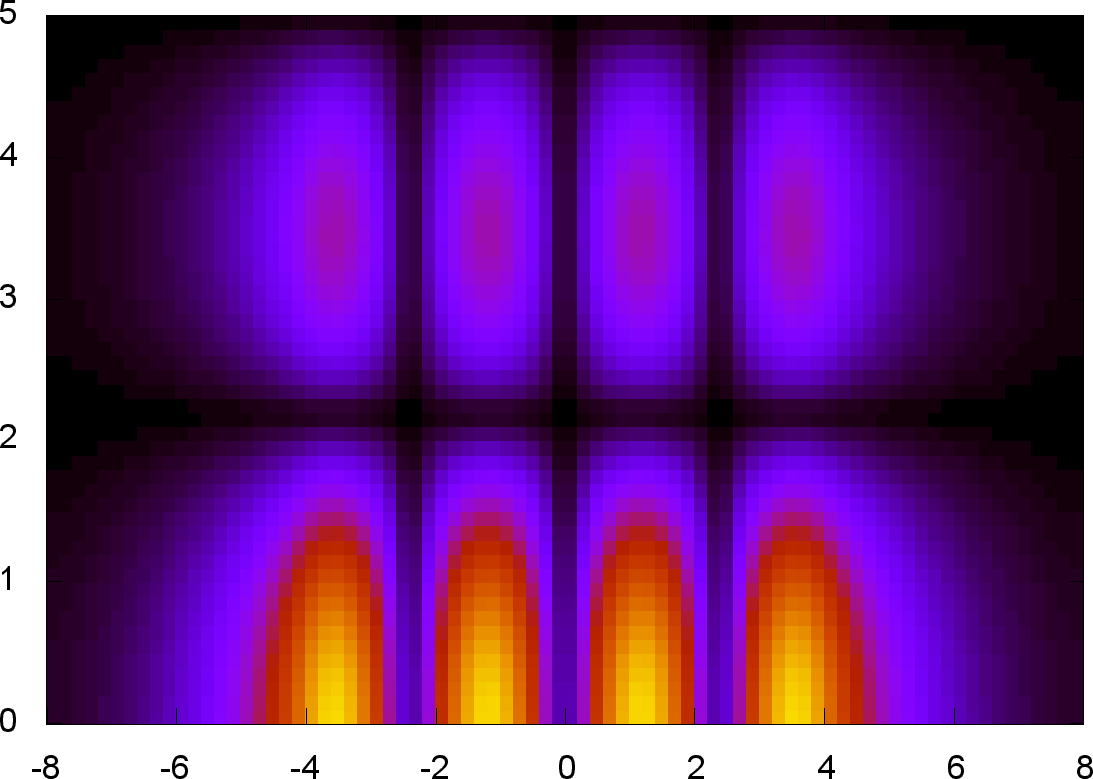} \\
\hline
 n=3 &\includegraphics[width=0.7in]{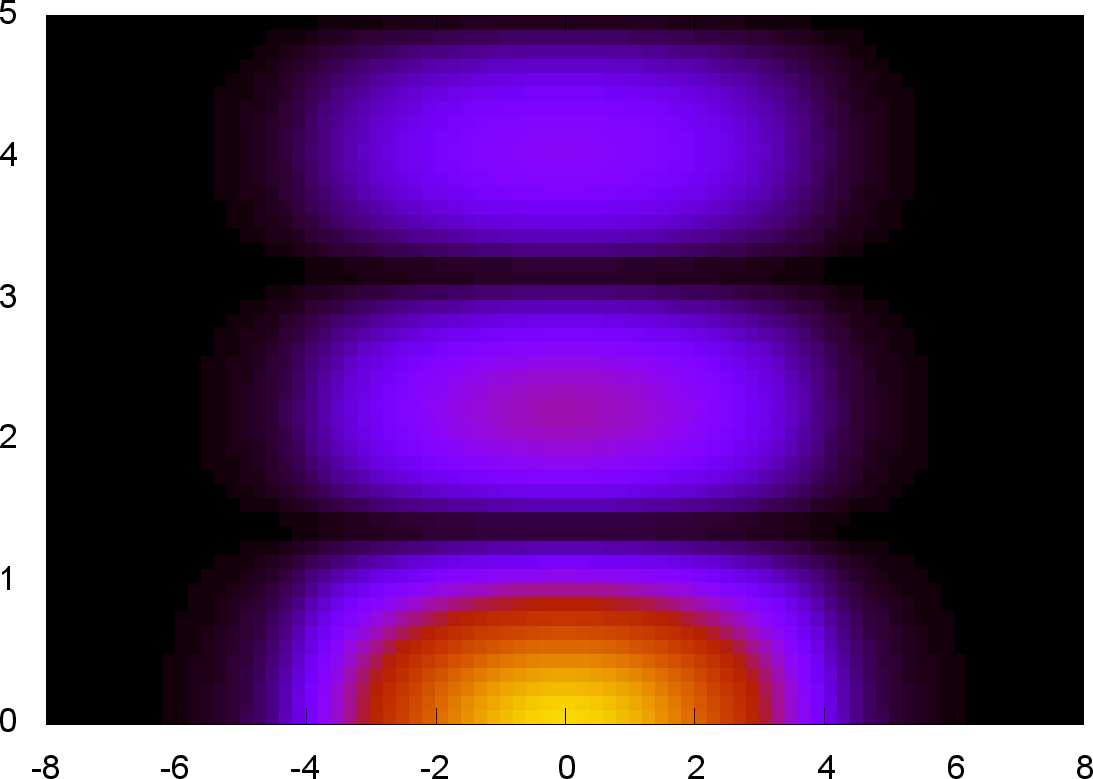}
     &\includegraphics[width=0.7in]{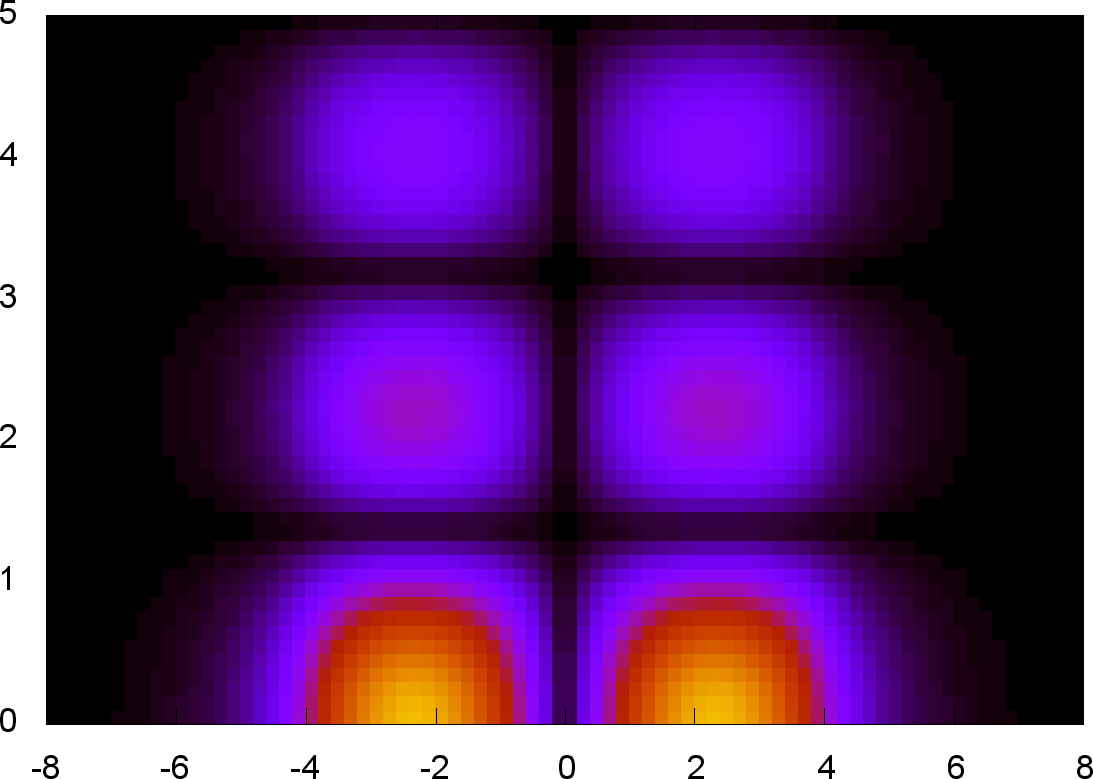}
     &\includegraphics[width=0.7in]{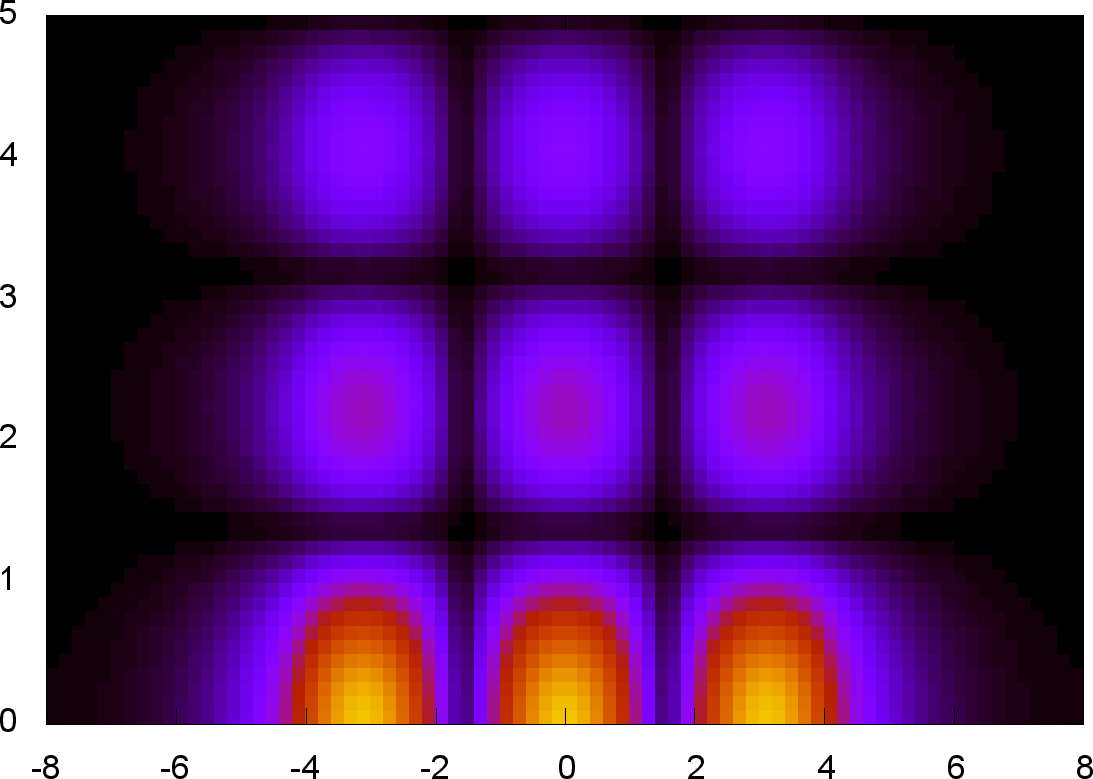}
     &\includegraphics[width=0.7in]{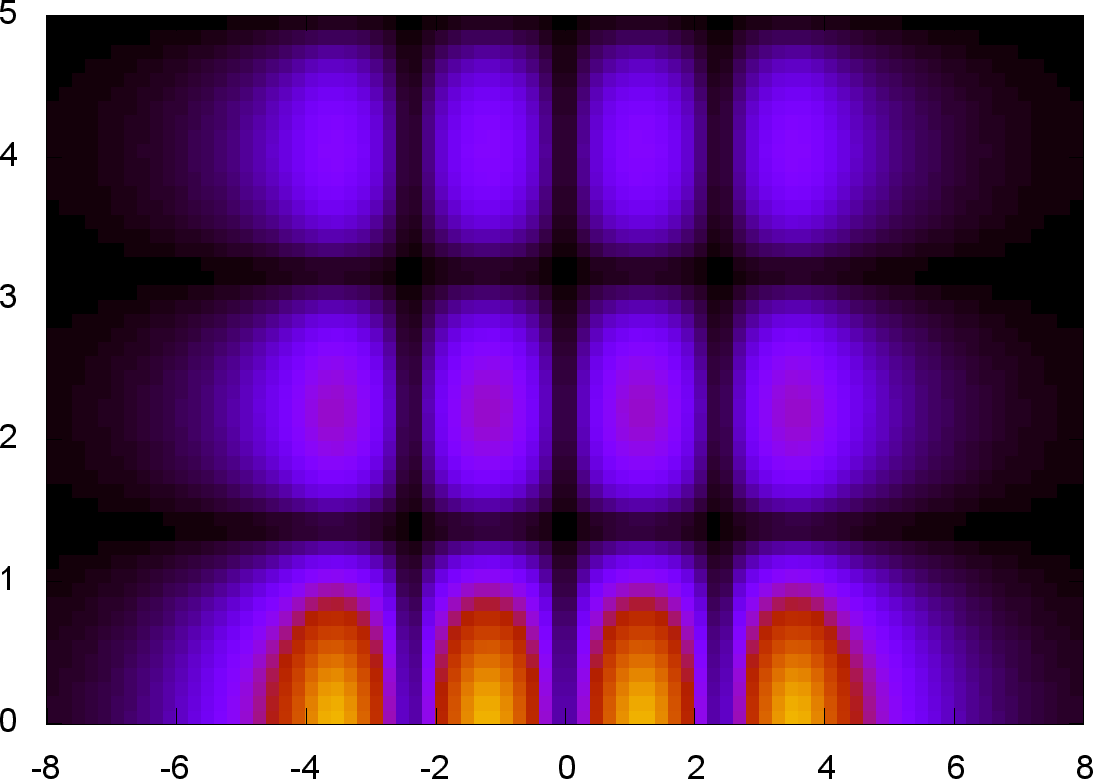} \\
\hline
\end{tabular}
\caption{Localization probability density $|\psi^{(1)}_n(E_{res,i},r,z)|^2$
for an electron with $m=0$ incident from reservoir $s=1$ into channel $n$ and 
corresponding to the resonance $i$ between the barriers.
The axis of abscissae is $z \in [-8,8]$nm, and 
the axis of ordinates is $r \in [0,5]$nm for all plots.}
\label{NWTDB_psim_m0}
\end{table}

In Table \ref{NWTDB_psim_m0} are represented the localization probability 
densities $|\psi^{(1)}_n(E_{res,i},r,z)|^2$
for the double-barriers region,
$z \in [-8,8]$nm  and $r \in [0,5]$nm,
for every indexed peak in Fig. \ref{NWTDB_T1_m0}. 
One can observe that the wave functions have pronounced maxima
between the barriers and decrease very quickly inside the barriers.
They are localized between the barriers, corresponding, indeed, to resonances
(quasi-bound states) between barriers and not to quasi-bound states
of evanescent channels.
Furthermore, the order $i$ of 
the resonance between the
barriers gives the number of nodes in the $z$-direction, namely $i-1$,
while the channel number $n$ gives the number of nodes in 
the $r$-direction, namely $n-1$.
This provides a picture of the orbitals of the "artificial atom",
which represents this quantum structure \cite{kouwenhoven01}.
\begin{figure}[hbt!]
\includegraphics[width=3.25in]{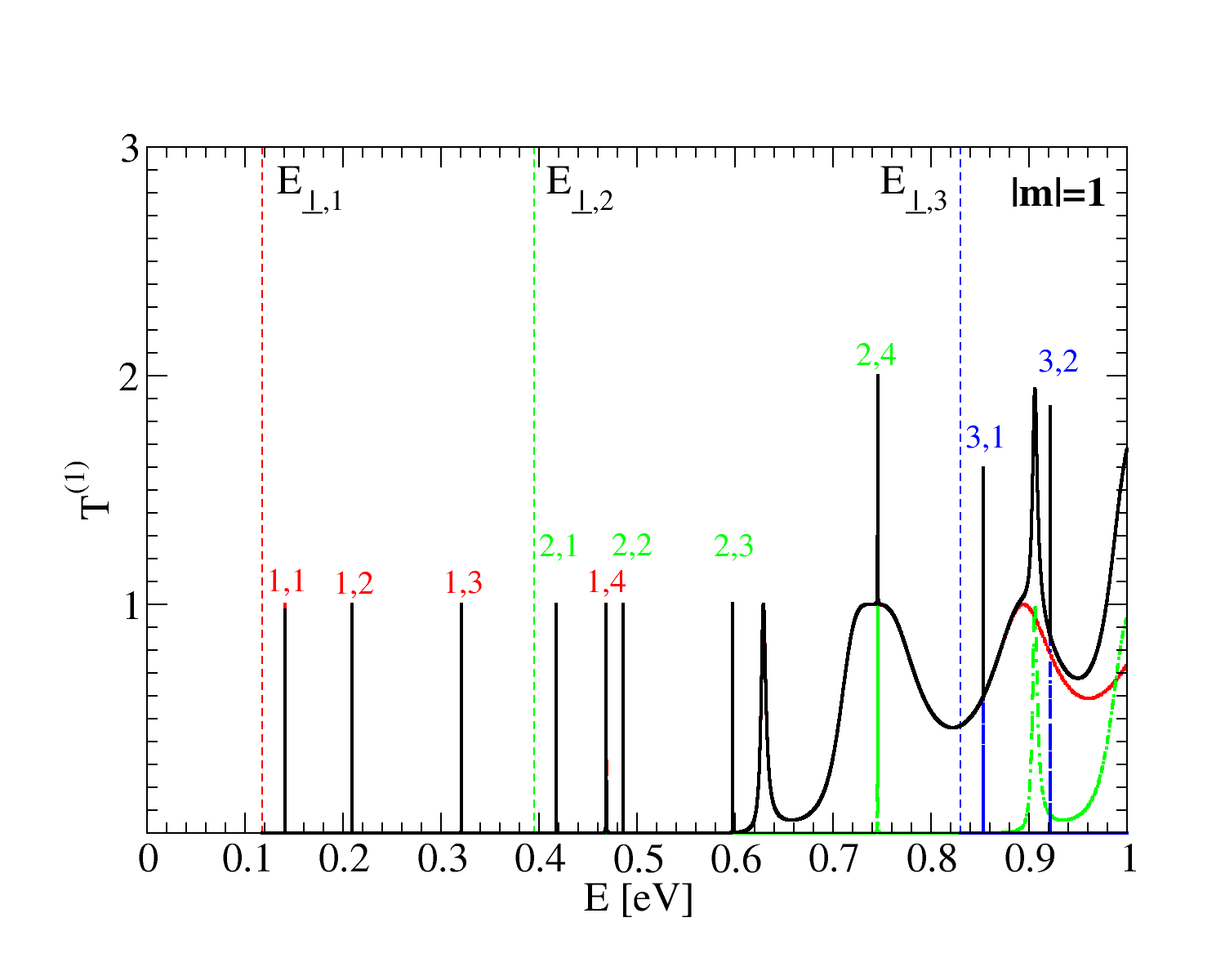}
\caption{(color online) Transmission coefficient $T^{(1)}$ for $|m|=1$
vs. total energy $E$ for
a double-barrier heterostructure along the nanowire
as depicted in Fig. \protect\ref{NWTDB_vb}. 
The peaks are indexed by $(n,i)$, where $n$ denotes the incident channel
and $i$ denotes the resonance between the barriers.
The same indexes are used in Table \protect\ref{NWTDB_psim_m1}.
}
\label{NWTDB_T1_m1}
\end{figure}

\begin{table}[b!]
\begin{tabular}{c|c|c|c|c|}
(n,i) & i=1 & i=2 & i=3 & i=4 \\
\hline
 n=1 &\includegraphics[width=0.7in]{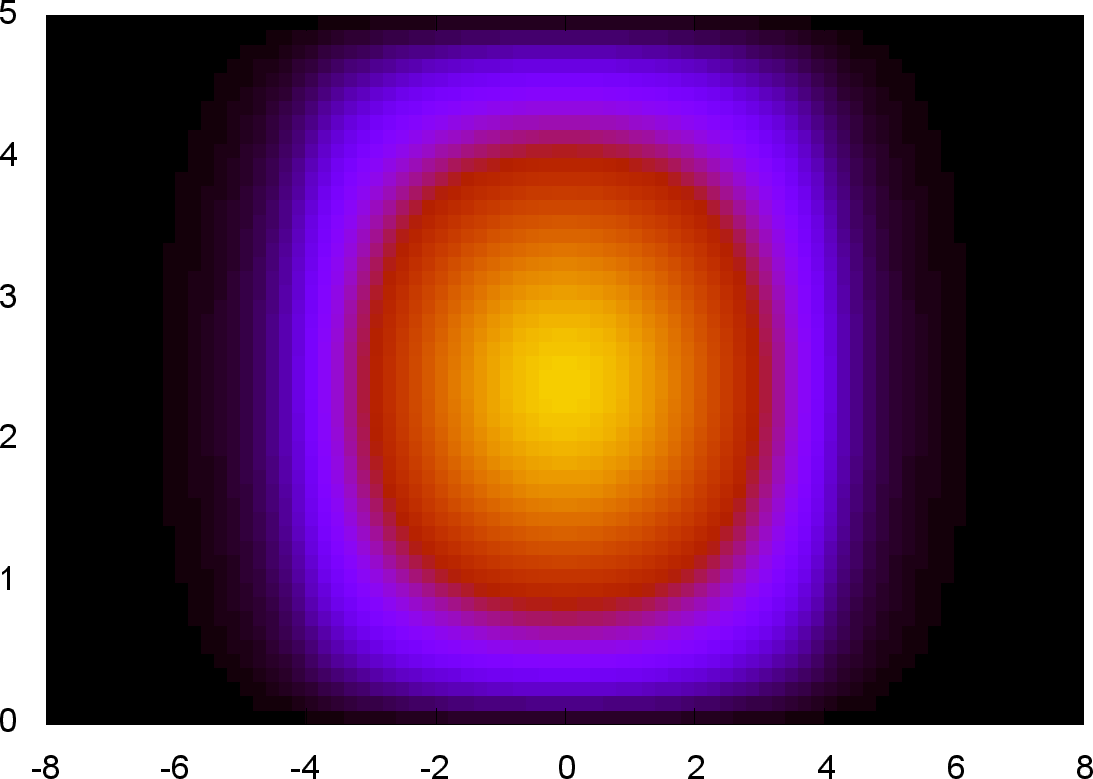}
     &\includegraphics[width=0.7in]{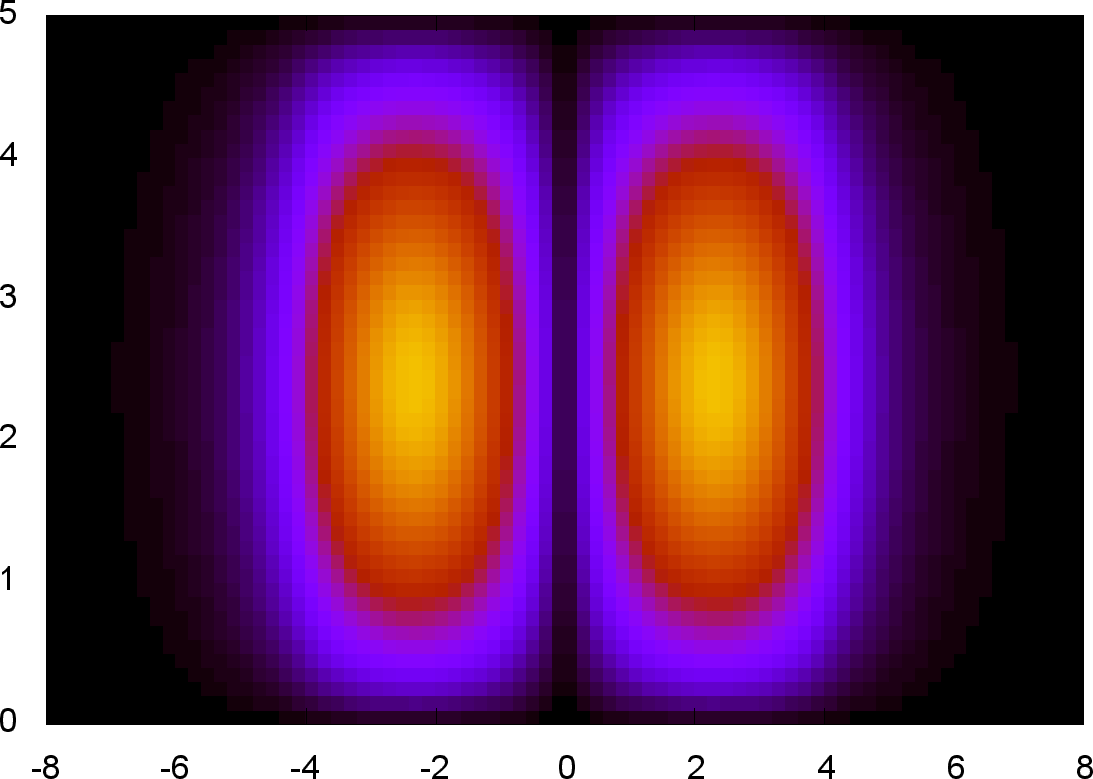}
     &\includegraphics[width=0.7in]{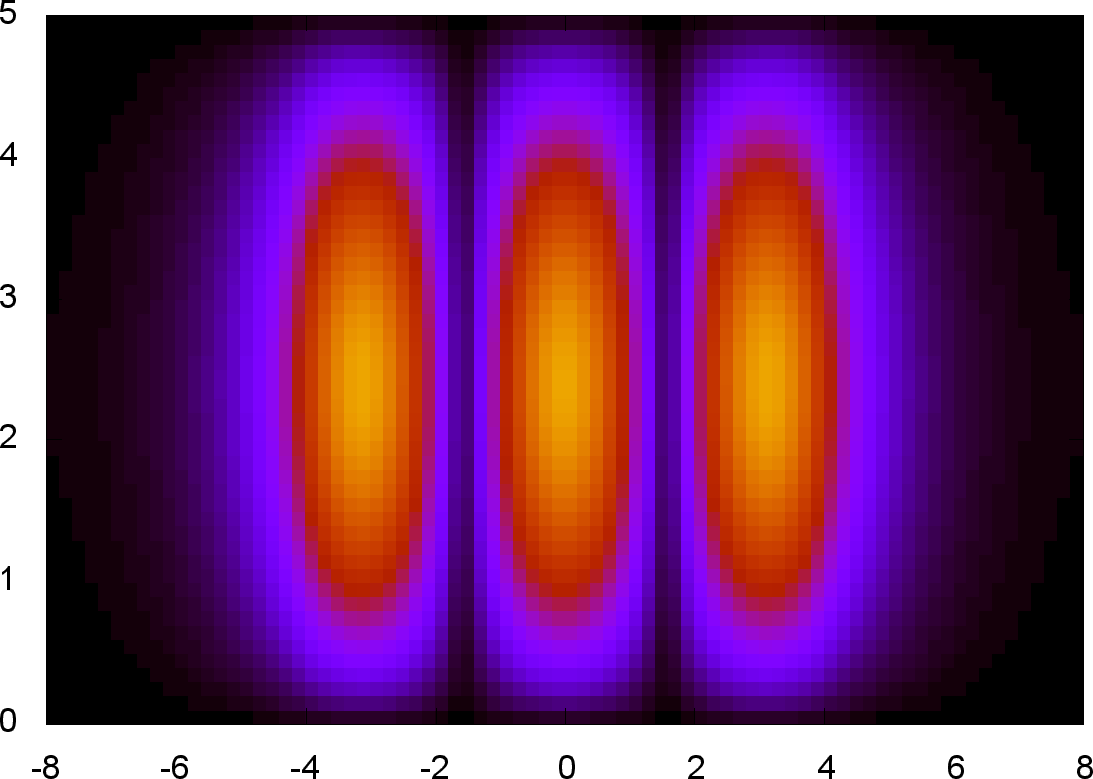}
     &\includegraphics[width=0.7in]{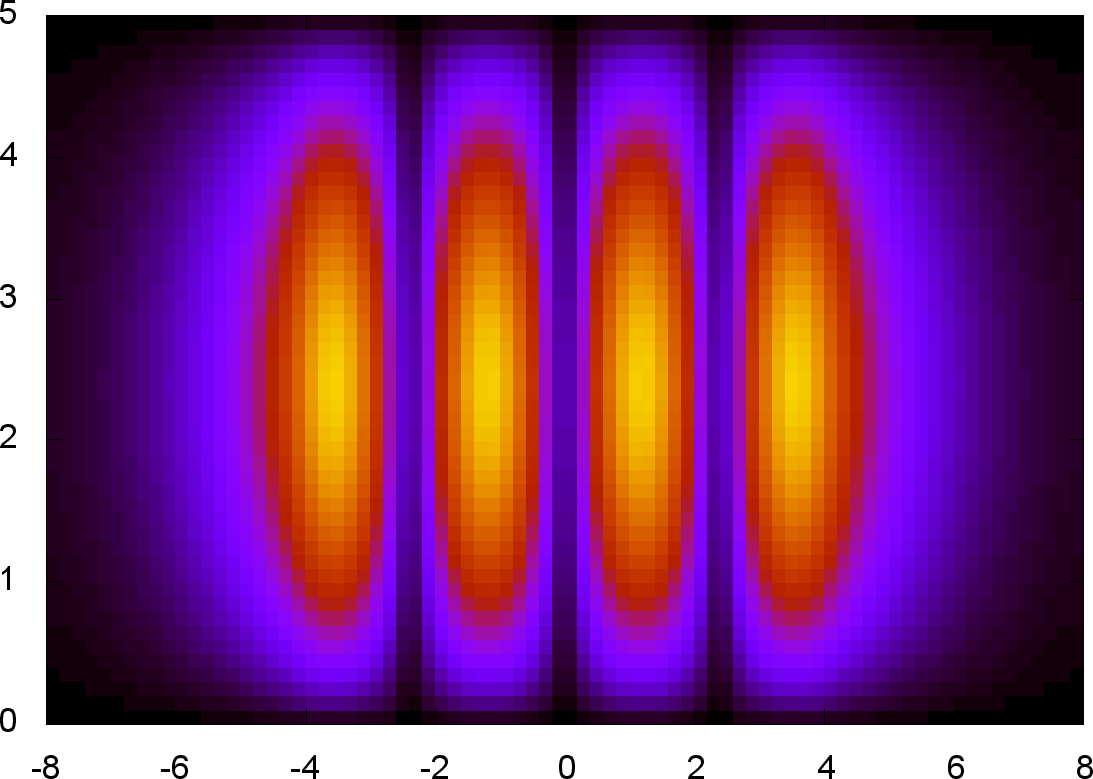} \\
\hline
 n=2 &\includegraphics[width=0.7in]{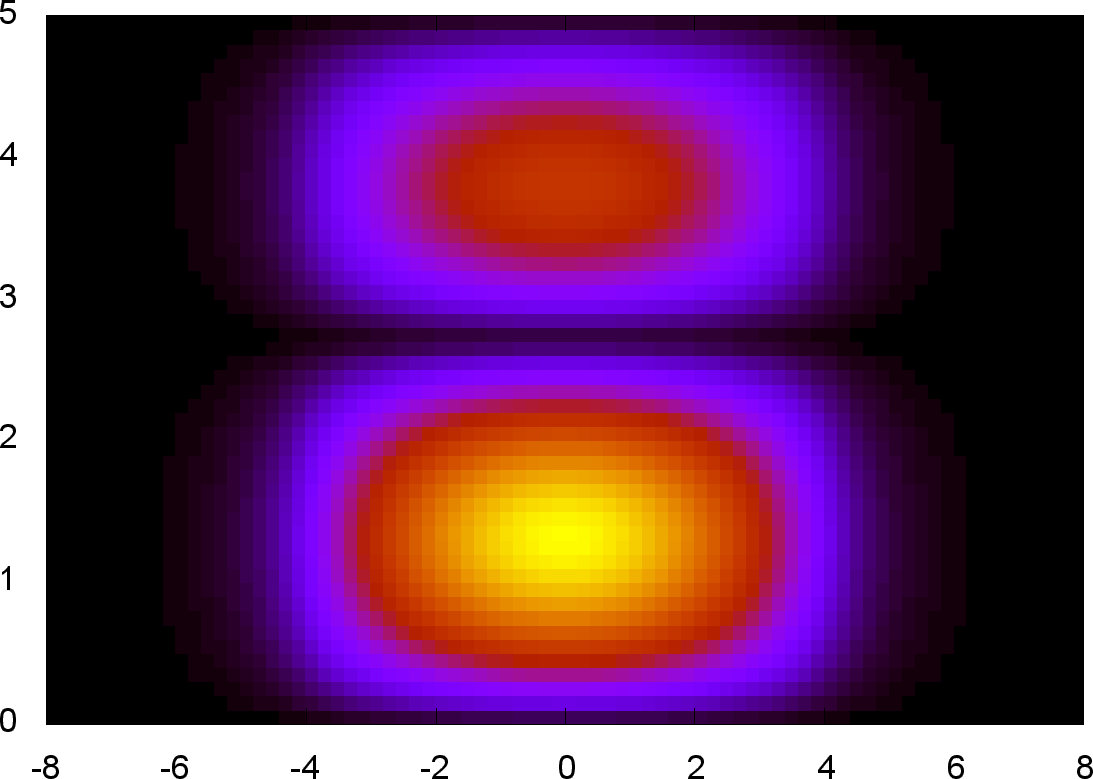}
     &\includegraphics[width=0.7in]{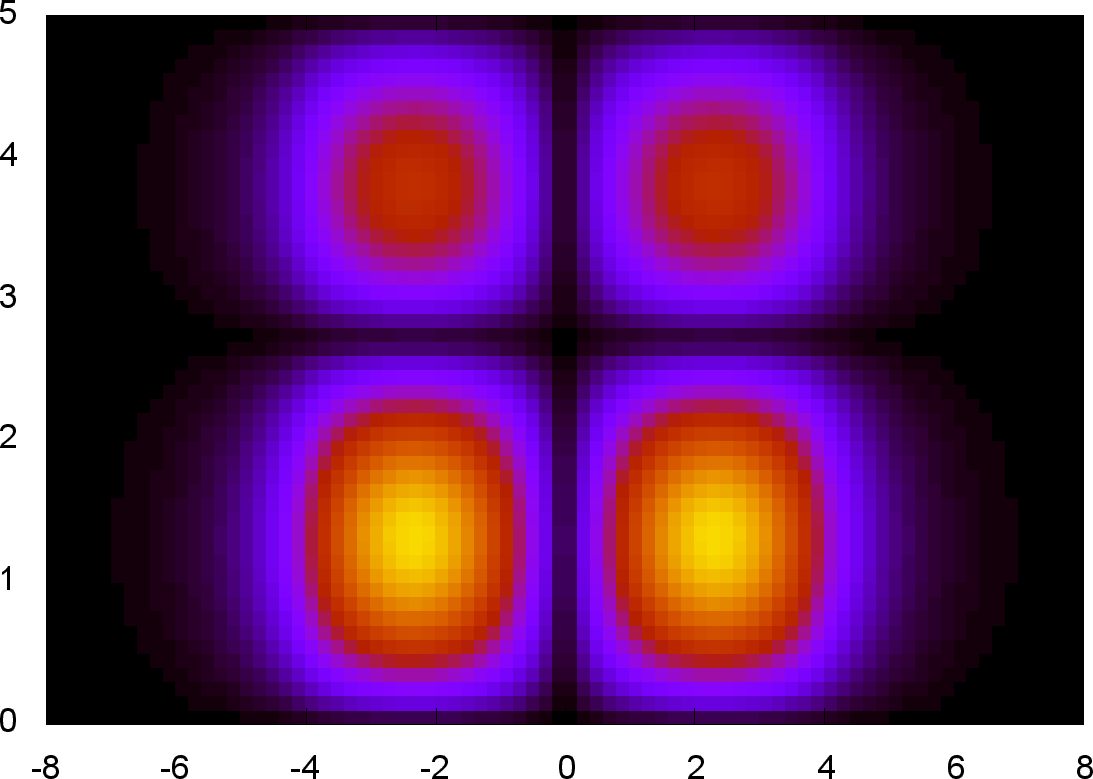}
     &\includegraphics[width=0.7in]{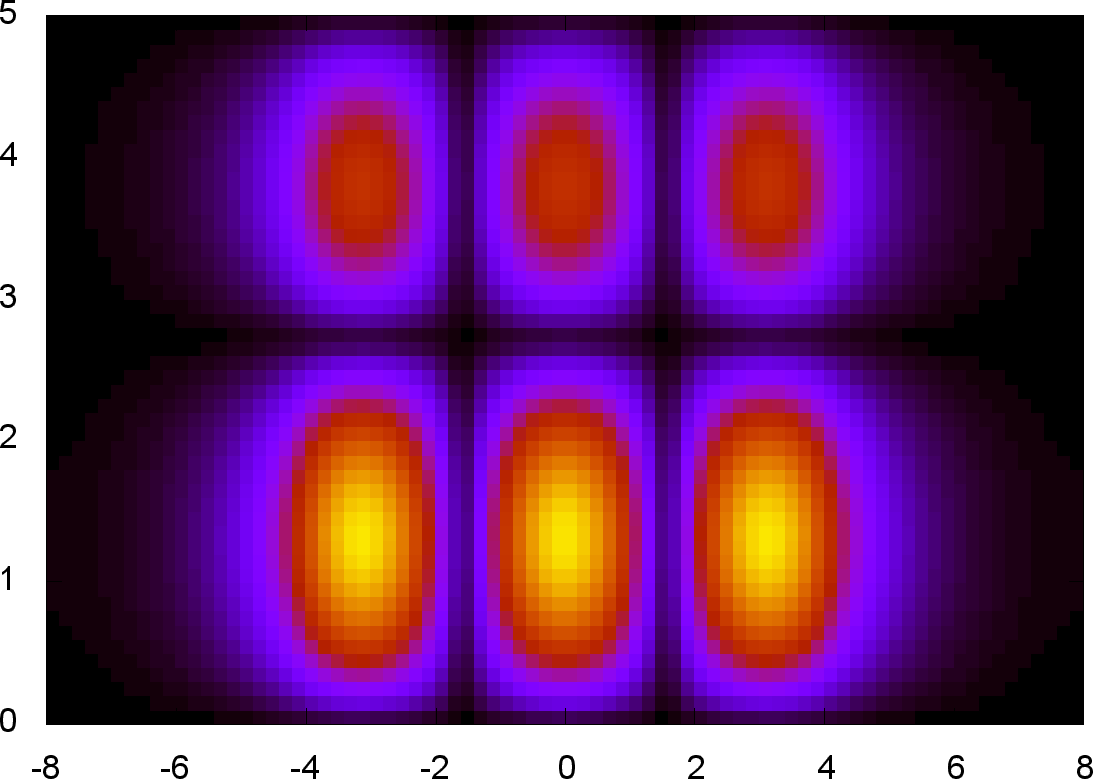}
     &\includegraphics[width=0.7in]{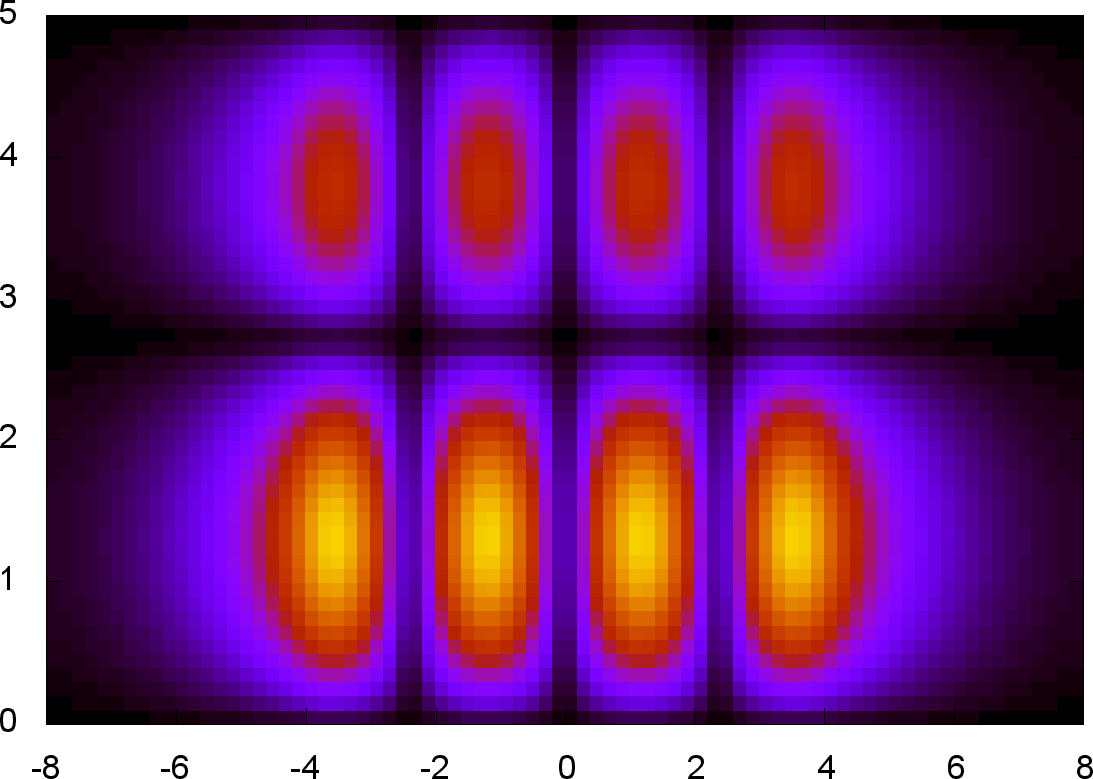} \\
\hline
 n=3 &\includegraphics[width=0.7in]{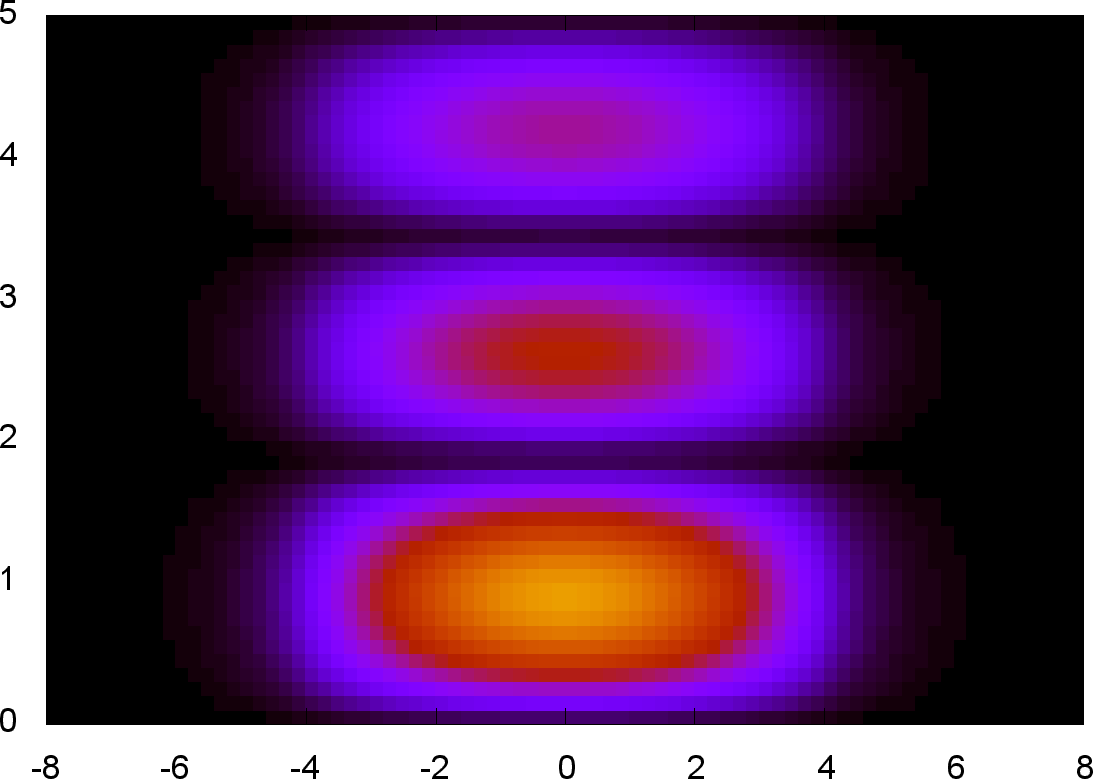}
     &\includegraphics[width=0.7in]{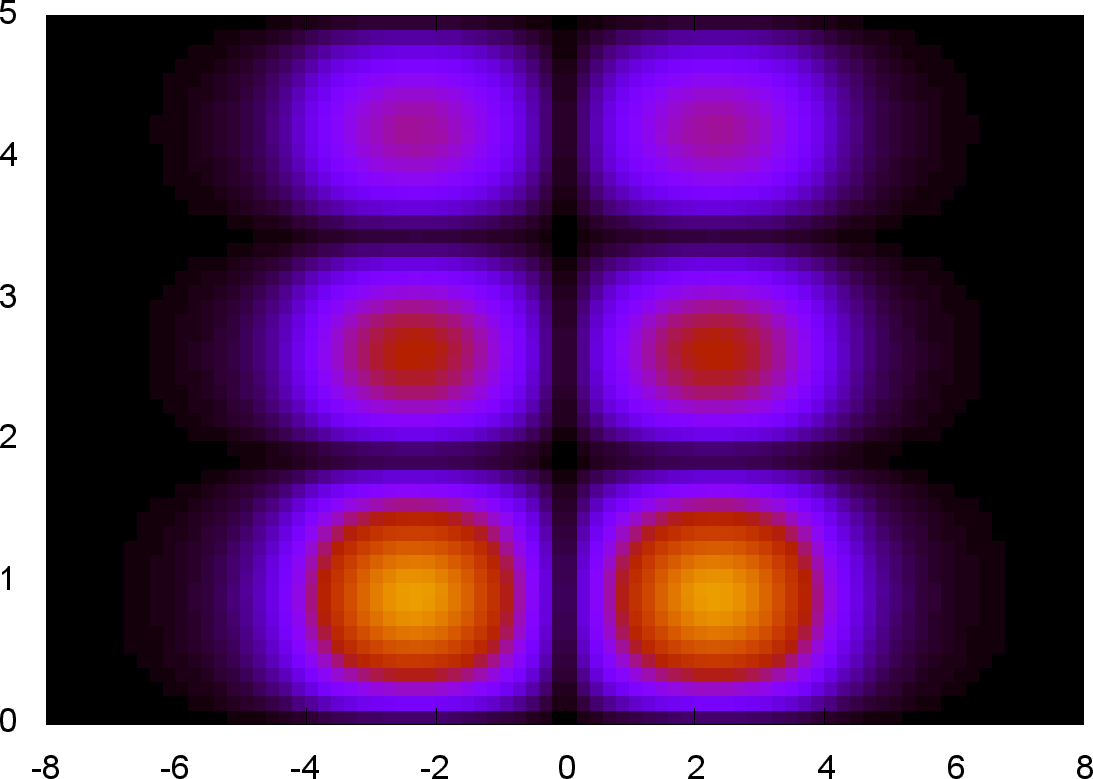}
     &
     & \\
\hline
\end{tabular}
\caption{(color online)
Localization probability density $|\psi^{(1)}_n(E_{res,i},r,z)|^2$
for an electron with $|m|=1$ 
incident from reservoir $s=1$ into channel $n$ and 
corresponding to the resonance $i$ between the barriers.
The axis of abscissae is $z \in [-8,8]$nm, and 
the axis of ordinates is $r \in [0,5]$nm for all plots.}
\label{NWTDB_psim_m1}
\end{table}

Similar behavior is observed for higher magnetic quantum numbers $m$.
In Fig. \ref{NWTDB_T1_m1} is represented
the total tunneling coefficient
and in Table \ref{NWTDB_psim_m1} the
localization probability
densities for the indexed peaks for the case $|m|=1$. 
The positions of the transmission peaks vary for different $m$-values
due to the dependence on $m$ of the
transversal energy channels $E^{(m)}_{\perp,n}$.
The scattering wave functions at resonances
for different $m$-values have different positions 
of nodes in the $r$-direction, and for any $m\not=0$
they are zero on the cylinder axis.

\section{Summary and discussion}

We have presented a general theory for computing the
scattering matrix and the scattering wave functions
for a general finite-range extended scattering potential
inside a cylindrical nanowire.
This formalism was applied to a variety of model systems,
like a quantum dot, a quantum ring and a double-barrier
heterostructure embedded  into the nanocylinder.
We have recovered the features for a nonseparable attractive
scattering potential in a multi-channel two-probe nanowire
taylored in the two-dimensional electron gas.
The difference to the Cartesian geometry is that every 
magnetic quantum number defines a two-dimensional scattering problem
with different structure of dips for the same scattering potential. 
How many of these problems have to be solved
depends on the further physical quantity calculation.
Furthermore, the cylindrical symmetry does not yield the same "selection rules"
for tunneling coefficient as the Cartesian symmetry,
so that dips could be observed in every subband. For stronger
attractive potential more than one dip can appear
due to the higher-order quasi-bound states of the next evanescent channel.
For quasi-bound states localized between barriers, it was possible
to compute the poles of the scattering matrix, which provide a quantitative
characterization of the resonances. 
Furthermore, the peaks of resonant tunneling can be indexed
by channel number and resonance index. 
Detailed maps of localization probability density sustain
the physical interpretation of the resonances (dips and peaks) 
found in the studied nanowire heterostructures.  

It will be the subject of next works to see, how the buildup of charge
around the scattering nonseparable attractive potential 
influences the overall electrical characteristics
of the nanowire-based devices.

\begin{acknowledgments}
It is a pleasure for us to acknowledge the fruitful discussions with 
Klaus G\"artner, Vidar Gudmundsson, Andrei Manolescu and Gheorghe Nenciu.
One of us (P.N.R.) also acknowledge partial support 
from German Research Foundation through SFB787 
and from the Romanian Ministry of Education and Research 
through the Program PNCDI2.
\end{acknowledgments}

\bibliography{nwt}

\begin{thebibliography}{36}
\expandafter\ifx\csname natexlab\endcsname\relax\def\natexlab#1{#1}\fi
\expandafter\ifx\csname bibnamefont\endcsname\relax
  \def\bibnamefont#1{#1}\fi
\expandafter\ifx\csname bibfnamefont\endcsname\relax
  \def\bibfnamefont#1{#1}\fi
\expandafter\ifx\csname citenamefont\endcsname\relax
  \def\citenamefont#1{#1}\fi
\expandafter\ifx\csname url\endcsname\relax
  \def\url#1{\texttt{#1}}\fi
\expandafter\ifx\csname urlprefix\endcsname\relax\def\urlprefix{URL }\fi
\providecommand{\bibinfo}[2]{#2}
\providecommand{\eprint}[2][]{\url{#2}}

\bibitem[{\citenamefont{Xiang et~al.}(2006)\citenamefont{Xiang, Lu, Hu, Hu,
  Yan, and Lieber}}]{Lieber_06}
\bibinfo{author}{\bibfnamefont{J.}~\bibnamefont{Xiang}},
  \bibinfo{author}{\bibfnamefont{W.}~\bibnamefont{Lu}},
  \bibinfo{author}{\bibfnamefont{Y.}~\bibnamefont{Hu}},
  \bibinfo{author}{\bibfnamefont{Y.}~\bibnamefont{Hu}},
  \bibinfo{author}{\bibfnamefont{H.}~\bibnamefont{Yan}}, \bibnamefont{and}
  \bibinfo{author}{\bibfnamefont{C.~M.} \bibnamefont{Lieber}},
  \bibinfo{journal}{Nature} \textbf{\bibinfo{volume}{441}},
  \bibinfo{pages}{489} (\bibinfo{year}{2006}).

\bibitem[{\citenamefont{Bryllert et~al.}(2006)\citenamefont{Bryllert,
  Wernersson, Loewgren, and Samuelson}}]{Samuelson_06}
\bibinfo{author}{\bibfnamefont{T.}~\bibnamefont{Bryllert}},
  \bibinfo{author}{\bibfnamefont{L.-E.} \bibnamefont{Wernersson}},
  \bibinfo{author}{\bibfnamefont{T.}~\bibnamefont{Loewgren}}, \bibnamefont{and}
  \bibinfo{author}{\bibfnamefont{L.}~\bibnamefont{Samuelson}},
  \bibinfo{journal}{Nanotechnology} \textbf{\bibinfo{volume}{17}},
  \bibinfo{pages}{S227} (\bibinfo{year}{2006}).

\bibitem[{\citenamefont{Yeo et~al.}(2006)}]{suk_iedm06}
\bibinfo{author}{\bibfnamefont{K.~H.} \bibnamefont{Yeo}} \bibnamefont{et~al.},
  \bibinfo{journal}{Tech. Dig. - Int. Electron Devices Meet.} p.
  \bibinfo{pages}{539} (\bibinfo{year}{2006}).

\bibitem[{\citenamefont{Cho et~al.}(2008)\citenamefont{Cho, Yeo, Yeoh, Suk, Li,
  Lee, Kim, Kim, Park, Hong et~al.}}]{cho_08}
\bibinfo{author}{\bibfnamefont{K.~H.} \bibnamefont{Cho}},
  \bibinfo{author}{\bibfnamefont{K.~H.} \bibnamefont{Yeo}},
  \bibinfo{author}{\bibfnamefont{Y.~Y.} \bibnamefont{Yeoh}},
  \bibinfo{author}{\bibfnamefont{S.~D.} \bibnamefont{Suk}},
  \bibinfo{author}{\bibfnamefont{M.}~\bibnamefont{Li}},
  \bibinfo{author}{\bibfnamefont{J.~M.} \bibnamefont{Lee}},
  \bibinfo{author}{\bibfnamefont{M.-S.} \bibnamefont{Kim}},
  \bibinfo{author}{\bibfnamefont{D.-W.} \bibnamefont{Kim}},
  \bibinfo{author}{\bibfnamefont{D.}~\bibnamefont{Park}},
  \bibinfo{author}{\bibfnamefont{B.~H.} \bibnamefont{Hong}},
  \bibnamefont{et~al.}, \bibinfo{journal}{Appl. Phys. Lett.}
  \textbf{\bibinfo{volume}{92}}, \bibinfo{eid}{052102}
  (pages~\bibinfo{numpages}{3}) (\bibinfo{year}{2008}).

\bibitem[{\citenamefont{M.T.Bjork et~al.}(2002)\citenamefont{M.T.Bjork,
  B.J.Ohlsson, Thelander, Persson, Deppert, Wallenberg, and
  Samuelson}}]{samuelson02}
\bibinfo{author}{\bibnamefont{M.T.Bjork}},
  \bibinfo{author}{\bibnamefont{B.J.Ohlsson}},
  \bibinfo{author}{\bibfnamefont{C.}~\bibnamefont{Thelander}},
  \bibinfo{author}{\bibfnamefont{A.}~\bibnamefont{Persson}},
  \bibinfo{author}{\bibfnamefont{K.}~\bibnamefont{Deppert}},
  \bibinfo{author}{\bibfnamefont{L.}~\bibnamefont{Wallenberg}},
  \bibnamefont{and}
  \bibinfo{author}{\bibfnamefont{L.}~\bibnamefont{Samuelson}},
  \bibinfo{journal}{Appl. Phys. Lett.} \textbf{\bibinfo{volume}{81}},
  \bibinfo{pages}{4458} (\bibinfo{year}{2002}).

\bibitem[{\citenamefont{Wensorra et~al.}(2005)\citenamefont{Wensorra,
  Indlekofer, Lepsa, Forster, and L{\"u}th}}]{wensorra}
\bibinfo{author}{\bibfnamefont{J.}~\bibnamefont{Wensorra}},
  \bibinfo{author}{\bibfnamefont{K.~M.} \bibnamefont{Indlekofer}},
  \bibinfo{author}{\bibfnamefont{M.~I.} \bibnamefont{Lepsa}},
  \bibinfo{author}{\bibfnamefont{A.}~\bibnamefont{Forster}}, \bibnamefont{and}
  \bibinfo{author}{\bibfnamefont{H.}~\bibnamefont{L{\"u}th}},
  \bibinfo{journal}{Nano Lett.} \textbf{\bibinfo{volume}{5}},
  \bibinfo{pages}{2470} (\bibinfo{year}{2005}).

\bibitem[{\citenamefont{Tian et~al.}(2007)\citenamefont{Tian, Zheng, Kempa,
  Fang, Yu, Yu, Huang, and Lieber}}]{Lieber_07}
\bibinfo{author}{\bibfnamefont{B.}~\bibnamefont{Tian}},
  \bibinfo{author}{\bibfnamefont{X.}~\bibnamefont{Zheng}},
  \bibinfo{author}{\bibfnamefont{T.~J.} \bibnamefont{Kempa}},
  \bibinfo{author}{\bibfnamefont{Y.}~\bibnamefont{Fang}},
  \bibinfo{author}{\bibfnamefont{N.}~\bibnamefont{Yu}},
  \bibinfo{author}{\bibfnamefont{G.}~\bibnamefont{Yu}},
  \bibinfo{author}{\bibfnamefont{J.}~\bibnamefont{Huang}}, \bibnamefont{and}
  \bibinfo{author}{\bibfnamefont{C.~M.} \bibnamefont{Lieber}},
  \bibinfo{journal}{Nature} \textbf{\bibinfo{volume}{449}},
  \bibinfo{pages}{885} (\bibinfo{year}{2007}).

\bibitem[{\citenamefont{Qian et~al.}(2008)\citenamefont{Qian, Li, Caronak,
  Park, Dong, Ding, Wang, and Lieber}}]{Lieber_08}
\bibinfo{author}{\bibfnamefont{F.}~\bibnamefont{Qian}},
  \bibinfo{author}{\bibfnamefont{Y.}~\bibnamefont{Li}},
  \bibinfo{author}{\bibfnamefont{S.~G.} \bibnamefont{Caronak}},
  \bibinfo{author}{\bibfnamefont{.~H.-G.} \bibnamefont{Park}},
  \bibinfo{author}{\bibfnamefont{Y.}~\bibnamefont{Dong}},
  \bibinfo{author}{\bibfnamefont{Y.}~\bibnamefont{Ding}},
  \bibinfo{author}{\bibfnamefont{.~Z.~L.} \bibnamefont{Wang}},
  \bibnamefont{and} \bibinfo{author}{\bibfnamefont{C.~M.}
  \bibnamefont{Lieber}}, \bibinfo{journal}{Nature Mater.}
  \textbf{\bibinfo{volume}{7}}, \bibinfo{pages}{701} (\bibinfo{year}{2008}).

\bibitem[{\citenamefont{Hu et~al.}(2007)\citenamefont{Hu, Churchill, Reilly,
  Xiang, Lieber, and Marcus}}]{Lieber_qubit}
\bibinfo{author}{\bibfnamefont{Y.}~\bibnamefont{Hu}},
  \bibinfo{author}{\bibfnamefont{H.~O.~H.} \bibnamefont{Churchill}},
  \bibinfo{author}{\bibfnamefont{D.~J.} \bibnamefont{Reilly}},
  \bibinfo{author}{\bibfnamefont{J.}~\bibnamefont{Xiang}},
  \bibinfo{author}{\bibfnamefont{C.~M.} \bibnamefont{Lieber}},
  \bibnamefont{and} \bibinfo{author}{\bibfnamefont{C.~M.}
  \bibnamefont{Marcus}}, \bibinfo{journal}{Nature Nanotechnol.}
  \textbf{\bibinfo{volume}{2}}, \bibinfo{pages}{622} (\bibinfo{year}{2007}).

\bibitem[{\citenamefont{Smr{\v{c}}ka}(1990)}]{smr}
\bibinfo{author}{\bibfnamefont{L.}~\bibnamefont{Smr{\v{c}}ka}},
  \bibinfo{journal}{Superlatt. and Microstruct.} \textbf{\bibinfo{volume}{8}},
  \bibinfo{pages}{221} (\bibinfo{year}{1990}).

\bibitem[{\citenamefont{Wulf et~al.}(1998)\citenamefont{Wulf, Ku\v{c}era,
  Racec, and Sigmund}}]{wulf98}
\bibinfo{author}{\bibfnamefont{U.}~\bibnamefont{Wulf}},
  \bibinfo{author}{\bibfnamefont{J.}~\bibnamefont{Ku\v{c}era}},
  \bibinfo{author}{\bibfnamefont{P.~N.} \bibnamefont{Racec}}, \bibnamefont{and}
  \bibinfo{author}{\bibfnamefont{E.}~\bibnamefont{Sigmund}},
  \bibinfo{journal}{Phys. Rev. B} \textbf{\bibinfo{volume}{58}},
  \bibinfo{pages}{16209} (\bibinfo{year}{1998}).

\bibitem[{\citenamefont{Onac et~al.}(2001)\citenamefont{Onac, Ku\v{c}era, and
  Wulf}}]{onac}
\bibinfo{author}{\bibfnamefont{E.}~\bibnamefont{Onac}},
  \bibinfo{author}{\bibfnamefont{J.}~\bibnamefont{Ku\v{c}era}},
  \bibnamefont{and} \bibinfo{author}{\bibfnamefont{U.}~\bibnamefont{Wulf}},
  \bibinfo{journal}{Phys. Rev. B} \textbf{\bibinfo{volume}{63}},
  \bibinfo{pages}{85319} (\bibinfo{year}{2001}).

\bibitem[{\citenamefont{Racec and Wulf}(2001)}]{roxana01}
\bibinfo{author}{\bibfnamefont{E.~R.} \bibnamefont{Racec}} \bibnamefont{and}
  \bibinfo{author}{\bibfnamefont{U.}~\bibnamefont{Wulf}},
  \bibinfo{journal}{Phys. Rev. B} \textbf{\bibinfo{volume}{64}},
  \bibinfo{pages}{115318} (\bibinfo{year}{2001}).

\bibitem[{\citenamefont{Racec et~al.}(2002)\citenamefont{Racec, Racec, and
  Wulf}}]{racec02}
\bibinfo{author}{\bibfnamefont{P.~N.} \bibnamefont{Racec}},
  \bibinfo{author}{\bibfnamefont{E.~R.} \bibnamefont{Racec}}, \bibnamefont{and}
  \bibinfo{author}{\bibfnamefont{U.}~\bibnamefont{Wulf}},
  \bibinfo{journal}{Phys. Rev. B} \textbf{\bibinfo{volume}{65}},
  \bibinfo{pages}{193314} (\bibinfo{year}{2002}).

\bibitem[{\citenamefont{Nemnes et~al.}(2004)\citenamefont{Nemnes, Wulf, and
  Racec}}]{nemnes04}
\bibinfo{author}{\bibfnamefont{G.~A.} \bibnamefont{Nemnes}},
  \bibinfo{author}{\bibfnamefont{U.}~\bibnamefont{Wulf}}, \bibnamefont{and}
  \bibinfo{author}{\bibfnamefont{P.~N.} \bibnamefont{Racec}},
  \bibinfo{journal}{J. Appl. Phys.} \textbf{\bibinfo{volume}{96}},
  \bibinfo{pages}{596} (\bibinfo{year}{2004}).

\bibitem[{\citenamefont{Nemnes et~al.}(2005)\citenamefont{Nemnes, Wulf, and
  Racec}}]{nemnes05}
\bibinfo{author}{\bibfnamefont{G.~A.} \bibnamefont{Nemnes}},
  \bibinfo{author}{\bibfnamefont{U.}~\bibnamefont{Wulf}}, \bibnamefont{and}
  \bibinfo{author}{\bibfnamefont{P.~N.} \bibnamefont{Racec}},
  \bibinfo{journal}{J. Appl. Phys.} \textbf{\bibinfo{volume}{98}},
  \bibinfo{pages}{084308} (\bibinfo{year}{2005}).

\bibitem[{\citenamefont{Wulf et~al.}(2007)\citenamefont{Wulf, Racec, and
  Racec}}]{07impe}
\bibinfo{author}{\bibfnamefont{U.}~\bibnamefont{Wulf}},
  \bibinfo{author}{\bibfnamefont{P.~N.} \bibnamefont{Racec}}, \bibnamefont{and}
  \bibinfo{author}{\bibfnamefont{E.~R.} \bibnamefont{Racec}},
  \bibinfo{journal}{Phys. Rev. B} \textbf{\bibinfo{volume}{75}},
  \bibinfo{pages}{075320} (\bibinfo{year}{2007}).

\bibitem[{\citenamefont{Behrndt et~al.}(2008)\citenamefont{Behrndt, Neidhardt,
  Racec, Racec, and Wulf}}]{hagen08}
\bibinfo{author}{\bibfnamefont{J.}~\bibnamefont{Behrndt}},
  \bibinfo{author}{\bibfnamefont{H.}~\bibnamefont{Neidhardt}},
  \bibinfo{author}{\bibfnamefont{E.~R.} \bibnamefont{Racec}},
  \bibinfo{author}{\bibfnamefont{P.~N.} \bibnamefont{Racec}}, \bibnamefont{and}
  \bibinfo{author}{\bibfnamefont{U.}~\bibnamefont{Wulf}}, \bibinfo{journal}{J.
  Differ. Equ.} \textbf{\bibinfo{volume}{244}}, \bibinfo{pages}{2545}
  (\bibinfo{year}{2008}).

\bibitem[{\citenamefont{Bagwell}(1990)}]{bagwell90}
\bibinfo{author}{\bibfnamefont{P.~F.} \bibnamefont{Bagwell}},
  \bibinfo{journal}{Phys. Rev. B} \textbf{\bibinfo{volume}{41}},
  \bibinfo{pages}{10354} (\bibinfo{year}{1990}).

\bibitem[{\citenamefont{Exner et~al.}(1996)\citenamefont{Exner, Gawlista, Seba,
  and Tater}}]{Exner96}
\bibinfo{author}{\bibfnamefont{P.}~\bibnamefont{Exner}},
  \bibinfo{author}{\bibfnamefont{R.}~\bibnamefont{Gawlista}},
  \bibinfo{author}{\bibfnamefont{P.}~\bibnamefont{Seba}}, \bibnamefont{and}
  \bibinfo{author}{\bibfnamefont{M.}~\bibnamefont{Tater}},
  \bibinfo{journal}{Ann. Phys.} \textbf{\bibinfo{volume}{252}},
  \bibinfo{pages}{133 } (\bibinfo{year}{1996}), ISSN \bibinfo{issn}{0003-4916}.

\bibitem[{\citenamefont{Gurvitz and Levinson}(1993)}]{levinson93}
\bibinfo{author}{\bibfnamefont{S.~A.} \bibnamefont{Gurvitz}} \bibnamefont{and}
  \bibinfo{author}{\bibfnamefont{Y.~B.} \bibnamefont{Levinson}},
  \bibinfo{journal}{Phys. Rev. B} \textbf{\bibinfo{volume}{47}},
  \bibinfo{pages}{10578} (\bibinfo{year}{1993}).

\bibitem[{\citenamefont{N\"ockel and Stone}(1994)}]{noeckel94}
\bibinfo{author}{\bibfnamefont{J.~U.} \bibnamefont{N\"ockel}} \bibnamefont{and}
  \bibinfo{author}{\bibfnamefont{A.~D.} \bibnamefont{Stone}},
  \bibinfo{journal}{Phys. Rev. B} \textbf{\bibinfo{volume}{50}},
  \bibinfo{pages}{17415} (\bibinfo{year}{1994}).

\bibitem[{\citenamefont{Bardarson et~al.}(2004)\citenamefont{Bardarson,
  Magnusdottir, Gudmundsdottir, Tang, Manolescu, and
  Gudmundsson}}]{gudmundsson04}
\bibinfo{author}{\bibfnamefont{J.~H.} \bibnamefont{Bardarson}},
  \bibinfo{author}{\bibfnamefont{I.}~\bibnamefont{Magnusdottir}},
  \bibinfo{author}{\bibfnamefont{G.}~\bibnamefont{Gudmundsdottir}},
  \bibinfo{author}{\bibfnamefont{C.-S.} \bibnamefont{Tang}},
  \bibinfo{author}{\bibfnamefont{A.}~\bibnamefont{Manolescu}},
  \bibnamefont{and}
  \bibinfo{author}{\bibfnamefont{V.}~\bibnamefont{Gudmundsson}},
  \bibinfo{journal}{Phys. Rev. B} \textbf{\bibinfo{volume}{70}},
  \bibinfo{pages}{245308} (\bibinfo{year}{2004}).

\bibitem[{\citenamefont{Gudmundsson et~al.}(2005)\citenamefont{Gudmundsson,
  Lin, Tang, Moldoveanu, Bardarson, and Manolescu}}]{gudmundsson05}
\bibinfo{author}{\bibfnamefont{V.}~\bibnamefont{Gudmundsson}},
  \bibinfo{author}{\bibfnamefont{Y.-Y.} \bibnamefont{Lin}},
  \bibinfo{author}{\bibfnamefont{C.-S.} \bibnamefont{Tang}},
  \bibinfo{author}{\bibfnamefont{V.}~\bibnamefont{Moldoveanu}},
  \bibinfo{author}{\bibfnamefont{J.~H.} \bibnamefont{Bardarson}},
  \bibnamefont{and}
  \bibinfo{author}{\bibfnamefont{A.}~\bibnamefont{Manolescu}},
  \bibinfo{journal}{Phys. Rev. B} \textbf{\bibinfo{volume}{71}},
  \bibinfo{eid}{235302} (pages~\bibinfo{numpages}{12}) (\bibinfo{year}{2005}).

\bibitem[{\citenamefont{Racec et~al.}()\citenamefont{Racec, Racec, and
  Wulf}}]{roxana08}
\bibinfo{author}{\bibfnamefont{E.~R.} \bibnamefont{Racec}},
  \bibinfo{author}{\bibfnamefont{P.~N.} \bibnamefont{Racec}}, \bibnamefont{and}
  \bibinfo{author}{\bibfnamefont{U.}~\bibnamefont{Wulf}},
  \bibinfo{note}{unpublished}.

\bibitem[{\citenamefont{Schanz and Smilansky}(1995)}]{schanz95}
\bibinfo{author}{\bibfnamefont{H.}~\bibnamefont{Schanz}} \bibnamefont{and}
  \bibinfo{author}{\bibfnamefont{U.}~\bibnamefont{Smilansky}},
  \bibinfo{journal}{Chaos Solitons \& Fractals} \textbf{\bibinfo{volume}{5}},
  \bibinfo{pages}{1289} (\bibinfo{year}{1995}).

\bibitem[{\citenamefont{Luisier et~al.}(2006)\citenamefont{Luisier, Schenk, and
  Fichtner}}]{luisier_06}
\bibinfo{author}{\bibfnamefont{M.}~\bibnamefont{Luisier}},
  \bibinfo{author}{\bibfnamefont{A.}~\bibnamefont{Schenk}}, \bibnamefont{and}
  \bibinfo{author}{\bibfnamefont{W.}~\bibnamefont{Fichtner}},
  \bibinfo{journal}{J. Appl. Phys.} \textbf{\bibinfo{volume}{100}},
  \bibinfo{pages}{043713} (\bibinfo{year}{2006}).

\bibitem[{\citenamefont{Wigner and Eisenbud}(1947)}]{wigeis}
\bibinfo{author}{\bibfnamefont{E.~P.} \bibnamefont{Wigner}} \bibnamefont{and}
  \bibinfo{author}{\bibfnamefont{L.}~\bibnamefont{Eisenbud}},
  \bibinfo{journal}{Phys. Rev.} \textbf{\bibinfo{volume}{72}},
  \bibinfo{pages}{29} (\bibinfo{year}{1947}).

\bibitem[{\citenamefont{Lane and Thomas}(1958)}]{lane}
\bibinfo{author}{\bibfnamefont{A.~M.} \bibnamefont{Lane}} \bibnamefont{and}
  \bibinfo{author}{\bibfnamefont{R.~G.} \bibnamefont{Thomas}},
  \bibinfo{journal}{Rev. Mod. Phys.} \textbf{\bibinfo{volume}{30}},
  \bibinfo{pages}{257} (\bibinfo{year}{1958}).

\bibitem[{\citenamefont{B{\"u}ttiker et~al.}(1985)\citenamefont{B{\"u}ttiker,
  Imry, Landauer, and Pinhas}}]{buett85}
\bibinfo{author}{\bibfnamefont{M.}~\bibnamefont{B{\"u}ttiker}},
  \bibinfo{author}{\bibfnamefont{Y.}~\bibnamefont{Imry}},
  \bibinfo{author}{\bibfnamefont{R.}~\bibnamefont{Landauer}}, \bibnamefont{and}
  \bibinfo{author}{\bibfnamefont{S.}~\bibnamefont{Pinhas}},
  \bibinfo{journal}{Phys. Rev. B} \textbf{\bibinfo{volume}{31}},
  \bibinfo{pages}{6207} (\bibinfo{year}{1985}).

\bibitem[{\citenamefont{Gudiksen et~al.}(2002)\citenamefont{Gudiksen, Lauhon,
  Wang, Smith, and Lieber}}]{lieber02}
\bibinfo{author}{\bibfnamefont{M.~S.} \bibnamefont{Gudiksen}},
  \bibinfo{author}{\bibfnamefont{L.~J.} \bibnamefont{Lauhon}},
  \bibinfo{author}{\bibfnamefont{J.}~\bibnamefont{Wang}},
  \bibinfo{author}{\bibfnamefont{D.~C.} \bibnamefont{Smith}}, \bibnamefont{and}
  \bibinfo{author}{\bibfnamefont{C.~M.} \bibnamefont{Lieber}},
  \bibinfo{journal}{Nature} \textbf{\bibinfo{volume}{415}},
  \bibinfo{pages}{617} (\bibinfo{year}{2002}).

\bibitem[{\citenamefont{Shen et~al.}(2008)\citenamefont{Shen, Chen, Bando, and
  Golberg}}]{shen08}
\bibinfo{author}{\bibfnamefont{G.}~\bibnamefont{Shen}},
  \bibinfo{author}{\bibfnamefont{D.}~\bibnamefont{Chen}},
  \bibinfo{author}{\bibfnamefont{Y.}~\bibnamefont{Bando}}, \bibnamefont{and}
  \bibinfo{author}{\bibfnamefont{D.}~\bibnamefont{Golberg}},
  \bibinfo{journal}{J. Mater. Sci. Technol.} \textbf{\bibinfo{volume}{24}},
  \bibinfo{pages}{541} (\bibinfo{year}{2008}).

\bibitem[{\citenamefont{van Wees et~al.}(1988)\citenamefont{van Wees, van
  Houten, Beenakker, Williamson, Kouwenhoven, van~der Marel, and
  Foxon}}]{vanwees88}
\bibinfo{author}{\bibfnamefont{B.~J.} \bibnamefont{van Wees}},
  \bibinfo{author}{\bibfnamefont{H.}~\bibnamefont{van Houten}},
  \bibinfo{author}{\bibfnamefont{C.~W.~J.} \bibnamefont{Beenakker}},
  \bibinfo{author}{\bibfnamefont{J.~G.} \bibnamefont{Williamson}},
  \bibinfo{author}{\bibfnamefont{L.~P.} \bibnamefont{Kouwenhoven}},
  \bibinfo{author}{\bibfnamefont{D.}~\bibnamefont{van~der Marel}},
  \bibnamefont{and} \bibinfo{author}{\bibfnamefont{C.~T.} \bibnamefont{Foxon}},
  \bibinfo{journal}{Phys. Rev. Lett.} \textbf{\bibinfo{volume}{60}},
  \bibinfo{pages}{848} (\bibinfo{year}{1988}).

\bibitem[{\citenamefont{Simon}(1976)}]{simon76}
\bibinfo{author}{\bibfnamefont{B.}~\bibnamefont{Simon}}, \bibinfo{journal}{Ann.
  Physics} \textbf{\bibinfo{volume}{97}}, \bibinfo{pages}{279}
  (\bibinfo{year}{1976}).

\bibitem[{\citenamefont{Klaus}(1977)}]{klaus77}
\bibinfo{author}{\bibfnamefont{M.}~\bibnamefont{Klaus}}, \bibinfo{journal}{Ann.
  Physics} \textbf{\bibinfo{volume}{108}}, \bibinfo{pages}{288}
  (\bibinfo{year}{1977}).

\bibitem[{\citenamefont{Kouwenhoven et~al.}(2001)\citenamefont{Kouwenhoven,
  Austing, and Tarucha}}]{kouwenhoven01}
\bibinfo{author}{\bibfnamefont{L.~P.} \bibnamefont{Kouwenhoven}},
  \bibinfo{author}{\bibfnamefont{D.~G.} \bibnamefont{Austing}},
  \bibnamefont{and} \bibinfo{author}{\bibfnamefont{S.}~\bibnamefont{Tarucha}},
  \bibinfo{journal}{Rep. Prog. Phys.} \textbf{\bibinfo{volume}{64}},
  \bibinfo{pages}{701} (\bibinfo{year}{2001}).

\end{thebibliography}

\printfigures

\printtables

\end{document}